\documentclass[seceq,paper,twocolumn,dvipdfmx]{jpsj3}
\usepackage{bm,amsmath,amssymb,graphicx,url,float,cases,fancybox}
\usepackage[dvipdfmx]{xcolor}

\definecolor{orange}{rgb}{1,0.5,0}
\definecolor{DG}{rgb}{0,0.65,0.2}

\renewcommand{\H}{{\mathcal H}}

\newcommand{\bS}{{\bm{S}}}
\newcommand{\bW}{{\bm{W}}}

\newcommand{\br}{{\bm{r}}}

\newcommand{\Sr}{{Sr$_2$CoSi$_2$O$_7$}}
\newcommand{\Ba}{{Ba$_2$CoGe$_2$O$_7$}}
\newcommand{\Tl}{{TlCuCl$_3$}}

\title{
Symmetry Analysis of Spin-Dependent Electric Dipole \\
and Its Application to Magnetoelectric Effects
}

\author{Masashige Matsumoto$^1$\thanks{E-mail address: matsumoto.masashige@shizuoka.ac.jp}, Kosuke Chimata$^1$, and Mikito Koga$^2$}

\inst{
$^1$Department of Physics, Faculty of Science, Shizuoka University, Shizuoka 422-8529, Japan \\
$^2$Department of Physics, Faculty of Education, Shizuoka University, Shizuoka 422-8529, Japan
}

\recdate{September 7, 2016}

\abst{
Spin-dependent electric dipole operators are investigated group-theoretically
for the emergence of an electric dipole induced by a single spin or by two spins,
where the spin dependences are completely classified up to the quadratic order.
For a single spin, a product of spin operators behaves as an even-parity electric quadrupole operator,
which differs from an odd-parity electric dipole.
The lack of the inversion symmetry allows the even- and odd-parity mixing,
which leads to the electric dipole described by the electric quadruple operators.
Point-group tables are given for classification of the possible spin-dependent electric dipoles
and for the qualitative analysis of multiferroic properties,
such as an emergent electric dipole moment coexisting with a magnetic moment,
electromagnon excitation, and directional dichroism.
The results can be applied to a magnetic ion in crystals
or embedded in molecules at a site without the inversion symmetry.
In the presence of an inversion symmetry, the electric dipole does not appear for a single spin.
This is not the case for the electric dipole induced by two spins with antisymmetric spin dependence,
which is known as vector spin chirality,
in the presence of the inversion center between the two spins.
In the absence of the inversion center, symmetric spin-dependent electric dipoles are also relevant.
The detailed analysis of various symmetries of two-spin states
is applied to spin dimer systems and the related multiferroic properties.
}

\begin{document}

\maketitle

\section{Introduction}

For conventional magnets, the magnetic dipole moment is usually controlled by the magnetic field.
On the other hand, there are unconventional magnets that reveal an electric dipole moment in the magnetically ordered phase.
\cite{Smolenskii-1982}
This is known as multiferroics and has attracted much attention in condensed matter physics,
especially regarding the cross-correlation between the electric and magnetic fields,
such as the control of an electric dipole moment by a magnetic field.
\cite{Tokura-2014}

In the multiferroic materials, the important point is that the electric dipole is related to spin operators.
In an ideal system, the spin space is independent of the real space and spins do not couple to the electric field.
There are two main microscopic origins connecting the two spaces.
One is known as an exchange striction effect with modulation in both the lattice and exchange interactions.
\cite{Picozzi-2007,Choi-2008}
The other is the spin-orbit interaction that transfers anisotropy in the real space into the spin space.
In both cases, the electric dipole is described by tensorial forms of the spin operators.
This spin-dependent electric dipole leads to various interesting multiferroic properties.

The spin-dependent electric dipole has been studied for a long time.
\cite{Tanabe-1965,Moriya-1966,Moriya-1968}
After the discovery of the giant magnetoelectric effect in TbMnO$_3$ and related compounds,
\cite{Kimura-2003,Kenzelmann-2005}
its mechanism within the spin-orbit interaction was investigated, focusing on the noncollinear (helical) ordered phase.
\cite{Katsura-2005,Mostovoy-2006,Sergienko-2006,Jia-2006,Jia-2007}
Katsura et al. found that the main source of the electric dipole is written in the following form:
\cite{Katsura-2005}
\begin{align}
\bm{p}_{ij}\propto \bm{e}_{ij}\times \bS_i \times \bS_j.
\label{eqn:P-Katsura}
\end{align}
Here, $\bS_i$ and $\bS_j$ represent the spin operators at the neighboring sites.
$\bm{e}_{ij}$ denotes the unit vector connecting the two spin sites.
$\bm{p}_{ij}$ is the electric dipole induced by the spin pair.
The outer spin product, $\bS_i \times \bS_j$, is termed the vector spin chirality.
This electric dipole is interpreted by spin current or inverse Dzyaloshinskii--Moriya mechanisms.
\cite{Katsura-2005,Sergienko-2006}
Since the helical magnetic structure breaks the inversion symmetry and it gives rise to a uniform vector spin chirality,
the ferroelectric properties in TbMnO$_3$ were successfully explained by Eq. (\ref{eqn:P-Katsura}).

Another intriguing multiferroic issue is the electric dipole moment observed in \Ba~and related compounds.
In the case of \Ba, the electric dipole is induced by a single $S=3/2$ spin of a Co$^{2+}$ ion surrounded by O$^{2-}$ ligand ions.
On the basis of the metal-ligand hybridization model with the spin-orbit interaction,
Arima found that the electric dipole is written in the following form:
\cite{Arima-2007}
\begin{align}
\bm{p}\propto \sum_i ( \bS \cdot \br_i )^2 \br_i.
\label{eqn:P-Arima}
\end{align}
Here, $\br_i$ denotes the position of the $i$th ligand ion relative to the Co$^{2+}$ site.
This electric dipole successfully explains the multiferroic properties of \Ba,
such as its magnetic-field-controlled ferroelectric polarization,
\cite{Murakawa-2010,Murakawa-2012,Akaki-2012}
electromagnon excitation,
\cite{Kezsmarki-2011,Miyahara-2011}
and directional dichroism.
\cite{Kezsmarki-2011,Miyahara-2011,Kezsmarki-2014}

As expressed by Eqs. (\ref{eqn:P-Katsura}) and (\ref{eqn:P-Arima}),
the electric dipole is described by the product of spin operators.
The electric dipole can be induced in both single-spin and two-spin systems.
For the electric dipole induced by a single spin,
the product of the spin operators can be interpreted as an electric quadrupole operator.
In the absence of the inversion symmetry at the spin site,
the even and odd parities are mixed up and the electric dipole operator
can be described by the electric quadrupole operator.
Romh\'{a}nyi and coworkers pointed out that in \Ba~the quadrupole operator of the $S^x S^y + S^y S^x$ type is
relevant to the ferroelectric moment parallel to the $z$-axis on the basis of a group theoretical discussion.
\cite{Romhanyi-2011,Penc-2012,Romhanyi-2012,Soda-2014}
As studied by Miyahara and Furukawa,
the same result can be obtained by Eq. (\ref{eqn:P-Arima}) from the metal-ligand hybridization model.
\cite{Miyahara-2011}
Group theoretical analysis is a powerful tool for the investigation of multiferroic materials.
For instance, the point-group symmetry of atomic positions surrounding a local spin
determines the possible spin dependence in the electric polarization without going into the microscopic origin.

So far, no systematic analysis on the electric polarization has yet been reported.
The purpose of this paper is to investigate the spin-dependent electric dipole operator from the viewpoint of symmetry
and to demonstrate what types of spin dependence are allowed or forbidden under various point-group symmetries.
We consider the product of spin operators up to the quadratic order as a minimum treatment
and the 32 point groups that are compatible with the space group.
Since the electric dipole induced by a single spin appears only in the absence of the inversion symmetry,
we focus on the 20 point groups having no inversion symmetry.

The electric dipole operator induced by two spins also plays an important role
even in the presence of the inversion center between the two spins,
where the spin dependence must be antisymmetric with respect to the inversion transformation.
Since the two spins are spatially separated, the symmetry around the two spins is not expressed by the point group.
Kaplan and Mahanti studied the antisymmetric spin-dependent electric dipole under various symmetries.
\cite{Kaplan-2011}
In this paper, we extend their work to the symmetric spin dependence
that is also allowed in the absence of the inversion center.

Our analysis can be used to study multiferroic properties of various types of magnetic materials.
As typical examples, we focus on $d$-electron systems with $S=1$, $S=3/2$, $S=2$, and $S=5/2$ spins in tetragonal symmetries
and discuss the emergent electric dipole moment, electromagnon excitation, and directional dichroism.
Since the result can also be applied to $f$-electron systems, we discuss the $J=5/2$ and $J=4$ cases in cubic point groups.
We classify the magnetic and electric dipole operators in the irreducible representation.
This helps us understand the selection rules of light absorption and the possibility of directional dichroism.
In the two-spin case, we analyze the emergent electric dipole moment in spin dimer systems
and discuss the expected multiferroic properties.

This paper is organized as follows.
In Sect. 2, we give a general formulation for the spin-dependent electric dipole operator induced by a single spin
and present the spin dependence for various point groups.
In Sect. 3, we study the symmetric spin-dependent electric dipole operators induced by two spins.
Sections 4 and 5 discuss possible applications of the electric dipole induced by a single spin and two spins, respectively.
The last section gives a summary and discussion of the results.
In Appendix \ref{appendix:polarization}, we present coefficient tensors of the spin-dependent electric dipoles
for various basal symmetry transformations to obtain the results in Sect. 2.
In Appendix \ref{appendix:parity-mix}, we present microscopic models for the spin-dependent electric dipoles
induced by parity mixing in the absence of the inversion symmetry.

\section{Electric Dipole by Single Spin}

We first study an electric dipole emerging at a single spin site in crystals or molecules,
where the spin-orbit interaction is required for the microscopic origin.
The electric dipole, namely, the polarized charge distribution, is related to the orbital of the electron
and it can be spin-dependent through the spin-orbit interaction.

We assume that the spin is located in an environment characterized by several symmetries, such as point-group representations.
Considering the symmetry properties of the electric dipole,
we can determine the spin dependences that are allowed under the symmetries
even though we do not discuss the microscopic origin within the spin-orbit interaction.

\subsection{General formulation}

We discuss the following spin-dependent electric dipole operator:
\begin{align}
p_{\rm S}^\alpha = K^\alpha_{\beta\gamma} S^\beta S^\gamma.
\label{eqn:P}
\end{align}
Here, $p_{\rm S}^\alpha$ and $S^\alpha$ ($\alpha=x,y,z$) are the $\alpha$ component
of the electric dipole and spin operators, respectively.
Since the spin operator has an even parity with respect to the spatial inversion transformation,
$S^\beta S^\gamma$ is symmetric for the inversion.
Here, the spin product is classified as an electric quadrupole
and Eq. (\ref{eqn:P}) means that electric dipoles can be induced by electric quadrupoles.
The subscript of $p_{\rm S}^\alpha$ represents the symmetric spin-dependent electric dipole operator,
while the asymmetric type, $p_{\rm A}^\alpha$, will be considered in Sect. 3 for two spins.
$K^\alpha_{\beta\gamma}$ is a coefficient tensor.
In Eq. (\ref{eqn:P}), summations of $\beta$ and $\gamma$ are implicitly taken over $x$, $y$, and $z$.
The electric dipole is invariant under the time-reversal transformation.
Since the product of the spin operators is invariant under this transformation,
the coefficient tensor $K^\alpha_{\beta\gamma}$ must be a real number.
In addition, the dipole operator must be Hermitian: $K^\alpha_{\beta\gamma}=K^\alpha_{\gamma\beta}$.
Thus, $K^\alpha_{\beta\gamma}$ is represented by a real symmetric tensor.

We assume that the spin is located in an environment represented by a point group.
In the presence of the spatial inversion symmetry,
the irreducible representations are classified by even and odd parities.
The electric dipole is for an odd parity.
In this case, the symmetric spin-dependent electric dipole vanishes.
In contrast, the dipole can appear in the absence of the inversion symmetry.
This is owing to the fact that the irreducible representations of the spatial inversion are meaningless,
namely, both even and odd parities are mixed.
The quadrupoles $(S^\beta S^\gamma$) in Eq. (\ref{eqn:P}) have an even parity; however,
an odd-parity component can be induced by the quadrupoles
when the inversion symmetry is broken in the environment (see Sect. \ref{sec:p-S}).
This is the reason why the symmetric spin-dependent component $p_{\rm S}^\alpha$ is regarded as an electric dipole.
More detailed investigation is required for the point groups without the inversion symmetry.

The electric dipole couples to the electric field, which is represented by the following Hamiltonian:
\begin{align}
\H = - \bm{p}_{\rm S} \cdot \bm{E}
= - K^\alpha_{\beta\gamma} E^\alpha S^\beta S^\gamma.
\label{eqn:H-p}
\end{align}
Here, $\bm{E}$ represents the electric field.
Let us discuss the symmetry properties of the coefficient tensor under symmetry transformations of a point group.
There are various symmetry operations in a point group.
Since we concentrate on point groups lacking the inversion symmetry,
the point-group symmetry operations consist of rotation and mirror operations.
The electric dipole $p_{\rm S}^\alpha$ and electric field $E^\alpha$ are polar vectors,
while the spin operators $S^\alpha$ are axial vectors.
The transformation can be expressed by a $3\times 3$ matrix
acting on $E^\alpha$, $S^\beta$, and $S^\gamma$ in Eq. (\ref{eqn:H-p}).
For the rotation, the matrix is the same for both types of vectors.
For the mirror operation, however, the matrices for the two types of vectors have opposite signs.
Since the right-hand side of Eq. (\ref{eqn:H-p}) is quadratic in the spin operators,
the matrix for the axial vector can be treated as that for the polar vector, owing to cancellation of the signs.
Therefore, it is appropriate to use only the matrices for the polar vector
in any symmetry operation (rotation or mirror), even for spin operators.
This means that $K^\alpha_{\beta\gamma}$ is a real-symmetric third-rank polar tensor.
After the transformation, the Hamiltonian given by Eq. (\ref{eqn:H-p}) is expressed as
\begin{align}
\H \rightarrow - K_{\beta\gamma}^\alpha (R_{\alpha\delta} E^\delta) (R_{\beta\mu} S^\mu) (R_{\gamma\nu} S^\nu).
\end{align}
Here, $R_{\alpha\beta}$ represents the $3\times 3$ matrix for the transformation.
Note that the same $R_{\alpha \beta}$ matrix can be used for both polar and axial vectors.
Since the Hamiltonian must be invariant under the symmetry transformation, i.e., Neumann's principal, we obtain
\begin{align}
K^\delta_{\mu\nu} = K_{\beta\gamma}^\alpha R_{\alpha\delta} R_{\beta\mu} R_{\gamma\nu}.
\end{align}
Using the properties of the orthonormal matrix of $R_{\alpha\beta}$,
i.e., $R_{\alpha\beta}R_{\beta\gamma}^T=\delta_{\alpha\gamma}$,
we arrive at the following equation to determine the coefficient tensor:
\begin{align}
R_{\alpha\beta} K^\beta_{\gamma\delta} = R_{\gamma\mu}^T K^\alpha_{\mu\nu} R_{\nu\delta}.
\label{eqn:K}
\end{align}
The coefficient tensor $K^\alpha_{\beta\gamma}$ in Eq. (\ref{eqn:P})
is determined so as to satisfy Eq. (\ref{eqn:K}) under all possible symmetry transformations in the point group.

\subsection{Examples of $C_2$ and $C_3$ point groups}

Let us show how to determine the coefficient tensor.
In general, the symmetric tensor can be expressed as
\begin{align}
K^\alpha =
\begin{pmatrix}
K^\alpha_{xx} & K^\alpha_{xy} & K^\alpha_{zx} \cr
K^\alpha_{xy} & K^\alpha_{yy} & K^\alpha_{yz} \cr
K^\alpha_{zx} & K^\alpha_{yz} & K^\alpha_{zz}
\end{pmatrix}.~~~(\alpha=x,y,z)
\end{align}
There are 18 $(=6\times 3$) degrees of freedom in total.
As a simple example, we consider the $C_2$ point group.
The symmetry operations are $E$ and $C_2$.
The former is the identity operation and we do not consider it.
The second is the $\pi$ rotation around the $z$-axis.
Since there is no inversion symmetry in the $C_2$ point group,
the symmetric spin-dependent electric dipole operator can be finite.
The $C_2$ operation is represented by the following $3\times 3$ matrix:
\begin{align}
C_2 =
\begin{pmatrix}
-1 & 0 & 0 \cr
0 & -1 & 0 \cr
0 & 0 & 1
\end{pmatrix}.
\end{align}
Equation (\ref{eqn:K}) is then written as
\begin{align}
\begin{pmatrix}
-1 & 0 & 0 \cr
0 & -1 & 0 \cr
0 & 0 & 1
\end{pmatrix}
\begin{pmatrix}
K^x \cr
K^y \cr
K^z
\end{pmatrix}
=
\begin{pmatrix}
C_2^T K^x C_2 \cr
C_2^T K^y C_2 \cr
C_2^T K^z C_2
\end{pmatrix}.
\label{eqn:C2}
\end{align}
Here, we omitted the subscripts of $K^\alpha_{\beta\gamma}$.
The matrices $C_2^T$ and $C_2$ in the right-hand side of Eq. (\ref{eqn:C2}) act on the subscripts of $K^\alpha_{\beta\gamma}$.
Equation (\ref{eqn:C2}) reduces the number of degrees of freedom of the coefficient tensors.
The tensors are determined as
\begin{align}
&K^x =
\begin{pmatrix}
0 & 0 & K^x_{zx} \cr
0 & 0 & K^x_{yz} \cr
K^x_{zx} & K^x_{yz} & 0
\end{pmatrix}, \cr
&K^y =
\begin{pmatrix}
0 & 0 & K^y_{zx} \cr
0 & 0 & K^y_{yz} \cr
K^y_{zx} & K^y_{yz} & 0
\end{pmatrix}, \cr
&K^z =
\begin{pmatrix}
K^z_{xx} & K^z_{xy} & 0 \cr
K^z_{xy} & K^z_{yy} & 0 \cr
0 & 0 & K^z_{zz}
\end{pmatrix}.
\end{align}
The spin-dependent electric dipole is then expressed as
\begin{align}
&p_{\rm S}^x = K^x_{yz} O_{yz} + K^x_{zx} O_{zx}, \cr
&p_{\rm S}^y = K^y_{yz} O_{yz} + K^y_{zx} O_{zx}, \label{eqn:P-C2} \\
&p_{\rm S}^z = K^z_{xx} O_{x^2} + K^z_{yy} O_{y^2} + K^z_{zz} O_{z^2} + K^z_{xy} O_{xy}. \nonumber
\end{align}
Here, we introduced the following operators:
\begin{align}
&O_{\alpha^2} = (S^\alpha)^2, \cr
&O_{\alpha\beta} = S^\alpha S^\beta + S^\beta S^\alpha~~~({\rm for}~\alpha\neq\beta).
\label{eqn:Gamma}
\end{align}
There are eight free parameters for the coefficients in Eq. (\ref{eqn:P-C2}).
The number of degrees of freedom is reduced from 18 to eight by Eq. (\ref{eqn:C2}).
Note that there is no additional possible spin dependence in the electric dipole.
A microscopic model with the spin-orbit interaction determines the values of the coefficients.

As another example, we next study the $C_3$ point group.
The possible symmetry operations are $E$ and $C_3$,
where the latter is the $2\pi/3$ rotation around the $z$-axis.
It is expressed by the following matrix:
\begin{align}
C_3 =
\begin{pmatrix}
-\frac{1}{2} & -\frac{\sqrt{3}}{2} & 0 \cr
\frac{\sqrt{3}}{2} & -\frac{1}{2} & 0 \cr
0 & 0 & 1
\end{pmatrix}.
\end{align}
Equation (\ref{eqn:K}) is written as
\begin{align}
\begin{pmatrix}
-\frac{1}{2} & -\frac{\sqrt{3}}{2} & 0 \cr
\frac{\sqrt{3}}{2} & -\frac{1}{2} & 0 \cr
0 & 0 & 1
\end{pmatrix}
\begin{pmatrix}
K^x \cr
K^y \cr
K^z
\end{pmatrix}
=
\begin{pmatrix}
C_3^T K^x C_3 \cr
C_3^T K^y C_3 \cr
C_3^T K^z C_3
\end{pmatrix}.
\label{eqn:C3}
\end{align}
Differing from the $C_2$ case, we can see that $K^x$ and $K^y$ are coupled in the left-hand side of Eq. (\ref{eqn:C3}).
The coefficient tensors are determined as
\begin{align}
&K^x =
\begin{pmatrix}
K^x_{xx} & K^x_{xy} & K^x_{zx} \cr
K^x_{xy} & -K^x_{xx} & K^x_{yz} \cr
K^x_{zx} & K^x_{yz} & 0
\end{pmatrix}, \cr
&K^y =
\begin{pmatrix}
K^x_{xy} & -K^x_{xx} & -K^x_{yz} \cr
-K^x_{xx} & -K^x_{xy} & K^x_{zx} \cr
-K^x_{yz} & K^x_{zx} & 0
\end{pmatrix}, \cr
&K^z =
\begin{pmatrix}
K^z_{xx} & 0 & 0 \cr
0 & K^z_{xx} & 0 \cr
0 & 0 & k^z_{zz}
\end{pmatrix}.
\end{align}
The symmetric spin-dependent electric dipole is then expressed as
\begin{align}
&p_{\rm S}^x = K^x_{xx} O_{x^2-y^2} + K^x_{xy} O_{xy} + K^x_{yz} O_{yz} + K^x_{zx} O_{zx}, \cr
&p_{\rm S}^y = K^x_{xy} O_{x^2-y^2} - K^x_{xx} O_{xy} + K^x_{zx} O_{yz} - K^x_{yz} O_{zx}, \cr
&p_{\rm S}^z = K^z_{xx} O_{x^2+y^2} + K^z_{zz} O_{z^2},
\label{eqn:P-C3}
\end{align}
where we introduced
\begin{align}
O_{x^2\pm y^2} = (S^x)^2 \pm (S^y)^2.
\end{align}
The $C_3$ operation mixes the $x$ and $y$ components
and the common coefficients appear in the $x$ and $y$ components in Eq. (\ref{eqn:P-C3}),
where the number of degrees of freedom is reduced from 18 to six by Eq. (\ref{eqn:C3}).

\subsection{Results for 32 point groups}

\begin{table}[t]\caption{
List of the general forms of the symmetric spin-dependent electric dipole operator
for various point groups compatible with the space group.
Among the 32 point groups, the lack of the inversion symmetry is required for the emergence of the electric dipole.
For monoclinic crystals, i.e., the $C_2$ and $C_s$ point-groups, the first setting is employed.
$K^\alpha_{\beta\gamma}$ is an arbitrary nonzero real coefficient.
The operators are defined as
$O_{\alpha^2} = (S^\alpha)^2$,
$O_{x^2\pm y^2} = (S^x)^2 \pm (S^y)^2$, and
$O_{\alpha\beta} = S^\alpha S^\beta + S^\beta S^\alpha~({\rm for}~\alpha\neq\beta)$.
When the dipole operator vanishes, we use ``$-$" in the list.
Note that all components vanish in case of the ``$O$" point group,
i.e., the dipole does not appear up to the quadratic order of the spin operators.
We add the result for the $C_{\infty v}$ point group for linear molecules.
}
\begin{tabular}{cccc}
\hline
Point group & $p_{\rm S}^x$ & $p_{\rm S}^y$ & $p_{\rm S}^z$ \cr
\hline
$C_2$ & $K^x_{yz} O_{yz}$ & $K^y_{yz} O_{yz}$ & $K^z_{xx} O_{x^2}$ \cr
      & $K^x_{zx} O_{zx}$ & $K^y_{zx} O_{zx}$ & $K^z_{yy} O_{y^2}$ \cr
      &                   &                   & $K^z_{zz} O_{z^2}$ \cr
      &                   &                   & $K^z_{xy} O_{xy}$ \cr
\hline
$C_s$ & $K^x_{xx} O_{x^2}$ & $K^y_{xx} O_{x^2}$ & $K^z_{yz} O_{yz}$ \cr
      & $K^x_{yy} O_{y^2}$ & $K^y_{yy} O_{y^2}$ & $K^z_{zx} O_{zx}$ \cr
      & $K^x_{zz} O_{z^2}$ & $K^y_{zz} O_{z^2}$ & \cr
      & $K^x_{xy} O_{xy}$  & $K^y_{xy} O_{xy}$  & \cr
\hline
$D_2$ & $K^x_{yz} O_{yz}$ & $K^y_{zx} O_{zx}$ & $K^z_{xy} O_{xy}$ \cr
\hline
$C_{2v}$ & $K^x_{zx} O_{zx}$ & $K^y_{yz} O_{yz}$ & $K^z_{xx} O_{x^2}$ \cr
         &                   &                   & $K^z_{yy} O_{y^2}$ \cr
         &                   &                   & $K^z_{zz} O_{z^2}$ \cr      
\hline
$C_4$ & $K^x_{yz} O_{yz}$ & $K^x_{zx} O_{yz}$  & $K^z_{xx} O_{x^2+y^2}$ \cr
      & $K^x_{zx} O_{zx}$ & $-K^x_{yz} O_{zx}$ & $K^z_{zz} O_{z^2}$ \cr
\hline
$S_4$ & $K^x_{yz} O_{yz}$ & $-K^x_{zx} O_{yz}$ & $K^z_{xx} O_{x^2-y^2}$ \cr
      & $K^x_{zx} O_{zx}$ & $K^x_{yz} O_{zx}$  & $K^z_{xy} O_{xy}$ \cr
\hline
$D_4$ & $K^x_{yz} O_{yz}$ & $-K^x_{yz} O_{zx}$ & $-$ \cr
\hline
$C_{4v}$ & $K^x_{zx} O_{zx}$ & $K^x_{zx} O_{yz}$ & $K^z_{xx} O_{x^2+y^2}$ \cr
         &                   &                   & $K^z_{zz} O_{z^2}$ \cr
\hline
$D_{2d}$ & $K^x_{yz} O_{yz}$ & $K^x_{yz} O_{zx}$ & $K^z_{xy} O_{xy}$ \cr
\hline
$C_3$ & $K^x_{xx} O_{x^2-y^2}$ & $K^x_{xy} O_{x^2-y^2}$ & $K^z_{xx} O_{x^2+y^2}$ \cr
      & $K^x_{xy} O_{xy}$      & $-K^x_{xx} O_{xy}$     & $K^z_{zz} O_{z^2}$ \cr
      & $K^x_{yz} O_{yz}$      & $K^x_{zx} O_{yz}$      & \cr
      & $K^x_{zx} O_{zx}$      & $-K^x_{yz} O_{zx}$     & \cr
\hline
$D_3$ & $K^x_{xx} O_{x^2-y^2}$ & $-K^x_{xx} O_{xy}$ & $-$ \cr
      & $K^x_{yz} O_{yz}$      & $-K^x_{yz} O_{zx}$ & \cr
\hline
$C_{3v}$ & $K^x_{xy} O_{xy}$ & $K^x_{xy} O_{x^2-y^2}$ & $K^z_{xx} O_{x^2+y^2}$ \cr
         & $K^x_{zx} O_{zx}$ & $K^x_{zx} O_{yz}$      & $K^z_{zz} O_{z^2}$ \cr
\hline
$C_6$ & $K^x_{yz} O_{yz}$ & $K^x_{zx} O_{yz}$  & $K^z_{xx} O_{x^2+y^2}$ \cr
      & $K^x_{zx} O_{zx}$ & $-K^x_{yz} O_{zx}$ & $K^z_{zz} O_{z^2}$ \cr
\hline
$C_{3h}$ & $K^x_{xx} O_{x^2-y^2}$ & $K^x_{xy} O_{x^2-y^2}$ & $-$ \cr
         & $K^x_{xy} O_{xy}$      & $-K^x_{xx} O_{xy}$  & \cr
\hline
$D_6$ & $K^x_{yz} O_{yz}$ & $-K^x_{yz} O_{zx}$ & $-$ \cr
\hline
$C_{6v}$ & $K^x_{zx} O_{zx}$ & $K^x_{zx} O_{yz}$ & $K^z_{xx} O_{x^2+y^2}$ \cr
         &                   &                   & $K^z_{zz} O_{z^2}$ \cr
\hline
$D_{3h}$ & $K^x_{xy} O_{xy}$ & $K^x_{xy} O_{x^2-y^2}$ & $-$ \cr
\hline
$T$ & $K^x_{yz} O_{yz}$ & $K^x_{yz} O_{zx}$ & $K^x_{yz} O_{xy}$ \cr
\hline
$O$ & $-$& $-$ & $-$ \cr
\hline
$T_d$ & $K^x_{yz} O_{yz}$ & $K^x_{yz} O_{zx}$ & $K^x_{yz} O_{xy}$ \cr
\hline
$C_{\infty v}$ & $K^x_{zx} O_{zx}$ & $K^x_{zx} O_{yz}$ & $K^z_{xx} O_{x^2+y^2}$ \cr
               &                   &                   & $K^z_{zz} O_{z^2}$ \cr
\hline
\end{tabular}
\label{table:P}
\end{table}

In this section, we focus on the 32 point groups that are compatible with the space group.
In the same way as in the previous subsection, we can determine the coefficient tensors
by solving Eq. (\ref{eqn:K}) under all the possible basal symmetry operations in the point group.
The lack of the inversion symmetry is required for a finite electric dipole operator.
Table \ref{table:P} shows finite spin-dependent components in various point groups.
In Appendix \ref{appendix:polarization},
we summarize the coefficient tensors for various basal symmetries used to obtain Table \ref{table:P} for convenience.
Up to the quadratic order of the spin operators, the spin dependences are completed by Table \ref{table:P}
irrespective of the microscopic origin.

The meaning of the several blanks in Table \ref{table:P} is that, for instance, $p_{\rm S}^z=0$ for the $D_4$ point group.
In the $D_4$ symmetry, two $C_2$ operations around the $x$- and $y$-axes and two $C_2(\pm\frac{\pi}{4})$ operations
(see Fig. \ref{fig:symmetry} in Appendix \ref{appendix:polarization}) are added to the $C_4$ case.
For these operations, $p_{\rm S}^z$ must change its sign.
However, the sign change is not allowed for $p_{\rm S}^z$ in $C_4$ since it is composed of $O_{x^2+y^2}$ and $O_{z^2}$.
Therefore, the coefficient of $p_{\rm S}^z$ must vanish for $D_4$.
The same argument is also applicable to $p_{\rm S}^z=0$ in the $D_3$, $D_{3h}$, and $D_6$ cases.
Adding the $C_2$ operations to $C_3$ and $C_6$ restricts the coefficients of $p_{\rm S}^z$ for the $D_3$, $D_{3h}$, and $D_6$ cases.
In the case of $C_{3h}$, the $\sigma_h$ operation is added to $C_3$.
The sign of $p_{\rm S}^z$ must be changed for this operation.
Since this is forbidden for $O_{x^2+y^2}$ and $O_{z^2}$ ($p_{\rm S}^z$ for $C_3$),
the coefficient of $p_{\rm S}^z$ vanishes in the case of $C_{3h}$.
In the $O$ point group, all components of the electric dipole vanish.
In this case, the $C_4$ operations are added to the $T$ point group.
From Table \ref{table:P}, we can see that the $C_4$ operation around the $z$-axis restricts the coefficient as $K^x_{yz}=-K^y_{zx}$.
When we add this restriction to the result of $T$,
we obtain $K^x_{yz}=0$ and all coefficients vanish in the $O$ point group.

Table \ref{table:P} is closely related to the basis functions of the irreducible representations.
For instance, the $z$ component of the dipole, $p_{\rm S}^z$, is classified in the irreducible representation
whose basis function contains $z$.
In the absence of the inversion symmetry, even and odd basis functions are mixed in the irreducible representations.
Let us explain this point for the $D_2$ point group as an example.
In the $D_2$ character table, we can see that the basis functions of $z$ and $xy$ (or $yx$) are classified
in the same irreducible representation.
\cite{Koster-1963}
This means that the two basis functions are transformed in the same way by the symmetry operations for $D_2$
and that they cannot be distinguished.
We next consider the spin product of $S^x S^y + S^y S^x$.
The $D_2$ point group consists of only rotational symmetry operations.
For these, $S^x S^y + S^y S^x$ behaves as $xy+yx$.
Therefore, the basis functions of $z$, $xy+yx$, and $S^x S^y + S^y S^x$
are classified in the same irreducible representation and they are mixed in the $D_2$ symmetry.
This leads to $p_{\rm S}^z=K^z_{xy} (S^x S^y + S^y S^x)$, as listed in Table \ref{table:P} for $D_2$.

We emphasize that this argument holds for other point-group symmetries.
In general, the point groups consist of rotation and mirror operations.
The $3\times 3$ matrices for the rotational operations are the same for the polar and axial vectors,
while the signs of their matrices are opposite for mirror operations.
Since $S^x S^y + S^y S^x$ is quadratic in the spin operators, it also behaves as $xy+yx$ for the mirror operations.
This means that the basis function of $\alpha\beta + \beta\alpha$
can be treated as $S^\alpha S^\beta + S^\beta S^\alpha$ in general.
Similarly, $\alpha^2$ can be treated as $(S^\alpha)^2$.

From the above discussion, we notice that most of the results listed in Table \ref{table:P}
can also be obtained by simply using the linear and quadratic basis functions of an irreducible representation
listed in the character tables of the point groups.
However, we have to be careful when the $x$ and $y$ basis functions
are classified in the same two-dimensional representation,
such as in the cases of $C_4$, $S_4$, $D_4$, $C_{4v}$, $D_{2d}$, $C_3$, and so on.
For these, the same coefficients appear in the $p_{\rm S}^x$ and $p_{\rm S}^y$ components, as shown in Table \ref{table:P}.
We have to pay attention to choose the correct sign of the coefficients,
which is determined by Eq. (\ref{eqn:K}).
Note that there are other ways to find the correct sign of the coefficient,
as described in Refs. \ref{ref:note-sign} and \ref{ref:note-piezo}.

\subsection{Microscopic origin of spin-dependent electric dipole}
\label{sec:p-S}

There are two possible microscopic origins of the electric dipole moment induced by the product of spin operators.
One is the metal-ligand hybridization model expressed by Eq. (\ref{eqn:P-Arima}),
where the energy levels of $d$ orbitals depend on the direction of the spin via the spin-orbit interaction
and the energy shift affects the metal-ligand hybridization.
Since the hybridization determines the charge distribution of the metal and ligand ions,
electric dipole moments are connected to the spin.
In the absence of the inversion symmetry, the electric dipole operator is described by products of spin operators.
\cite{Arima-2007}

The other microscopic origin is parity mixing between even- and odd-parity orbitals.
In the absence of the inversion symmetry at the metal-ion site,
even-parity $d$ orbitals mix with odd-parity $p$-orbitals.
In this case, the charge density of the mixed orbital has both even- and odd-parity components
and the latter leads to an electric dipole moment.
In the presence of the spin-orbit interaction, the orbitals are connected to the spin
and the electric dipoles are described by the product of spin operators.
The details are given in Appendix \ref{appendix:parity-mix},
where both $d$-$p$ and $f$-$d$ orbital hybridizations are discussed, focusing on the $T_d$ point-group symmetry.

Both origins lead to the same spin dependences in the electric dipole
and it is difficult to distinguish them by experiments.
We emphasize here that the spin dependences in the electric dipole are precisely determined group-theoretically
on the basis of the local symmetry at the metal-ion site, regardless of its microscopic origin.

\section{Electric Dipole by Two Spins}

\begin{figure}[t]
\begin{center}
\includegraphics[width=5cm,clip]{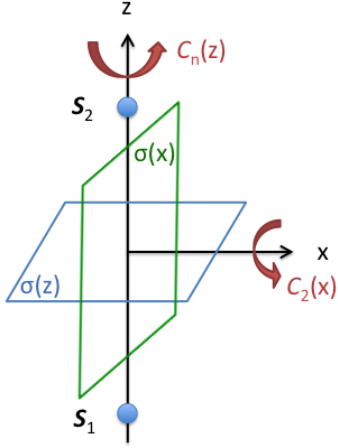}
\end{center}
\caption{
(Color online)
Schematic of two spins ($\bS_1$ and $\bS_2$) aligned along the $z$-axis.
$\sigma(z)$ and $\sigma(x)$ represent the mirrors whose normal vectors are along the $z$- and $x$-axes, respectively.
$C_2(x)$ and $C_n(z)$ are $\pi$ and $2\pi/n$ rotations around the $x$- and $z$-axes, respectively.
}
\label{fig:spin}
\end{figure}

In this section, we study the electric dipole operator generated by a pair of spins
represented by $\bS_1$ and $\bS_2$ at different sites, as shown in Fig. \ref{fig:spin}.
Up to the quadratic order of the spin operators,
the general form of the spin-dependent electric dipole can be expressed as
\begin{align}
&p^\alpha = p_{\rm S}^\alpha + p_{\rm A}^\alpha, \cr
&p_{\rm S}^\alpha = A^\alpha_{\beta\gamma} S_1^\beta S_2^\gamma
            + D^\alpha_{\beta\gamma} ( S_1^\beta S_1^\gamma + S_2^\beta S_2^\gamma ), \cr
&p_{\rm A}^\alpha = C^\alpha_{\beta\gamma} S_1^\beta S_2^\gamma
            + B^\alpha_{\beta\gamma} ( S_1^\beta S_1^\gamma - S_2^\beta S_2^\gamma ).
\label{eqn:P2}
\end{align}
Here, $p_{\rm S}^\alpha$ and $p_{\rm A}^\alpha$ are symmetric and antisymmetric spin-dependent components
with respect to the interchange of the two spins, respectively.
$A^\alpha_{\beta\gamma}$ and $C^\alpha_{\beta\gamma}$ are real coefficient tensors
for the product of spin operators at different sites,
while $D^\alpha_{\beta\gamma}$ and $B^\alpha_{\beta\gamma}$ are those for the same sites.
They have the following relations:
\begin{align}
&A^\alpha_{\beta\gamma} = A^\alpha_{\gamma\beta},~~~
C^\alpha_{\beta\gamma} = - C^\alpha_{\gamma\beta}, \cr
&D^\alpha_{\beta\gamma} = D^\alpha_{\gamma\beta},~~~
B^\alpha_{\beta\gamma} = B^\alpha_{\gamma\beta}.
\label{eqn:ABDE}
\end{align}
The first two relations are required to classify the even and odd spin dependences,
while the latter two ensure the Hermitian nature of the dipole operator.
Owing to the antisymmetric property of $C^\alpha_{\beta\gamma}$ in Eq. (\ref{eqn:ABDE}),
the first term of $p_{\rm A}^\alpha$ in Eq. (\ref{eqn:P2}) can be written as
$p_{\rm A}^\alpha=C^\alpha_\beta (\bS_1\times\bS_2)_\beta$.
\cite{Moriya-1968,Kaplan-2011}
Here, $C^\alpha_\beta$ is an arbitrary constant and $(\cdots)_\beta$ represents the $\beta$ component of the vector.

Let us discuss the symmetry property of the electric dipole.
When there is an inversion center between the two spins,
the inversion, $I$, is one of the possible symmetry operations.
For the inversion transformation, we obtain
\begin{align}
I p^\alpha I^{-1} = I ( p_{\rm S}^\alpha + p_{\rm A}^\alpha ) I^{-1} = p_{\rm S}^\alpha - p_{\rm A}^\alpha.
\label{eqn:IP}
\end{align}
Since the electric dipole is transformed as a polar vector by the inversion operation,
the sign changes in the left-hand side of Eq. (\ref{eqn:IP}).
This indicates that $p_{\rm S}^\alpha$ disappears and that $p_{\rm A}^\alpha$ is the only possible electric dipole.
In the absence of the inversion center, however, the irreducible representations are not classified by this symmetry
and the even and odd parities are mixed.
In this case, the symmetric spin-dependent component of the electric dipole operator can be finite.

Kaplan and Mahanti focussed on the antisymmetric component
and classified the electric dipole under various symmetries.
\cite{Kaplan-2011}
Their result holds for cases both with and without the inversion center.
In this paper, we extend their study to the symmetric component
and investigate the possible forms of the spin-dependent electric dipole operator
that can appear in the absence of the inversion center.

The symmetry operations we consider here are shown in Fig. \ref{fig:spin}.
Among them, $\sigma(x)$ and $C_n(z)$ operations do not interchange the two spin sites,
while $\sigma(z)$ and $C_2(x)$ interchange them.
Since $p_{\rm S}^\alpha$ is the symmetric component, all the operations behave the same with respect to the spin interchange.
Therefore, we arrive at the following relations to determine the coefficient tensors:
\begin{align}
&R_{\alpha\beta} A^\beta_{\gamma\delta} = R_{\gamma\mu}^T A^\alpha_{\mu\nu} R_{\nu\delta}, \cr
&R_{\alpha\beta} D^\beta_{\gamma\delta} = R_{\gamma\mu}^T D^\alpha_{\mu\nu} R_{\nu\delta}.
\label{eqn:AD}
\end{align}
Here, $R_{\alpha\beta}$ represents the $3\times 3$ matrix for the symmetry operation.
Equation (\ref{eqn:AD}) is essentially the same as Eq. (\ref{eqn:K}).
The coefficient tensors, $A^\alpha_{\beta\gamma}$ and $D^\alpha_{\beta\gamma}$,
are determined so as to satisfy Eq. (\ref{eqn:AD}).
In Table \ref{table:PS}, we list the possible forms of the symmetric spin-dependent electric dipole operator
under various symmetries for the two spins.

\begin{table}[t]
\caption{
List of the general forms of the symmetric spin-dependent electric dipole operator
generated by two spins for various symmetries.
The symmetry operations we consider here are shown in Fig. \ref{fig:spin}.
The coefficient $A^\alpha_{\beta\gamma}$ is an arbitrary, nonzero, and real value for the spin product at the different sites.
The operators are defined as
$F_{\alpha^2} = S^\alpha_1 S^\alpha_2$,
$F_{x^2\pm y^2} = (S^x_1 S^x_2) \pm (S^y_1 S^y_2)$, and
$F_{\alpha\beta} = S^\alpha_1 S^\beta_2 + S^\beta_1 S^\alpha_2$ for $\alpha\neq\beta$.
Note that the results are the same for the $C_4(z)$, $C_6(z)$, and $C_\infty(z)$ symmetries.
In the case of spin products at the same site, the result can be obtained by replacing
$A^\alpha_{\beta\gamma}\rightarrow D^\alpha_{\beta\gamma}$ and $F\rightarrow G$.
Here, the operator $G$ is written as
$G_{\alpha^2} = (S^\alpha_1)^2 + (S^\alpha_2)^2$,
$G_{x^2\pm y^2} = [(S^x_1)^2 \pm (S^y_1)^2] + [(S^x_2)^2 \pm (S^y_2)^2]$, and
$G_{\alpha\beta} = (S^\alpha_1 S^\beta_1 + S^\beta_1 S^\alpha_1) + (S^\alpha_2 S^\beta_2 + S^\beta_2 S^\alpha_2)$
for $\alpha\neq\beta$.
}
\begin{tabular}{cccc}
\hline
Symmetry & $p_{\rm S}^x$ & $p_{\rm S}^y$ & $p_{\rm S}^z$ \cr
\hline
$\sigma(z)$ & $A^x_{xx} F_{x^2}$ & $A^y_{xx} F_{x^2}$ & $A^z_{yz} F_{yz}$ \cr
            & $A^x_{yy} F_{y^2}$ & $A^y_{yy} F_{y^2}$ & $A^z_{zx} F_{zx}$ \cr
            & $A^x_{zz} F_{z^2}$ & $A^y_{zz} F_{z^2}$ & \cr
            & $A^x_{xy} F_{xy}$  & $A^y_{xy} F_{xy}$  & \cr
\hline
$\sigma(x)$ & $A^x_{zx} F_{zx}$ & $A^y_{xx} F_{x^2}$ & $A^z_{xx} F_{x^2}$ \cr
            & $A^x_{xy} F_{xy}$ & $A^y_{yy} F_{y^2}$ & $A^z_{yy} F_{y^2}$ \cr
            &                   & $A^y_{zz} F_{z^2}$ & $A^z_{zz} F_{z^2}$ \cr
            &                   & $A^y_{yz} F_{yz}$  & $A^z_{yz} F_{yz}$ \cr
\hline
$C_2(x)$ & $A^x_{xx} F_{x^2}$ & $A^y_{zx} F_{zx}$ & $A^z_{zx} F_{zx}$ \cr
         & $A^x_{yy} F_{y^2}$ & $A^y_{xy} F_{xy}$ & $A^z_{xy} F_{xy}$ \cr
         & $A^x_{zz} F_{z^2}$ &                   & \cr
         & $A^x_{yz} F_{yz}$  &                   & \cr
\hline
$C_2(z)$ & $A^x_{yz} F_{yz}$ & $A^y_{yz} F_{yz}$ & $A^z_{xx} F_{x^2}$ \cr
         & $A^x_{zx} F_{zx}$ & $A^y_{zx} F_{zx}$ & $A^z_{yy} F_{y^2}$ \cr
         &                   &                   & $A^z_{zz} F_{z^2}$ \cr
         &                   &                   & $A^z_{xy} F_{xy}$ \cr
\hline
$C_3(z)$ & $A^x_{xx} F_{x^2-y^2}$ & $A^x_{xy} F_{x^2-y^2}$ & $A^z_{xx} F_{x^2+y^2}$ \cr
         & $A^x_{xy} F_{xy}$      & $-A^x_{xx} F_{xy}$     & $A^z_{zz} F_{z^2}$ \cr
         & $A^x_{yz} F_{yz}$      & $A^x_{zx} F_{yz}$      & \cr
         & $A^x_{zx} F_{zx}$      & $-A^x_{yz} F_{zx}$     & \cr
\hline
$C_4(z)$, $C_6(z)$ & $A^x_{yz} F_{yz}$ & $A^x_{zx} F_{yz}$  & $A^z_{xx} F_{x^2+y^2}$ \cr
$C_\infty(z)$      & $A^x_{zx} F_{zx}$ & $-A^x_{yz} F_{zx}$ & $A^z_{zz} F_{z^2}$ \cr
\hline
\end{tabular}
\label{table:PS}
\end{table}

In case of the antisymmetric component, the same formulation can be used.
The coefficient tensors are determined to satisfy the following relations:
\begin{align}
&R_{\alpha\beta} C^\beta_{\gamma\delta} = \eta_{1\leftrightarrow 2} R_{\gamma\mu}^T C^\alpha_{\mu\nu} R_{\nu\delta}, \cr
&R_{\alpha\beta} B^\beta_{\gamma\delta} = \eta_{1\leftrightarrow 2} R_{\gamma\mu}^T B^\alpha_{\mu\nu} R_{\nu\delta}.
\label{eqn:BE}
\end{align}
Here, $\eta_{1\leftrightarrow 2}$ is a factor for the interchange of the two spins.
When the symmetry operation interchanges the spins, $\eta_{1\leftrightarrow 2}=-1$,
while $\eta_{1\leftrightarrow 2}=1$ for no spin interchange.
In the same way as for the symmetric component, we can determine the coefficient tensors by solving Eq. (\ref{eqn:BE}).
For completeness, we also show the results for the antisymmetric case in Table \ref{table:PA},
which were reported by Kaplan and Mahanti.
\cite{Kaplan-2011}

\begin{table}[t]
\caption{
List of the general forms of the antisymmetric spin-dependent electric dipole operator
generated by two spins with various symmetries,
which were obtained by Kaplan and Mahanti.
\cite{Kaplan-2011}
The symmetry operations we consider here are listed in Fig. \ref{fig:spin}.
In Ref. \ref{ref:Kaplan-2011}, they took the $x$-axis along the two spins.
In this paper, we choose the $z$-axis along the two spins
for easy comparison with the results summarized in Table \ref{table:P}.
$C^\alpha_\beta$ are arbitrary nonzero real coefficients for the spin product at the different sites.
Following Kaplan and Mahanti,
\cite{Kaplan-2011}
they are defined as
$C^\alpha_x=C^\alpha_{yz}$, $C^\alpha_y=C^\alpha_{zx}$, and $C^\alpha_z=C^\alpha_{xy}$,
where $C^\alpha_{\beta\gamma}$ were introduced in Eq. (\ref{eqn:P2}).
$W_\alpha$ is defined as $W_\alpha = ( \bS_1 \times \bS_2 )_\alpha$.
Here, $(\cdots)_\alpha$ represents the $\alpha$ component.
For spin products at the same site, $B^\alpha_{\beta\gamma}$ are arbitrary nonzero coefficients.
The operators are defined as
$V_{\alpha^2} = O_{\alpha^2,1} - O_{\alpha^2,2}$,
$V_{x^2\pm y^2} = O_{x^2 \pm y^2,1} - O_{x^2 \pm y^2,2}$, and
$V_{\alpha\beta} = O_{\alpha\beta,1} - O_{\alpha\beta,2}$ for $\alpha\neq\beta$.
Here, $O_{m,1}$ and $O_{m,2}$ are the quadrupole operators defined by the spin operators of $\bS_1$ and $\bS_2$, respectively.
Note that the results are the same for the $C_4(z)$, $C_6(z)$, and $C_\infty(z)$ symmetries.
}
\begin{tabular}{cccc}
\hline
Symmetry & $p_{\rm A}^x$ & $p_{\rm A}^y$ & $p_{\rm A}^z$ \cr
\hline
$\sigma(z)$ & $C^x_x W_x$ & $C^y_x W_x$ & $C^z_z W_z$ \cr
            & $C^x_y W_y$ & $C^y_y W_y$ & \cr
            & $B^x_{yz} V_{yz}$ & $B^y_{yz} V_{yz}$ & $B^z_{xx} V_{x^2}$ \cr
            & $B^x_{zx} V_{zx}$ & $B^y_{zx} V_{zx}$ & $B^z_{yy} V_{y^2}$ \cr
            &                   &                   & $B^z_{zz} V_{z^2}$ \cr
            &                   &                   & $B^z_{xy} V_{xy}$ \cr
\hline
$\sigma(x)$ & $C^x_y W_y$ & $C^y_x W_x$ & $C^z_x W_x$ \cr
            & $C^x_z W_z$ &             & \cr
            & $B^x_{zx} V_{zx}$ & $B^y_{xx} V_{x^2}$ & $B^z_{xx} V_{x^2}$ \cr
            & $B^x_{xy} V_{xy}$ & $B^y_{yy} V_{y^2}$ & $B^z_{yy} V_{y^2}$ \cr
            &                   & $B^y_{zz} V_{z^2}$ & $B^z_{zz} V_{z^2}$ \cr
            &                   & $B^y_{yz} V_{yz}$  & $B^z_{yz} V_{yz}$ \cr
\hline
$C_2(x)$ & $C^x_y W_y$ & $C^y_x W_x$ & $C^z_x W_x$ \cr
         & $C^x_z W_z$ &             & \cr
         & $B^x_{zx} V_{zx}$ & $B^y_{xx} V_{x^2}$ & $B^z_{xx} V_{x^2}$ \cr
         & $B^x_{xy} V_{xy}$ & $B^y_{yy} V_{y^2}$ & $B^z_{yy} V_{y^2}$ \cr
         &                   & $B^y_{zz} V_{z^2}$ & $B^z_{zz} V_{z^2}$ \cr
         &                   & $B^z_{yz} V_{yz}$  & $B^z_{yz} V_{yz}$ \cr
\hline
$C_2(z)$ & $C^x_x W_x$ & $C^y_x W_x$ & $C^z_z W_z$ \cr
         & $C^x_y W_y$ & $C^y_y W_y$ & \cr
         & $B^x_{yz} V_{yz}$ & $B^y_{yz} V_{yz}$ & $B^z_{xx} V_{x^2}$ \cr
         & $B^x_{zx} V_{zx}$ & $B^y_{zx} V_{zx}$ & $B^z_{yy} V_{y^2}$ \cr
         &                   &                   & $B^z_{zz} V_{z^2}$ \cr
         &                   &                   & $B^z_{xy} V_{xy}$ \cr
\hline
$C_3(z)$ & $C^x_x W_x$ & $-C^x_y W_x$ & $C^z_z W_z$ \cr
         & $C^x_y W_y$ & $C^x_x W_y$  & \cr
         & $B^x_{xx} V_{x^2-y^2}$ & $B^x_{xy} V_{x^2-y^2}$ & $B^z_{xx} V_{x^2+y^2}$ \cr
         & $B^x_{xy} V_{xy}$      & $-B^x_{xx} V_{xy}$     & $B^z_{zz} V_{z^2}$ \cr
         & $B^x_{yz} V_{yz}$      & $B^x_{zx} V_{yz}$      & \cr
         & $B^x_{zx} V_{zx}$      & $-B^x_{yz} V_{zx}$     & \cr
\hline
$C_4(z)$, $C_6(z)$ & $C^x_x W_x$ & $-C^x_y W_x$ & $C^z_z W_z$ \cr
$C_\infty(z)$      & $C^x_y W_y$ & $C^x_x W_y$  & \cr
                   & $B^x_{yz} V_{yz}$ & $B^x_{zx} V_{yz}$ & $B^z_{xx} V_{x^2+y^2}$ \cr
                   & $B^x_{zx} V_{zx}$ & $-B^x_{yz} V_{zx}$  & $B^z_{zz} V_{z^2}$ \cr
\hline
\end{tabular}
\label{table:PA}
\end{table}

\section{Magnetoelectric Effect by Single Spin}

In this section, we study magnetoelectric effects caused by a spin-dependent electric dipole.
We discuss the case of a dipole generated by a single spin,
where the antisymmetric spin-dependent component vanishes $(p_{\rm A}=0)$.
There are the following three expected effects:
(1) The emergence of a static electric dipole in magnetic systems.
(2) Electromagnon excitation.
(3) Directional dichroism.
We discuss these points in this section.

We assume that a single spin is located in an environment represented by a point group.
Since the dipole is given by a product of spin operators, as shown in Table \ref{table:P},
$S=1/2$ is irrelevant and we only consider $S\ge 1$ cases.
As an example, we mainly study the tetragonal point groups of $D_4$, $C_{4v}$, and $D_{2d}$.
The classification of the spin and electric dipole is given in Table \ref{table:list-tetra}.
The energy eigenstates of a single spin in the tetragonal system are classified in Table \ref{table:spin-tetra}.
For later convenience, we show the multiplication table in Table \ref{table:product-tetra}.

\subsection{Emergence of static electric dipole}

We study a static electric dipole moment emerging in magnetic systems.
To have a finite expectation value of the electric dipole operator,
$\langle{\rm GS}|p_{\rm S}^\alpha|{\rm GS}\rangle$ must contain the $\Gamma_1$ representation,
where $|{\rm GS}\rangle$ represents a groundstate.
We discuss this from the group-theoretical point of view.

\subsubsection{$S=1$}

We first discuss the $S=1$ case.
As in Table \ref{table:spin-tetra}, the energy eigenstates are classified in the $\Gamma_2$ and $\Gamma_5$ representations
in the tetragonal symmetry.
This is consistent with the energy eigenstates of the spin Hamiltonian of $D(S^z)^2$.
The groundstate is $\Gamma_2$ for the easy-plane ($D>0$) case, whereas it is $\Gamma_5$ for the easy-axis ($D<0$) case.

The $z$ component of the dipole operator is nonzero for $C_{4v}$ and $D_{2d}$.
As shown in Table \ref{table:P}, it is expressed as
\begin{align}
p_{\rm S}^z =
\begin{cases}
  K^z_{xx} [(S^x)^2+(S^y)^2] + K^z_{zz} (S^z)^2~~~({\rm for}~C_{4v}) \\
  K^z_{xy} (S^xS^y+S^yS^x)~~~~~~~~~~~~~~~~~~~({\rm for}~D_{2d})
\end{cases}.
\end{align}
For $C_{4v}$, $p_{\rm S}^z$ is classified in the $\Gamma_1$ representation (see Table \ref{table:list-tetra}).
Since the $\Gamma_1$ operator can always have a finite expectation value for any state,
we concentrate our attention on $p_{\rm S}^z$ for $D_{2d}$ in the following discussion.

In the case of $D_{2d}$, $p_{\rm S}^z$ is classified in the $\Gamma_4$ representation (see Table \ref{table:list-tetra}).
Since only $\Gamma_5\otimes\Gamma_5$ contains the $\Gamma_4$ representation
among the basis $\Gamma_2$ and $\Gamma_5$ states (see Table \ref{table:product-tetra}),
$p_{\rm S}^z$ can be finite only for the $\Gamma_5$ groundstate.
For the $D_{2d}$ symmetry, the electric dipole operator is expressed as
\begin{align}
p_{\rm S}^z = K^z_{xy} O_{xy} = K^z_{xy}
\begin{pmatrix}
0 & 0 & -i \cr
0 & 0 & 0 \cr
i & 0 & 0
\end{pmatrix},
\label{eqn:Pz-d2d}
\end{align}
where the matrix elements are given on the basis of the $|1\rangle$, $|0\rangle$, and $|-1\rangle$ states.
To have a finite expectation value, the groundstate must be a linear combination
of the $|1\rangle$ and $|-1\rangle$ states ($\Gamma_5$ states).
In general, note that the quadrupole operator $O_{xy}$ has finite matrix elements
between the $|m\rangle$ and $|m\pm 2\rangle$ states.

\begin{table}[t]
\caption{
Classification of spin and electric dipole for $D_4$, $C_{4v}$, and $D_{2d}$ point groups.
$S^\alpha$ and $p_{\rm S}^\alpha$ represent the $\alpha~(=x,y,z)$ component of the spin and electric dipole, respectively.
The classification of $p_{\rm S}^z$ depends on the point-group symmetry.
It vanishes for $D_4$, as shown in Table \ref{table:P}.
}
\begin{tabular}{ccccccc}
\hline
Tetragonal ($D_4,C_{4v},D_{2d}$) & Magnetic dipole & Electric dipole \cr
\hline
$A_1$ $\Gamma_1$ &             & $p_{\rm S}^z~(C_{4v})$ \cr
$A_2$ $\Gamma_2$ & $S^z$       &                \cr
$B_1$ $\Gamma_3$ &             &               \cr
$B_2$ $\Gamma_4$ &             & $p_{\rm S}^z~(D_{2d})$ \cr
$E$~~$\Gamma_5$  & ($S^x,S^y$) & ($p_{\rm S}^x,p_{\rm S}^y$)   \cr
\hline
\end{tabular}
\label{table:list-tetra}
\end{table}

\begin{table}[t]
\caption{
Classification of single-spin states for the tetragonal ($D_4$, $C_{4v}$, and $D_{2d}$) point groups.
$|m\rangle$ represents the spin state for $S^z=m$.
$a$ and $b$ are arbitrary constants.
}
\begin{tabular}{ccccc}
\hline
Tetragonal                   & $S=1$           & $S=3/2$                   & $S=2$                  & $S=5/2$ \cr
\hline
$A_1$ $\Gamma_1$             &                 &                           & $|0\rangle$            & \cr
$A_2$ $\Gamma_2$             & $|0\rangle$     &                           &                        & \cr
$B_1$ $\Gamma_3$             &                 &                           & $|2\rangle+|-2\rangle$ & \cr
$B_2$ $\Gamma_4$             &                 &                           & $|2\rangle-|-2\rangle$ & \cr
$E$~~$\Gamma_5$              & $|\pm 1\rangle$ &                           & $|\pm 1\rangle$        & \cr
\hline
$E_{\frac{1}{2}}$ $\Gamma_6$ &                 & $|\pm \frac{1}{2}\rangle$ &                        & $|\pm\frac{1}{2}\rangle$ \cr
$E_{\frac{3}{2}}$ $\Gamma_7$ &                 & $|\pm \frac{3}{2}\rangle$ &                        & $a|\pm\frac{3}{2}\rangle+b|\mp\frac{5}{2}\rangle$ \cr
                             &                 &                           &                        & $a|\pm\frac{5}{2}\rangle-b|\mp\frac{3}{2}\rangle$ \cr
\hline
\end{tabular}
\label{table:spin-tetra}
\end{table}

\begin{table}[t]
\caption{
Multiplication table for $D_4$, $C_{4v}$, and $D_{2d}$ point groups.
\cite{Koster-1963}
$\Gamma_{i+j}$ represents $\Gamma_i\oplus\Gamma_j$.
}
\begin{tabular}{cccccccc}
\hline
$\otimes$  & $\Gamma_1$ & $\Gamma_2$ & $\Gamma_3$ & $\Gamma_4$ & $\Gamma_5$ & $\Gamma_6$ & $\Gamma_7$ \cr
\hline
$\Gamma_1$ & $\Gamma_1$ & $\Gamma_2$ & $\Gamma_3$ & $\Gamma_4$ & $\Gamma_5$ & $\Gamma_6$ & $\Gamma_7$ \cr
$\Gamma_2$ & $\Gamma_2$ & $\Gamma_1$ & $\Gamma_4$ & $\Gamma_3$ & $\Gamma_5$ & $\Gamma_6$ & $\Gamma_7$ \cr
$\Gamma_3$ & $\Gamma_3$ & $\Gamma_4$ & $\Gamma_1$ & $\Gamma_2$ & $\Gamma_5$ & $\Gamma_7$ & $\Gamma_6$ \cr
$\Gamma_4$ & $\Gamma_4$ & $\Gamma_3$ & $\Gamma_2$ & $\Gamma_1$ & $\Gamma_5$ & $\Gamma_7$ & $\Gamma_6$ \cr
$\Gamma_5$ & $\Gamma_5$ & $\Gamma_5$ & $\Gamma_5$ & $\Gamma_5$ & $\Gamma_{1+2+3+4}$ & $\Gamma_{6+7}$ & $\Gamma_{6+7}$ \cr
$\Gamma_6$ & $\Gamma_6$ & $\Gamma_6$ & $\Gamma_7$ & $\Gamma_7$ & $\Gamma_{6+7}$ & $\Gamma_{1+2+5}$ & $\Gamma_{3+4+5}$ \cr
$\Gamma_7$ & $\Gamma_7$ & $\Gamma_7$ & $\Gamma_6$ & $\Gamma_6$ & $\Gamma_{6+7}$ & $\Gamma_{3+4+5}$ & $\Gamma_{1+2+5}$ \cr
\hline
\end{tabular}
\label{table:product-tetra}
\end{table}

To induce the quadrupole moment, it is necessary to lift the threefold degeneracy of the $S=1$ states.
A simple way is to apply an external magnetic field, where the time-reversal symmetry is broken.
A typical example is given by the following Hamiltonian:
\begin{align}
\H = D(S^z)^2 - h ( \cos\phi S^x + \sin\phi S^y ).
\label{eqn:H-D-h}
\end{align}
Here, $h=g\mu_{\rm B}H$ represents an effective magnetic field applied in the $xy$-plane
that mixes the $|\pm 1\rangle$ states.
$\phi$ is the angle of the magnetic field measured from the $x$-axis.
Note that $h>0$ and the direction of $\bm{h}=h(\cos\phi,\sin\phi)$ is represented by the angle $\phi$.
In both the easy-plane and easy-axis cases, the groundstate is given by
\begin{align}
|{\rm GS}\rangle &= \frac{1}{N_0} \left[
                   e^{-i\phi} \sqrt{2}h |1\rangle
                 + \left(D+\sqrt{D^2+4h^2}\right) |0\rangle \right. \cr
&~~~~~~~~~\left.
                 + e^{i\phi} \sqrt{2}h |-1\rangle \right],
\end{align}
with
$N_0 = [ 4h^2 + (D+\sqrt{D^2+4h^2})^2 ]^{1/2}$.
We can see that the superposition of the $|\pm 1\rangle$ states is realized by the applied field.
The expectation values of the spin and electric dipole operators are given by
\begin{align}
&
\begin{pmatrix}
\langle S^x \rangle \cr
\langle S^y \rangle
\end{pmatrix}
= \frac{2h}{\sqrt{D^2+4h^2}}
\begin{pmatrix}
x \cr
y
\end{pmatrix}, \cr
&~~~
\langle p_{\rm S}^z \rangle = K^z_{xy} \frac{8h^2}{N_0^2} xy,
\label{eqn:expect}
\end{align}
where $(x,y)=(\cos\phi,\sin\phi)$.
In Eq. (\ref{eqn:expect}), $\langle\cdots\rangle$ represents the expectation value for the groundstate $|{\rm GS}\rangle$.
Note that $\langle S^z\rangle=\langle p_{\rm S}^x\rangle=\langle p_{\rm S}^y\rangle=0$,
while $\langle S^x\rangle\propto h$, $\langle S^y\rangle\propto h$, and $\langle p_{\rm S}^z\rangle\propto h^2$.
We can see that $p_{\rm S}^z\propto O_{xy}$ behaves as $xy$ with respect to the direction of the magnetic moment,
reflecting the quadrupole nature.

For the $x$ and $y$ components, the electric dipole operators are classified in the $\Gamma_5$ representation,
as shown in Table \ref{table:list-tetra}.
Using Table \ref{table:product-tetra},
we can check that the expectation values of these operators are zero for both the $\Gamma_2$ and $\Gamma_5$ groundstates.
Since $p_{\rm S}^x\propto O_{yz}$ and $p_{\rm S}^y\propto O_{zx}$,
their expectation values can be finite when the magnetic field direction is inclined from the $xy$-plane.

In a cubic symmetry, the situation becomes simple since the spin Hamiltonian becomes isotropic for $S\le 3/2$,
i.e., $D=0$.
Under the magnetic field $\bm{H}=H(\sin\theta\cos\phi,\sin\theta\sin\phi,\cos\theta)$,
the spin Hamiltonian is expressed as
\begin{align}
\H = - g\mu_{\rm B}\bm{H}\cdot\bS.
\end{align}
The groundstate is given by
\begin{align}
|{\rm GS}\rangle &= e^{i\phi}\cos^2\left(\frac{\theta}{2}\right) |1\rangle
                 + \frac{1}{\sqrt{2}}\sin\theta |0\rangle \cr
&~~~
                 + e^{-i\phi}\sin^2\left(\frac{\theta}{2}\right) |-1\rangle.
\end{align}
For the cubic ($T$ and $T_d$) point groups,
the dipole operators are classified in the same three-dimensional representation.
The expectation values of the spin and electric dipole operators are given by
\begin{align}
\begin{pmatrix}
\langle S^x \rangle \cr
\langle S^y \rangle \cr
\langle S^z \rangle
\end{pmatrix}
=
\begin{pmatrix}
x \cr
y \cr
z
\end{pmatrix},~~~
\begin{pmatrix}
\langle p_{\rm S}^x \rangle \cr
\langle p_{\rm S}^y \rangle \cr
\langle p_{\rm S}^z \rangle
\end{pmatrix}
=K
\begin{pmatrix}
yz \cr
zx \cr
xy
\end{pmatrix},
\end{align}
where $(x,y,z)=(\sin\theta\cos\phi,\sin\theta\sin\phi,\cos\theta)$ and $K$ is an arbitrary constant.
In the cubic symmetry, there is no magnetic anisotropy for $S\le 3/2$.
Then, the local moment is smoothly rotated by the external field.
This gives rise to the quadrupole moment and results in a finite electric dipole.
Thus, both the spin and the electric dipole are induced instantaneously
and are controlled simultaneously by the external magnetic field.
It will be interesting to check this point by experiments on real materials.

The above magnetoelectric effect is discussed on the basis of a local spin;
however, note that it can also be applied to interacting spin systems.
The spontaneous magnetic moments at neighboring sites can give rise to a molecular field,
which plays the role of an external magnetic field at the local spin site.

\subsubsection{$S=3/2$}

For $S=3/2$, the energy eigenstates are classified in the $\Gamma_6$ and $\Gamma_7$ representations,
as shown in Table \ref{table:spin-tetra}.
This is realized by the spin Hamiltonian of $D(S^z)^2$,
where the $|\pm\frac{1}{2}\rangle$ ($\Gamma_6$) states are separated from the $|\pm\frac{3}{2}\rangle$ ($\Gamma_7$) states.
The $p_{\rm S}^z$ operator for $D_{2d}$ has the $\Gamma_4$ character
and its expectation value is zero for both the $\Gamma_6$ and $\Gamma_7$ states,
since $\Gamma_6\otimes\Gamma_6$ and $\Gamma_7\otimes\Gamma_7$ do not contain the $\Gamma_4$ representation,
as shown in Table \ref{table:product-tetra}.
On the other hand, we can see that $\Gamma_6\otimes\Gamma_7$ contains $\Gamma_4$.
This means that mixing of the $\Gamma_6$ ($|\pm\frac{1}{2}\rangle$) and $\Gamma_7$ ($|\mp\frac{3}{2}\rangle$) states
is essential for a finite expectation value of $p_{\rm S}^z$.
This can be realized by applying an external magnetic field in the $xy$-plane
and we consider the Hamiltonian given by Eq. (\ref{eqn:H-D-h}).
For $S=3/2$, the groundstate is given by
\begin{align}
|{\rm GS}\rangle &= \frac{1}{N_0} \left[
  e^{-i\frac{3}{2}\phi} \sqrt{3}h \left|\frac{3}{2}\right\rangle
+ e^{-i\frac{1}{2}\phi} f(D,h) \left|\frac{1}{2}\right\rangle \right. \cr
&~~~\left.
+ e^{i\frac{1}{2}\phi} f(D,h) \left|\frac{-1}{2}\right\rangle
+ e^{i\frac{3}{2}\phi} \sqrt{3}h \left|-\frac{3}{2}\right\rangle \right]
\end{align}
with
$f(D,h)=2D+h+2\sqrt{D^2+Dh+h^2}$ and $N_0=[ 6h^2 + 2f^2(D,h) ]^{1/2}$.
We can see that the superposition of the $|\pm\frac{1}{2}\rangle$ and $|\mp\frac{3}{2}\rangle$ states
is realized by the field.
The expectation values of the spin and electric dipole operators are expressed as
\begin{align}
&
\begin{pmatrix}
\langle S^x \rangle \cr
\langle S^y \rangle
\end{pmatrix}
= \frac{D+2h+\sqrt{{D^2+Dh+h^2}}}{2\sqrt{{D^2+Dh+h^2}}}
\begin{pmatrix}
x \cr
y
\end{pmatrix}, \cr
&~~~
\langle p_{\rm S}^z \rangle = K^z_{xy} \frac{3h}{\sqrt{D^2+Dh+h^2}} xy,
\end{align}
where $(x,y)=(\cos\phi,\sin\phi)$.
Note that $\langle S^z\rangle=\langle p_{\rm S}^x\rangle=\langle p_{\rm S}^y\rangle=0$.
The result is essentially the same as that in the $S=1$ case.

It was reported that finite electric polarization was observed in multiferroic systems of akermanite compounds,
such as \Ba~and \Sr,
\cite{Murakawa-2010,Murakawa-2012,Akaki-2012}
where the Co$^{2+}$ ion has the $S=3/2$ spin and is located at the center of a tetrahedron.
The point group at the Co$^{2+}$ site is $D_{2d}$.
These compounds show a long-range ordering at the N\'{e}el temperature accompanied by a collinear antiferromagnetic (AF) structure.
Owing to the easy-plane ($D>0$) single-ion anisotropy, the energy level scheme is $\Gamma_6$-$\Gamma_7$
and the magnetic moment aligns in the $xy$-plane.
The AF moment tends to align perpendicular to the external magnetic field applied in the $xy$-plane.
In this case, the magnetic field $\bm{h}=h(\cos\phi,\sin\phi,0)$ in Eq. (\ref{eqn:H-D-h}) is understood
as an effective field originating from both the molecular field and the external field.
By changing the direction of the external magnetic field,
it was observed that the polarization (electric dipole) actually behaves as $\langle p_{\rm S}^z\rangle\propto xy\propto\sin(2\phi)$,
and it was successfully explained by the metal-ligand hybridization model.
\cite{Murakawa-2010,Murakawa-2012,Akaki-2012}
Our study understands this group-theoretically.

\subsubsection{$S=2$ and $S=5/2$}

The energy eigenstates for $S=2$ and $S=5/2$ spins are listed in Table \ref{table:spin-tetra}.
For $S=2$, there are $\Gamma_1$, $\Gamma_3$, and $\Gamma_4$ states in addition to the $\Gamma_5$ state.
However, the $\Gamma_3$ and $\Gamma_4$ states do not contribute to the appearance of $p_{\rm S}^z$,
since the products of $\Gamma_3\otimes\Gamma_4$, $\Gamma_3\otimes\Gamma_5$, and $\Gamma_4\otimes\Gamma_5$
do not contain the $\Gamma_4$ representation for the $p_{\rm S}^z$ operator (see Table \ref{table:product-tetra}).
As in the $S=1$ case, the $\Gamma_5$ state ($|\pm 1\rangle$) results in a finite $\langle p_{\rm S}^z \rangle$ for $D_{2d}$ symmetry.
Another possibility is a mixed state between the $\Gamma_1$ and $\Gamma_4$ states.

For $S=5/2$, there are two $\Gamma_7$ states.
As in the $S=3/2$ case, $\Gamma_6$ ($|\pm\frac{1}{2}\rangle$) and
$\Gamma_7$ ($a|\mp\frac{3}{2}\rangle+b|\pm\frac{5}{2}\rangle$ and $a|\pm\frac{5}{2}\rangle-b|\mp\frac{3}{2}\rangle$) mixing
is essential for a finite value of $\langle p_{\rm S}^z\rangle$.

In the $D_{2d}$ point group, the electric dipole operators are described by quadrupole operators
such as $p_{\rm S}^x\propto O_{yz}$, $p_{\rm S}^y\propto O_{zx}$, and $p_{\rm S}^z\propto O_{xy}$.
Their expectation values can be finite when the magnetic moment has $yz$, $zx$, and $xy$ components, respectively.

\subsubsection{$J=5/2$ and $J=4$ in cubic symmetry}

In the above discussion, we focused on $d$-electron systems
and studied the electric dipole operator described by the spin operators.
Note that the formulation can also be applied to $f$-electron systems
by replacing the spin operators with total angular momentum operators, i.e., $\bm{S}\rightarrow \bm{J}$,
where the spin-orbit interaction is strong and the total angular momentum is a good quantum number.
In this subsection, we discuss the $J=5/2$ and $J=4$ cases for Ce$^{3+}$ and Pr$^{3+}$ ions, respectively.

In a cubic system without the inversion symmetry, for the $T_d$ and $T$ point groups,
the electric dipole operators are expressed by the quadrupole operators as (see Table \ref{table:P})
\begin{align}
\begin{pmatrix}
p_{\rm S}^x \cr
p_{\rm S}^y \cr
p_{\rm S}^z
\end{pmatrix}
= K
\begin{pmatrix}
O_{yz} \cr
O_{zx} \cr
O_{xy}
\end{pmatrix}
= K
\begin{pmatrix}
J^y J^z + J^z J^y \cr
J^z J^x + J^x J^z \cr
J^x J^y + J^y J^x
\end{pmatrix}.
\end{align}
This indicates that the electric dipole can emerge in the quadrupole ordered phase
of the ($O_{yz},O_{zx},O_{xy}$) type.

In the $T_d$ point group, the electric dipole is classified in the $\Gamma_5$ representation,
as shown in Table \ref{table:list-Td}.
The crystal-field energy levels are classified as in Table \ref{table:spin-Td}.
For $J=5/2$, the electric dipole moment can emerge for the $\Gamma_8$ groundstate,
since $\Gamma_8\otimes\Gamma_8$ contains the $\Gamma_5$ representation, as shown in the multiplication table for $T_d$.
\cite{Koster-1963}
For $J=4$, note that this can be realized by both the $\Gamma_4$ and $\Gamma_5$ groundstates.
\cite{Koster-1963}
Another possibility for $J=4$ is a $\Gamma_1$-$\Gamma_5$ low-energy level scheme.
When the energy splitting of the two levels is much smaller than
the intersite quadrupole interaction of the $\Gamma_5$ type,
the system can be regarded as a pseudoquartet system
and the quadrupole long-range ordered phase can be stabilized at low temperatures.
In this case, we can expect that the electric dipole moment is also induced spontaneously by the quadrupole moment
at the Pr$^{3+}$ site in the $T_{d}$ point-group symmetry.

\begin{table}[t]
\caption{
Classification of magnetic dipole, electric dipole, and electric quadrupole operators for $T_d$ point group.
$J^\alpha$ and $p_{\rm S}^\alpha$ represent the $\alpha~(=x,y,z)$ component
of the total angular momentum and electric dipole operators, respectively.
The quadrupole operators in the same representation of the electric dipole are also shown,
which are defined as $O_{\alpha\beta}=J^\alpha J^\beta + J^\beta J^\alpha$.
Note that the $\Gamma_5$ electric dipole and quadrupole are indistinguishable
in the $T_d$ point group without the inversion symmetry.
The electric dipole is then expressed by a linear combination of the quadrupole operators in the same representation.
The coefficients of the linear combination are listed in Table I.
}
\begin{tabular}{ccccc}
\hline
$T_d$ & Magnetic dipole & Electric dipole & Electric quadrupole \cr
\hline
$T_1$ $\Gamma_4$ & ($J^x,J^y,J^z$) & \cr
$T_2$ $\Gamma_5$ &                 & ($p_{\rm S}^x,p_{\rm S}^y,p_{\rm S}^z$) & ($O_{yz},~O_{zx},~O_{xy}$) \cr
\hline
\end{tabular}
\label{table:list-Td}
\end{table}

\begin{table}[t]
\caption{
Classification of the local energy eigenstates for the $T_d$ point group.
$|m\rangle$ represents the spin state for $J^z=m$.
}
{\tiny
\begin{tabular}{ccc}
\hline
$T_d$                        & $J=5/2$ & $J=4$ \cr
\hline
$A_1$ $\Gamma_1$             & & $\sqrt{\frac{5}{24}}|4\rangle+\sqrt{\frac{7}{12}}|0\rangle+\sqrt{\frac{5}{24}}|-4\rangle$ \cr
$A_2$ $\Gamma_2$             & \cr
$E$~~$\Gamma_3$              & & $\sqrt{\frac{7}{24}}|4\rangle-\sqrt{\frac{5}{12}}|0\rangle+\sqrt{\frac{7}{24}}|-4\rangle$ \cr
                             & & $\sqrt{\frac{1}{2}}|2\rangle+\sqrt{\frac{1}{2}}|-2\rangle$ \cr
$T_1$ $\Gamma_4$             & & $\sqrt{\frac{1}{8}}|\pm 3\rangle+\sqrt{\frac{7}{8}}|\mp 1\rangle$ \cr
                             & & $\sqrt{\frac{1}{2}}|4\rangle-\sqrt{\frac{1}{2}}|-4\rangle$ \cr
$T_2$ $\Gamma_5$             & & $\sqrt{\frac{7}{8}}|\pm 3\rangle-\sqrt{\frac{1}{8}}|\mp 1\rangle$ \cr
                             & & $\sqrt{\frac{1}{2}}|2\rangle-\sqrt{\frac{1}{2}}|-2\rangle$ \cr
\hline
$E_{\frac{1}{2}}$ $\Gamma_6$             & \cr
$E_{\frac{5}{2}}$ $\Gamma_7$ & $\sqrt{\frac{1}{6}}|\pm\frac{5}{2}\rangle-\sqrt{\frac{5}{6}}|\mp\frac{3}{2}\rangle$ \cr
$G_{\frac{3}{2}}$ $\Gamma_8$ & $\pm\sqrt{\frac{5}{6}}|\pm\frac{5}{2}\rangle\pm\sqrt{\frac{1}{6}}|\mp\frac{3}{2}\rangle$ \cr
                             & $\pm|\pm\frac{1}{2}\rangle$ \cr
\hline
\end{tabular}}
\label{table:spin-Td}
\end{table}

The important point is that the electric quadrupole and dipole are categorized
in the same representation of the $T_d$ point group.
There is no symmetry transformation to distinguish them in the absence of the inversion symmetry.
For the $T$ point group, the magnetic dipole also belongs to the same representation
in addition to the electric dipole and quadrupole.
Since the electric field couples to the electric dipole,
applying an external electric field can control multipole orderings,
such as the quadrupole ordering in $f$-electron systems.
A larger coefficient $K_{\beta \gamma}^\alpha$ for an electric dipole is expected in $f$-electron systems
with a rather strong spin-orbit interaction.
We emphasize that appropriate $f$-electron systems can be good candidates for the detection of an emergent electric dipole
and thus have great potential for future application to electromagnetic control.

\subsection{Electromagnon excitation}

In conventional magnets, light absorption is caused by the magnetic field component.
In contrast, magnetic excitation caused by the electric field component is termed electromagnon excitation.
When an electric dipole is described by the spin operators, the electric field is connected to spin systems.
This is the origin of the electromagnon excitation.
In this subsection, we study the selection rule of the electromagnon excitation, focusing on the $D_{2d}$ point group.

\subsubsection{$S=1$}

In the case of $S=1$, the energy levels split into the $\Gamma_2$ and $\Gamma_5$ states, as shown in Table \ref{table:spin-tetra}.
In the multiplication table, we can see that the $\Gamma_5$ operator connects the $\Gamma_2$ and $\Gamma_5$ states.
Since $p_{\rm S}^x$ and $p_{\rm S}^y$ are classified in the $\Gamma_5$ representation,
an electromagnon can be excited when an alternating electric field is applied in the $xy$-plane.
In addition, $S^x$ and $S^y$ are also classified in the same $\Gamma_5$ representation (see Table \ref{table:list-tetra}).
Therefore, the $\Gamma_2$-$\Gamma_5$ transition can be realized by both magnetic and electric dipole operators.
This leads to a cross-correlation and results in directional dichroism,
\cite{Miyahara-2011}
as will be discussed in the next subsection.

In the case of the $D_{2d}$ point group, $p_{\rm S}^z$ is classified in the $\Gamma_4$ irreducible representation
(see Table \ref{table:list-tetra}).
Since $\Gamma_5\otimes\Gamma_5$ contains $\Gamma_4$, as shown in Table \ref{table:product-tetra},
the $\Gamma_4$ operator can connect the $\Gamma_5$ states.
This means that the $\Gamma_5$-$\Gamma_5$ transition is an electromagnon excitation caused by $p_{\rm S}^z$.
The specific form of $p_{\rm S}^z$ is given by Eq. (\ref{eqn:Pz-d2d}) for $S=1$.
We can see that it actually connects the $|\pm 1\rangle$ states.
The $\Gamma_2$-$\Gamma_5$ excitation is possible by the $\Gamma_5$ operators, as mentioned above.

\subsubsection{$S \ge 3/2$}

For $S=3/2$, the energy levels split into $\Gamma_6$ and $\Gamma_7$ (see Table \ref{table:spin-tetra}).
The $\Gamma_5$ ($p_{\rm S}^x,p_{\rm S}^y)$ operators can have finite matrix elements for all combinations of the two states,
while the $\Gamma_4$ ($p_{\rm S}^z$) operator is finite only between $\Gamma_6$ and $\Gamma_7$.

\begin{figure}[t]
\begin{center}
\includegraphics[width=6.5cm,clip]{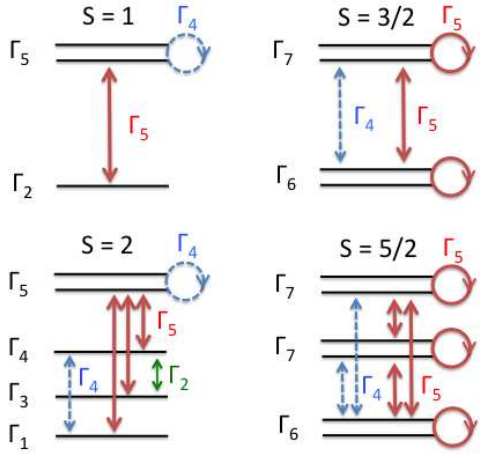}
\end{center}
\caption{
(Color online)
Schematic of selection rules of $S=1$, $S=3/2$, $S=2$, and $S=5/2$ states
for tetragonal symmetry ($D_4$, $C_{4v}$, $D_{2d}$).
The wave functions of the energy levels are listed in Table \ref{table:spin-tetra}.
Note that the energy positions are arbitrary and are determined by the spin Hamiltonian of the single-ion anisotropy.
The following are the irreducible representations of the operators causing the transitions
(see Table \ref{table:list-tetra}).
$\Gamma_2$: $S^z$ (thin green arrow),
$\Gamma_4$: $p_{\rm S}^z$ (dashed blue arrow),
$\Gamma_5$: ($S^x,S^y)$, ($p_{\rm S}^x,p_{\rm S}^y$) (thick red arrow).
Here, $S^\alpha$ is a magnetic dipole operator, while $p_{\rm S}^\alpha$ is an electric dipole operator.
The $\Gamma_4$ ($p_{\rm S}^z$) operator only appears for the $D_{2d}$ point group.
Since the magnetic and electric dipoles are classified in the same $\Gamma_5$ representation,
we can expect cross-correlation effects in the transition caused by the $\Gamma_5$ operator.
Directional dichroism is discussed in Sect. \ref{sec:dichroism} as an example.
For integer spins, the degenerate $\Gamma_5$ states are connected by the $\Gamma_4$ operator.
For half-integer spins, the degenerate $\Gamma_6$ states are connected by the $\Gamma_5$ operator.
This also holds for the degenerate $\Gamma_7$ states (see Table \ref{table:product-tetra}).
In the presence of an external magnetic field, the degenerate $\Gamma_5$, $\Gamma_6$, and $\Gamma_7$ states are split.
In such a case, the $\Gamma_4$ and $\Gamma_5$ operators determine the selection rule
inside the $\Gamma_5$, $\Gamma_6$, and $\Gamma_7$ states.
}
\label{fig:e-magnon}
\end{figure}

In Fig. \ref{fig:e-magnon}, we summarize the selection rule of the electromagnon excitation for $S\ge 1$.
For $S=2$, the $\Gamma_4$ ($|2\rangle - |-2\rangle$) state cannot be excited
from the $\Gamma_1$ ($|0\rangle$) state by a magnetic dipole,
while it can be excited by the $\Gamma_4$ electric dipole ($p_{\rm S}^z$) in $D_{2d}$ symmetry (see Table \ref{table:list-tetra}).
Under a finite field along the $z$-direction, the $\Gamma_4$ and $\Gamma_5$ states are mixed
and both can be excited by the electric field component.
Thus, the transition forbidden by the magnetic dipole can be excited by the electric dipole.
This indicates that we have to be careful when we analyze the intensity of electron spin resonance
in the absence of the inversion symmetry,
where the conventional selection rule by the magnetic dipole cannot be applied.

For $S=5/2$, we discuss the electromagnon excitation observed in \Ba~in connection with the directional dichroism
in the last part of the following subsection.

\subsection{Directional dichroism}
\label{sec:dichroism}

In conventional materials, an electromagnetic wave propagates in the same way when its propagating direction is reversed.
In multiferroic materials, it can propagate differently.
This is known as directional dichroism and is one of the typical signatures of magnetoelectric effects.

As in the above discussion, an electromagnon can be excited between the $\Gamma_2$ and $\Gamma_5$ states for $S=1$.
As shown in Table \ref{table:list-tetra},
($S^x,S^y)$ and ($p_{\rm S}^x,p_{\rm S}^y$) are classified in the same $\Gamma_5$ representation.
This indicates that the transition between the $\Gamma_2$ and $\Gamma_5$ states can be induced
by both electric and magnetic field components of light.
We can expect an interference effect between the two origins.
This appears as the directional dichroism.
\cite{Miyahara-2011}

As a fundamental example, we study the $S=1$ case in the $D_{2d}$ point group.
The spin and electric dipole operators are expressed by the following matrix forms:
\begin{align}
&S^x =
\begin{pmatrix}
0 & \frac{1}{\sqrt{2}} & 0 \cr
\frac{1}{\sqrt{2}} & 0 & \frac{1}{\sqrt{2}} \cr
0 & \frac{1}{\sqrt{2}} & 0
\end{pmatrix},~~~
S^y =
\begin{pmatrix}
0 & \frac{-i}{\sqrt{2}} & 0 \cr
\frac{i}{\sqrt{2}} & 0 & \frac{-i}{\sqrt{2}} \cr
0 & \frac{i}{\sqrt{2}} & 0
\end{pmatrix}, \cr
&p_{\rm S}^x =K
\begin{pmatrix}
0 & \frac{-i}{\sqrt{2}} & 0 \cr
\frac{i}{\sqrt{2}} & 0 & \frac{i}{\sqrt{2}} \cr
0 & \frac{-i}{\sqrt{2}} & 0
\end{pmatrix},~
p_{\rm S}^y =K
\begin{pmatrix}
0 & \frac{1}{\sqrt{2}} & 0 \cr
\frac{1}{\sqrt{2}} & 0 & \frac{-1}{\sqrt{2}} \cr
0 & \frac{-1}{\sqrt{2}} & 0
\end{pmatrix}.
\end{align}
Here, $K$ is a constant.
We assume $K>0$ for a simple discussion below.
We consider the following Hamiltonian:
\begin{align}
\H = D (S^z)^2 - h_z S^z + \H'.
\end{align}
Here, $D~(>0)$ represents the easy-plane single-ion anisotropy.
$h_z=g\mu_{\rm B} H_z(>0)$ represents a static magnetic field applied along the positive $z$-direction.
$\H'$ is the perturbation Hamiltonian given by
\begin{align}
\H' = - \bm{E}^\omega \cdot \bm{p}_{\rm S} - g\mu_{\rm B} \bm{H}^\omega \cdot \bm{S}.
\end{align}
Here, $\bm{E}^\omega = (E_x^\omega,E_y^\omega,0)$ and $\bm{H}^\omega = (H_x^\omega,H_y^\omega,0)$
represent the alternating electric and magnetic fields, respectively.
We introduce the following effective fields for simplicity:
\begin{align}
&\bm{e}^\omega=(e_x^\omega,e_x^\omega,0)=K(E_x^\omega,E_x^\omega,0), \cr
&\bm{h}^\omega=(h_x^\omega,h_x^\omega,0)=g\mu_{\rm B}(H_x^\omega,H_x^\omega,0).
\end{align}
The energy eigenstates of the unperturbed Hamiltonian are given by
\begin{align}
&\omega_0 = 0,~~~~~~~~~~~~~|0\rangle, \cr
&\omega_1 = D - h_z,~~~~~|1\rangle, \cr
&\omega_{-1} = D + h_z,~~~|-1\rangle.
\end{align}
The matrix elements between the $|0\rangle$ and $|\pm 1\rangle$ states are calculated as
\begin{align}
&\langle 1| \H' |0\rangle
= - \frac{1}{\sqrt{2}} [ ( e_y^\omega + h_x^\omega ) - i ( e_x^\omega + h_y^\omega ) ], \cr
&\langle-1| \H' |0\rangle
= - \frac{1}{\sqrt{2}} [ ( - e_y^\omega + h_x^\omega ) - i ( e_x^\omega - h_y^\omega ) ].
\end{align}
The transition probability, i.e., the absorption rate of light, is given by
\begin{align}
P(\omega) &= I_1 \delta( \omega - \omega_1 ) + I_{-1} \delta( \omega - \omega_{-1} ),
\end{align}
where $I_{\pm 1}$ represent the intensities for the $|\pm 1\rangle$ states, respectively.
They are given by
\begin{align}
&I_{\pm 1}
= \pi [ (\bm{e}^\omega)^2 + (\bm{h}^\omega)^2 \pm 2 ( e_y^\omega h_x^\omega + e_x^\omega h_y^\omega ) ].
\label{eqn:I-pm}
\end{align}

\begin{figure}[t]
\begin{center}
\includegraphics[width=8cm,clip]{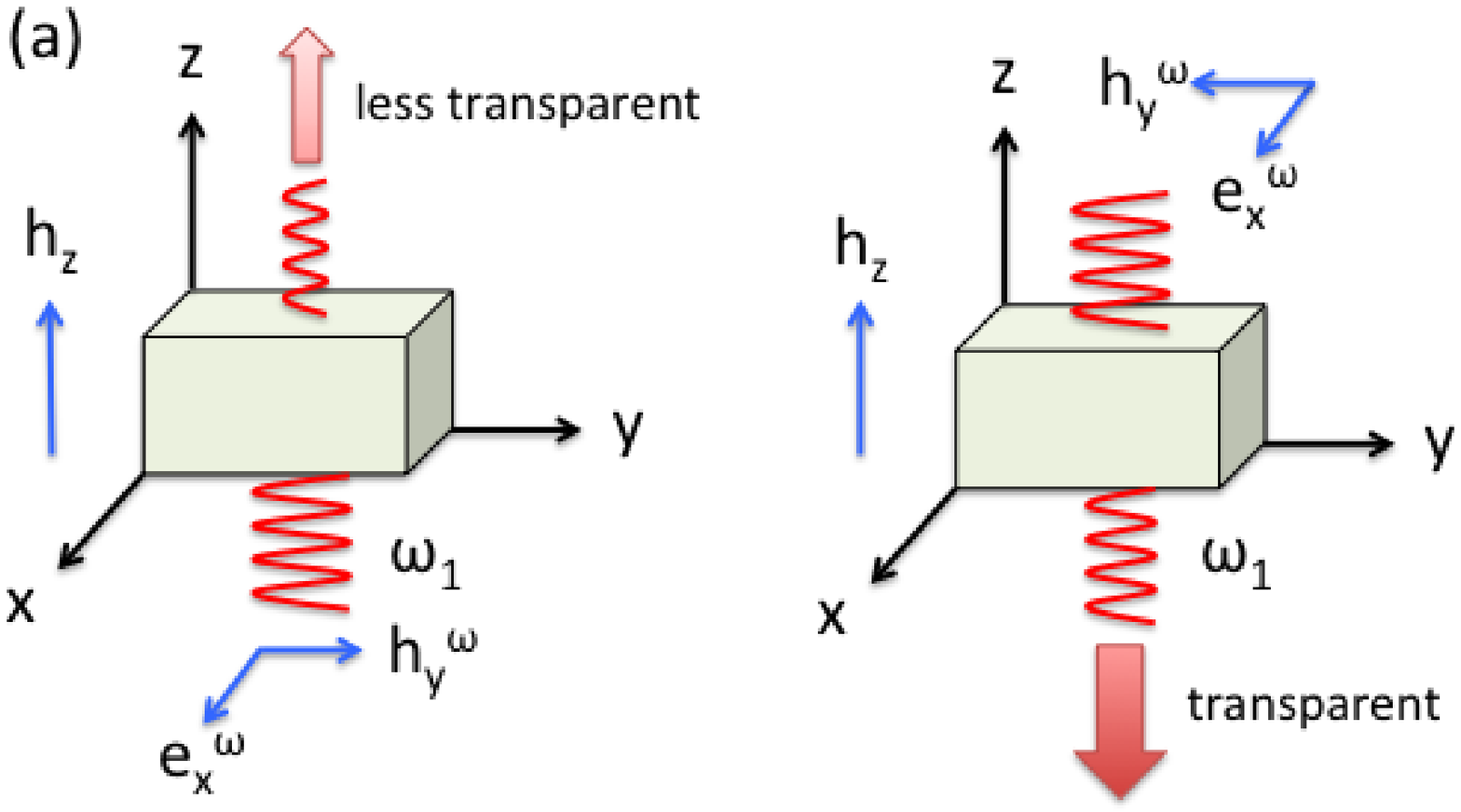}
\includegraphics[width=8.5cm,clip]{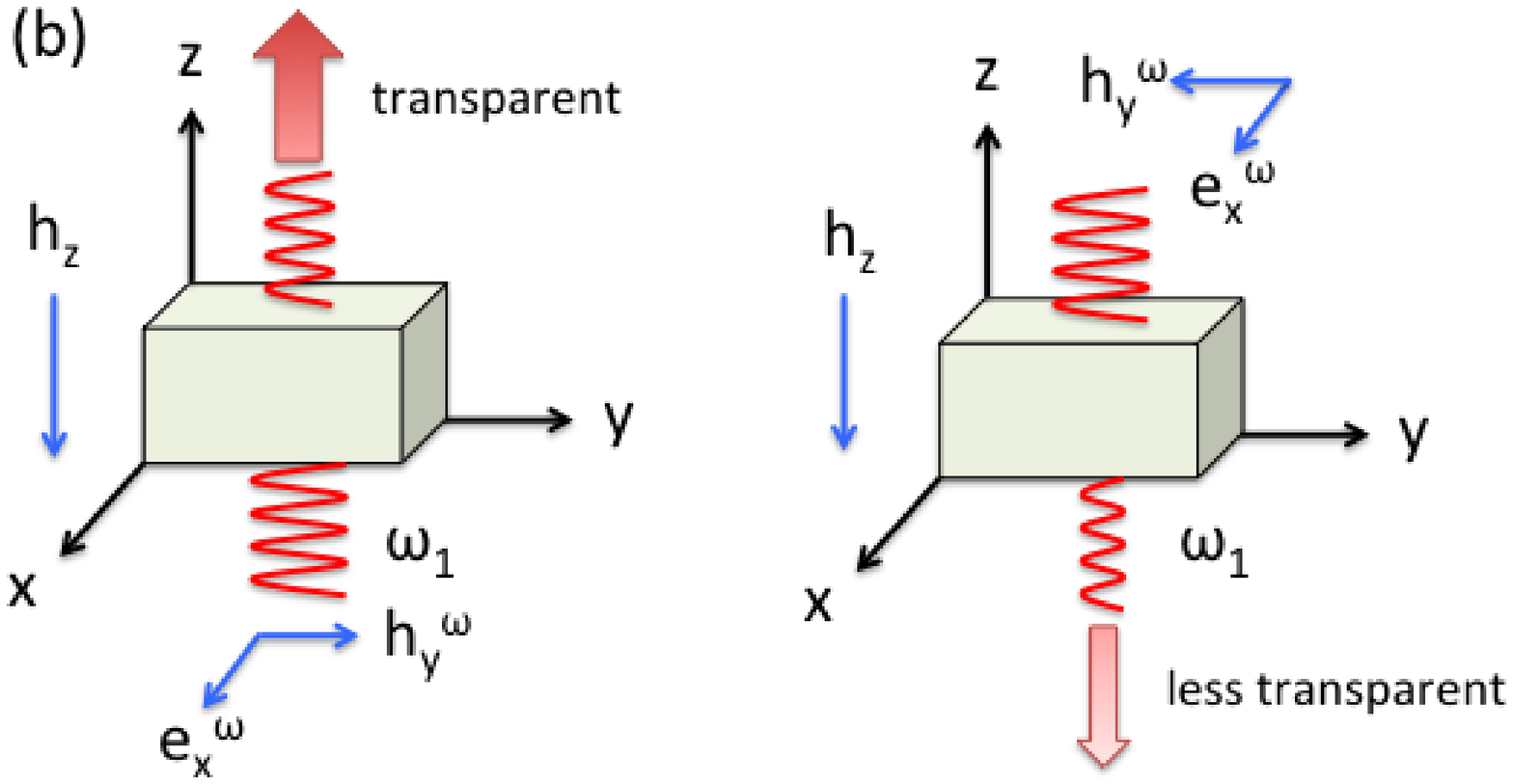}
\includegraphics[width=5cm,clip]{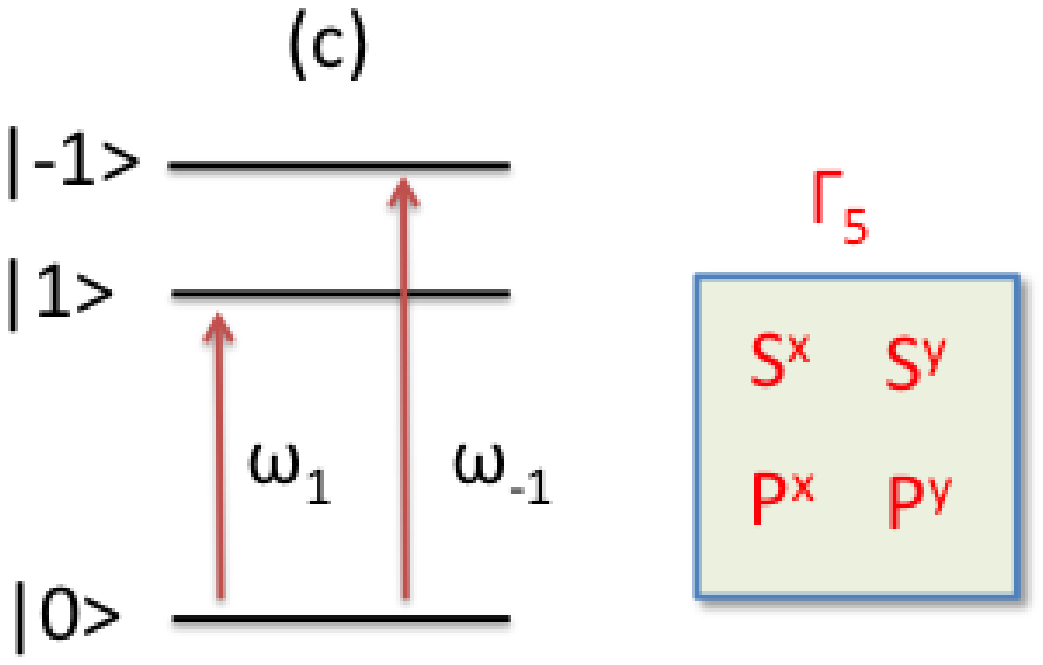}
\end{center}
\caption{
(Color online)
Schematic of directional dichroism for tetragonal ($D_4$, $C_{4v}$, $D_{2d}$) $S=1$ systems.
A static magnetic field is applied along the $z$-direction.
(a) For $\omega_1=D-h_z$.
The lower-lying $|1\rangle$ state is excited by both electric and magnetic fields of light.
The light propagating in the negative $z$-direction is more transparent.
(b) When the direction of the static magnetic field is reversed,
the light propagating in the positive $z$-direction is more transparent.
(c) Schematic of magnetoelectric excitations.
The magnetic ($S^x,S^y$) and electric ($p_{\rm S}^x,p_{\rm S}^y$) dipoles are classified in the same representation.
They have finite matrix elements between the $|0\rangle$ and $|\pm 1\rangle$ states.
This is the origin of the cross-correlation for the directional dichroism.
}
\label{fig:dichroism}
\end{figure}

Let us consider light propagating parallel to the $z$-direction.
We assume that the direction of the electric field is fixed in the $x$-direction,
i.e., $e_x^\omega>0$, $e_y^\omega=0$, and $h_x^\omega=0$.
The sign of the magnetic field is $h_y^\omega>0$ ($h_y^\omega<0$) for the light
propagating in the positive (negative) $z$-direction.
When the frequency of light is tuned to $\omega=\omega_1=D-h_z$, the light is absorbed
accompanied by a $|0\rangle\rightarrow |1\rangle$ transition.
In Eq. (\ref{eqn:I-pm}), we can see that the intensity for this transition, $I_1$, depends on the sign of $h_y^\omega$.
Therefore, the transparency of light depends on the propagating direction.
The light propagating in the positive direction is absorbed strongly (less transparent).
When the propagating direction is reversed, it is less absorbed (more transparent).
Therefore, directional dichroism appears, as shown in Fig. \ref{fig:dichroism}(a).
The asymmetry of the absorption rate is given by
\begin{align}
\frac{I_1(h_y^\omega>0) - I_1(h_y^\omega<0)}{I_1(h_y^\omega>0) + I_1(h_y^\omega<0)}
\simeq \frac{2 e_x^\omega}{h_y^\omega}
= \frac{2 K E_x^\omega}{g\mu_{\rm B} H_y^\omega},
\end{align}
where we used $e_x^\omega / h_y^\omega \ll 1$ assuming a small $K$ value.
This expresses the strength of the directional dichroism.
A larger $K$ (stronger spin-orbit interaction) is favorable for dichroism.
When the direction of the external magnetic field is reversed ($h_z\rightarrow -h_z$),
the absorption rate changes from $I_1$ to $I_{-1}$ with the same $\omega_1=D-|h_z|$.
The directional dichroism changes its direction, as shown in Fig. \ref{fig:dichroism}(b).
This phenomenon can be used as an ``optical diode" with which we can control the direction of the transparent light.
In the present study, we can control the easy-transparent direction
by changing the direction of the external magnetic field.
By tuning the frequency $\omega=\omega_{-1}=D+h_z$,
the direction is reversed in the dichroism compared with the $\omega_1$ case,
since the $|-1\rangle$ state is excited instead of the $|1\rangle$ state.

The above result is owing to the fact that the excitation is caused by both the magnetic and electric dipoles,
which are classified in the same irreducible representation in the present case.
Their cross-correlation leads to the directional dichroism, as summarized in Fig. \ref{fig:dichroism}(c).

In the case of $S=3/2$, we can also expect directional dichroism, as in the $S=1$ case.
The local energy levels split into the $\Gamma_6$ and $\Gamma_7$ states, as shown in Fig. \ref{fig:e-magnon} for $S=3/2$.
The transition between them is caused by both the $\Gamma_4$ ($p_{\rm S}^z$) and $\Gamma_5$ ($S^x,S^y,p_{\rm S}^x,p_{\rm S}^y$) operators.
Therefore, the transition can be induced by both magnetic and electric components
and we can expect directional dichroism in $S=3/2$ systems.
Directional dichroism was actually observed in \Ba.
\cite{Kezsmarki-2011,Kezsmarki-2014}
Details of the microscopic theory for the directional dichroism in \Ba~were presented convincingly by Miyahara and Furukawa
on the basis of the metal-ligand hybridization model.
\cite{Miyahara-2011}
As pointed out in Ref. \ref{ref:Miyahara-2011}, this idea can also be applied to other symmetries.
The spin-dependent electric dipole operator listed in Table \ref{table:P} for various point-group symmetries
is very useful for deducing relevant spin dependences in specific magnetic materials.

\section{Magnetoelectric Effect by Two Spins}

In this section, we study an electric dipole generated by a pair of spins.
As shown in Fig. \ref{fig:spin}, the $z$-axis is chosen along the two spins.
We first discuss a case with the inversion symmetry between the two spins in Sect. \ref{sec:presence}
and apply the result to a typical example of an interacting spin dimer system, \Tl.
A case without the inversion symmetry is also discussed in Sect. \ref{sec:absence}.

\subsection{In the presence of inversion symmetry}
\label{sec:presence}

When there is an inversion center between the two spins,
the electric dipole operator must have antisymmetric spin dependence.
Therefore, the symmetric spin-dependent component vanishes $(p_{\rm S}=0)$
and only the antisymmetric one ($p_{\rm A}^\alpha$) remains in Eq. (\ref{eqn:P2}).
Since $C^\alpha_{\beta\gamma}=-C^\alpha_{\gamma\beta}$,
the term described by $C^\alpha_{\beta\gamma}$ is expressed by $\bW=\bS_1\times\bS_2$,
\cite{Moriya-1968,Katsura-2005,Kaplan-2011}
which leads to
\begin{align}
p_{\rm A}^\alpha = C^\alpha_\beta W^\beta + B^\alpha_{\beta\gamma} ( S_1^\beta S_1^\gamma - S_2^\beta S_2^\gamma ),
\label{eqn:p-W}
\end{align}
where the coefficients $C^\alpha_\beta$ and $B^\alpha_{\beta\gamma}$
are listed in Table \ref{table:PA} for various symmetries.

\subsubsection{Under high symmetry}

First, we consider a high-symmetry case,
where the two spins have all the symmetries listed in Table \ref{table:PA} in addition to the inversion symmetry.
Retaining the common quadrupole operators in Table \ref{table:PA},
we can express the electric polarization operator in the following form:
\begin{align}
&\begin{pmatrix}
p_{\rm A}^x \cr
p_{\rm A}^y \cr
p_{\rm A}^z
\end{pmatrix}
= -C
\begin{pmatrix}
( \bm{e}_{12}\times\bS_1\times\bS_2 )_x \cr
( \bm{e}_{12}\times\bS_1\times\bS_2 )_y \cr
( \bm{e}_{12}\times\bS_1\times\bS_2 )_z \cr
\end{pmatrix} \label{eqn:P-A-1} \\
&~+
\begin{pmatrix}
B_1 ( O_{zx,1} - O_{zx,2} ) \cr
B_1 ( O_{yz,1} - O_{yz,2} ) \cr
B_2 ( O_{x^2+y^2,1} - O_{x^2+y^2,2} ) + B_3 ( O_{z^2,1} - O_{z^2,2} )
\end{pmatrix}.
\nonumber
\end{align}
Here, $C$, $B_1$, $B_2$, and $B_3$ are arbitrary constants.
$\bm{e}_{12}$ denotes a unit vector connecting the two spins of $\bS_1$ and $\bS_2$,
which is parallel to the $z$-axis, as shown in Fig. \ref{fig:spin}.
Note that the $z$ component vanishes, i.e., $( \bm{e}_{12}\times\bS_1\times\bS_2 )_z = 0$.
The first term in Eq. (\ref{eqn:P-A-1}) represents the fact that $\bm{p}_{\rm A} \propto \bm{e}_{12}\times\bS_1\times\bS_2$.
This coincides with the result obtained by Katsura et al. on the basis of the spin current mechanism,
\cite{Katsura-2005}
since the result was derived under a high symmetry, as pointed out by Kaplan and Mahanti.
\cite{Kaplan-2011}
The second term in Eq. (\ref{eqn:P-A-1}) is written by the antisymmetric form of the quadrupole operators at each site.
The transverse ($B_1$) and longitudinal ($B_2$ and $B_3$) terms of the electric dipole
coincide with those obtained by Jia et al,
\cite{Jia-2006,Jia-2007,note-B3}
since a microscopic model with the spin-orbit interaction was employed under a high symmetry.
\cite{Kaplan-2011}
When the symmetry of the two spins becomes lower, we still retain the electric dipole in Eq. (\ref{eqn:P-A-1}),
although other terms arise in addition, as shown in Table \ref{table:PA}.
This indicates the robustness of Eq. (\ref{eqn:P-A-1}) in general cases.

\subsubsection{Spin dimer system}

\begin{table}[t]
\caption{
Symmetry properties of a local Hamiltonian $J\bS_1\cdot\bS_2$, singlet state, triplet state, $\bS_\pm$, and $\bW$ operators
with respect to the spatial inversion ($I$) and time-reversal ($T$) transformations.
The even and odd characters are represented by $\pm$, respectively.
In terms of the $T$ transformation, $\bm{S}_\pm$ are magnetic, while $\bm{W}$ and $F_{\alpha\beta}$ are electric.
Since the magnetic dipole is staggered on the left and right sides of a dimer,
$\bS_-$ can be interpreted as a magnetic quadrupole (or multipole) (see text below Fig. \ref{fig:multipole}).
Since $\bW$ has an odd character for $I$, it can be interpreted as an electric dipole operator.
In the absence of the inversion center between the two spins, the even and odd parities are mixed.
The symmetric spin-dependent electric dipole operator remains,
which is described by electric quadrupole operators.
}
\begin{tabular}{lcccc}
\hline
 & $I$ & $T$ & $IT$ & Classification \cr
\hline
$J\bS_1\cdot\bS_2$ & $+$ & $+$ & $+$ \cr

\hline
$|s\rangle=\frac{1}{\sqrt{2}}(|\uparrow\downarrow\rangle-|\downarrow\uparrow\rangle)$    & $-$ & $+$ & $-$ \cr
$|t_x\rangle=\frac{-1}{\sqrt{2}}(|\uparrow\uparrow\rangle-|\downarrow\downarrow\rangle)$ & $+$ & $-$ & $-$\cr
$|t_y\rangle=\frac{i}{\sqrt{2}}(|\uparrow\uparrow\rangle+|\downarrow\downarrow\rangle)$  & $+$ & $-$ & $-$\cr
$|t_z\rangle=\frac{1}{\sqrt{2}}(|\uparrow\downarrow\rangle+|\downarrow\uparrow\rangle)$  & $+$ & $-$ & $-$\cr
\hline
$\bS_+=\bS_1 + \bS_2$     & $+$ & $-$ & $-$ & magnetic dipole \cr
$\bS_-=\bS_1 - \bS_2$     & $-$ & $-$ & $+$ & magnetic quadrupole \cr
                          &     &     &     & (magnetic multipole) \cr
\hline
$\bW=\bS_1 \times \bS_2$  & $-$ & $+$ & $-$ & electric dipole \cr
$F_{\alpha\beta}=S_1^\alpha S_2^\beta+S_1^\beta S_2^\alpha$ & $+$ & $+$ & $+$ & electric quadrupole \cr
\hline
\end{tabular}
\label{table:dimer}
\end{table}

Next, we examine the electric dipole moment originating from the vector spin chirality.
As recently revealed by Kimura et al.,
a spin dimer system is a typical system showing the magnetoelectric effect,
\cite{Kimura-2016}
where $S=1/2$ spins are strongly coupled by the exchange interaction $J\bS_1\cdot\bS_2$.
We assume an inversion center between the two spins, such as in \Tl.
In this case, the symmetric spin-dependent electric dipole operator disappears
and only the antisymmetric one remains.
For $S=1/2$, the $B^\alpha_{\beta\gamma}$ term in Eq. (\ref{eqn:p-W}) is irrelevant
since it is proportional to the product of the spin operators at the same spin site.
Therefore, the vector spin chirality is the only source of the electric dipole
for the $S=1/2$ spin dimer with an inversion center.

The energy eigenstates split into singlet and triplet states.
Their characters with respect to the spatial inversion ($I$) and time-reversal ($T$) transformations
are listed in Table \ref{table:dimer},
where we employ the $x$, $y$, and $z$ representations for the triplet states.
In this table, $\bS_\pm=\bS_1\pm\bS_2$ operators are introduced
to represent the uniform and staggered components, respectively.
Since $\bS_\pm$ are odd for the $T$ transformation, they are magnetic, while $\bW$ is nonmagnetic.
$\bS_+$ has an even parity, while $\bS_-$ and $\bW$ have an odd parity.
Thus, the three kinds of operators are distinguishable by the $I$ and $T$ symmetries.
Since $\bW$ has the same character as the electric dipole,
it can be regarded as an electric dipole.
This is the reason why the electric dipole operator is described by the vector spin chirality.
Note that this also holds even in the absence of the inversion center between the two spins.
In this case, the symmetric component of the electric polarization operator
coexists with the antisymmetric component, as expressed in Eq. (\ref{eqn:P2}).

For a weak interdimer interaction, the singlet groundstate is stabilized
and there is no long-range order down to zero temperature.
In the disordered phase, the local Hamiltonian of a dimer is invariant under the inversion transformation.
The energy eigenstates are then classified by even ($|t_x\rangle,|t_y\rangle, |t_z\rangle$) and odd ($|s\rangle$) parities.
The Hamiltonian is invariant when the $I$ and $T$ transformations are performed simultaneously,
i.e., under the $IT$ transformation.
As in Table \ref{table:dimer}, both the singlet and triplet states have an odd character for $IT$.

Next, we consider the $\bS_\pm$ and $\bW$ operators.
Their specific forms are expressed as
\begin{align}
&S_+^\alpha = -i \epsilon_{\alpha\beta\gamma} |t_\beta\rangle\langle t_\gamma|, \cr
&S_-^\alpha = (|t_\alpha\rangle\langle s|) + (|s\rangle\langle t_\alpha|), \cr
&W^\alpha = -i \frac{1}{2} \left[ (|t_\alpha\rangle\langle s|) - (|s\rangle\langle t_\alpha|) \right].
\label{eqn:operator}
\end{align}
Here, $\epsilon_{\alpha\beta\gamma}$ represents the antisymmetric tensor.
Both $\bS_-$ and $\bW$ have finite matrix elements between the singlet and triplet states,
while $\bS_+$ is finite only between the triplet states.

When the interdimer interaction is increased, spin dimer systems show a long-range order
with a staggered magnetic moment on a dimer, as in the pressure-induced ordered phase of \Tl.
\cite{Oosawa-2004,Ruegg-2008,Matsumoto-2004}
In this case, the order parameter is represented by $\bS_-$.
The local Hamiltonian of a dimer is then expressed as
\begin{align}
\H_{\rm dimer} = J\bS_1\cdot\bS_2 - \bm{h}_{\rm AF} \cdot \bS_-.
\label{eqn:H-dimer}
\end{align}
Here, $\bm{h}_{\rm AF}$ denotes the staggered magnetic field from the neighboring sites, which couples to $\bS_-$.
The inversion symmetry is broken in $\H_{\rm dimer}$ owing to the $\bm{h}_{\rm AF}\cdot\bS_-$ term.
The singlet and triplet states are mixed up and the energy eigenstates
are given by the superposition of these states with real coefficients,
since the matrix elements of $\bS_-$ between the singlet and triplet states are real numbers,
as shown in Eq. (\ref{eqn:operator}).
In this case, it may be thought that $\bm{W}$ is induced because of the broken inversion symmetry.
However, it is not, since the matrix elements of $\bW$ between the singlet and triplet states
are pure imaginary numbers, as shown in Eq. (\ref{eqn:operator}).

This is also understood from a symmetrical point of view.
Although the inversion symmetry is broken,
$\H_{\rm dimer}$ given by Eq. (\ref{eqn:H-dimer}) is invariant for the $IT$ transformation
since both $J\bS_1\cdot\bS_2$ and $\bS_-$ are invariant (see Table \ref{table:dimer}).
Thus, the $IT$ transformation is the remaining symmetry of $\H_{\rm dimer}$ in Eq. (\ref{eqn:H-dimer})
and the $\bS_-$ operator is distinguishable from the $\bS_+$ and $\bW$ operators.
The expectation value of $\bS_-$ can be finite, while those of $\bS_+$ and $\bW$ vanish.
The reason for the latter case is explained by the fact that
\begin{align}
\langle {\rm GS}|\bW|{\rm GS}\rangle &= \langle {\rm GS}|(IT)^{-1} (IT) \bW (IT)^{-1} (IT) |{\rm GS}\rangle \cr
&= - \langle {\rm GS}| \bW |{\rm GS}\rangle.
\label{eqn:W}
\end{align}
Here, $|{\rm GS}\rangle$ represents the groundstate.
Note that both $|{\rm GS}\rangle$ and $\bW$ have an odd character for the $IT$ transformation.
Equation (\ref{eqn:W}) indicates that the expectation values of both $\bW$ and $\bS_+$ vanish
because of their odd character for the $IT$ transformation.
Therefore, the electric dipole moment (or $\bW$) is not induced
in the pressure-induced ordered phase of interacting spin dimer systems.

A long-range ordered phase can also be stabilized by an external magnetic field.
This is termed field-induced magnetic order and is interpreted to be a consequence of Bose--Einstein condensation of a magnon.
\cite{Oosawa-1999,Nikuni-2000,Tanaka-2001,Matsumoto-2002,Ruegg-2003}
In the ordered phase, the local Hamiltonian of a dimer is expressed as
\begin{align}
\H_{\rm dimer} = J\bS_1\cdot\bS_2 - \bm{h}_{\rm AF} \cdot \bS_- - \bm{h}_{\rm ex} \cdot \bS_+.
\label{eqn:H-dimer-h}
\end{align}
Here, $\bm{h}_{\rm ex}$ denotes the external magnetic field, which couples to $\bS_+$.
Owing to the $\bm{h}_{\rm ex}\cdot\bS_+$ term, the Hamiltonian is no longer invariant under the $IT$ transformation.
The groundstate is then given by a superposition of the odd and even states for $IT$,
where the even component can be obtained by taking new linear combinations
of the $|t_x\rangle$, $|t_y\rangle$, and $|t_z\rangle$ triplet states with complex coefficients.
The expectation values of $\bS_+$ and $\bW$ can be finite since they can connect the even and odd states for $IT$.
As a result, the electric dipole moment is induced in the field-induced ordered phase.
This was actually observed recently by Kimura et al. in \Tl~under a finite magnetic field.
\cite{Kimura-2016}
In a pyroelectric current measurement, they revealed that an electric polarization is induced
in the Bose--Einstein condensation phase of the magnon above the critical field.
In \Tl, a dimer is located at a low-symmetry position, i.e., it only has the inversion symmetry.
The electric dipole operator is then expressed as
$p_{\rm S}^\alpha = C^\alpha_\beta W^\beta$.
It was confirmed that the induced electric polarization in \Tl~
is proportional to the expectation value of the magnitude of the vector spin chirality, $\langle |\bW| \rangle$.
\cite{Kimura-2016}

\begin{figure}[t]
\begin{center}
\includegraphics[width=8cm,clip]{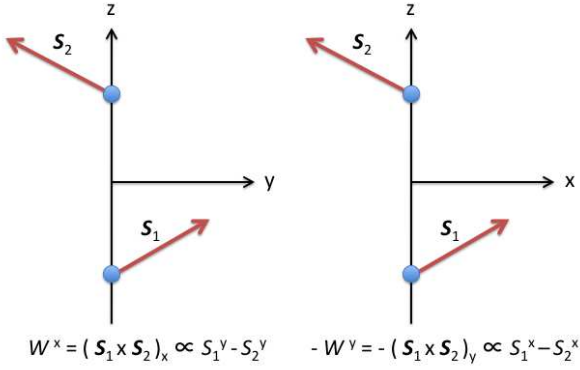}
\end{center}
\caption{
(Color online)
Interpretation of $(W^x,W^y)\propto(S_-^y,-S_-^x)$.
There is a uniform magnetic moment in the $z$-direction.
The staggered component aligns in the $xy$-plane.
The $x$ and $y$ components of $\bW=\bS_1\times\bS_2$ are proportional to
$(S_1^y-S_2^y,-S_1^x+S_2^x)=(S_-^y,-S_-^x)$.
}
\label{fig:W}
\end{figure}

Under a magnetic field parallel to the $z$-direction,
a staggered moment appears in the $xy$-plane in the field-induced ordered phase.
Under the field, the $\bS_-$ and $\bW$ operators are expressed as
\begin{align}
&S_-^x = \frac{1}{\sqrt{2}}[ - (|t_1\rangle\langle s|) + (|t_{-1}\rangle\langle s|) ] + {\rm h.c.}, \cr
&S_-^y = i\frac{1}{\sqrt{2}}[ (|t_1\rangle\langle s|) + (|t_{-1}\rangle\langle s|) ] + {\rm h.c.}, \cr
&W^x = i\frac{1}{2\sqrt{2}}[ (|t_1\rangle\langle s|) - (|t_{-1}\rangle\langle s|) ] + {\rm h.c.}, \cr
&W^y = \frac{1}{2\sqrt{2}}[ (|t_1\rangle\langle s|) + (|t_{-1}\rangle\langle s|) ] + {\rm h.c.},
\end{align}
where
$|t_1\rangle=|\uparrow\uparrow\rangle$,
$|t_0\rangle=\frac{1}{\sqrt{2}}(|\uparrow\downarrow\rangle+|\downarrow\uparrow\rangle)$, and
$|t_{-1}\rangle=|\downarrow\downarrow\rangle$.
Under the field, the $|t_1\rangle$ state is stabilized.
The groundstate is then expected to be a superposition mainly of the $|s\rangle$ and $|t_1\rangle$ states.
When we restrict ourselves to the lowest-lying two states, the $\bW$ operator is written as
\begin{align}
\begin{pmatrix}
W^x \cr
W^y
\end{pmatrix}
=\frac{1}{2}
\begin{pmatrix}
S_-^y \cr
-S_-^x
\end{pmatrix}.
\label{eqn:WS}
\end{align}
This restriction becomes valid for a strong intradimer interaction,
where a large splitting of the singlet-triplet state is realized at zero field.
In this case, the $\bW$ operator becomes equivalent to the $\bS_-$ operator in the $x$ and $y$ components.
A schematic of the magnetic moment in the ordered phase is shown in Fig. \ref{fig:W}
with the interpretation of Eq. (\ref{eqn:WS}).

In general cases, the groundstate of Eq. (\ref{eqn:H-dimer-h}) is expressed in the following form:
\cite{Kimura-2016}
\begin{align}
|{\rm GS}\rangle = a_s|s\rangle - a_1 e^{-i\phi} |t_1\rangle + a_{-1} e^{i\phi} |t_{-1}\rangle.
\label{eqn:GS-H}
\end{align}
Here, $a_s$ and $a_{\pm1}$ are real coefficients
and $\phi$ represents the angle of the staggered moment in the $xy$-plane measured from the $x$-axis.
The expectation values of the operators for the groundstate are given by
\cite{Kimura-2016}
\begin{align}
&
\begin{pmatrix}
\langle S_-^x \rangle \cr
\langle S_-^y \rangle \cr
\end{pmatrix}
=\sqrt{2} a_s ( a_1 + a_{-1} )
\begin{pmatrix}
x \cr
y
\end{pmatrix}, \cr
&~~~\langle S_+^z \rangle = a_1^2 - a_{-1}^2, \cr
&
\begin{pmatrix}
\langle W^x \rangle \cr
\langle W^y \rangle
\end{pmatrix}
= \frac{1}{\sqrt{2}} a_s ( a_1 - a_{-1} )
\begin{pmatrix}
y \cr
-x
\end{pmatrix},
\label{eqn:WS2}
\end{align}
where $(x,y)=(\cos\phi,\sin\phi)$.
Note that
$\langle S_-^z \rangle
=\langle S_+^x \rangle
=\langle S_+^y \rangle
=\langle W^z \rangle
=0$.
When $a_{-1}=0$, Eq. (\ref{eqn:WS2}) reduces to Eq. (\ref{eqn:WS}).
We can see in Eq. (\ref{eqn:WS2}) that $\langle\bW\rangle$ is spontaneously induced
together with the staggered moment in the $xy$-plane in the field-induced ordered phase.
The electric dipole moment is also induced accordingly.
In the absence of the magnetic field, $a_1=a_{-1}$ and the electric dipole moment disappears even in the ordered phase.

For a spin dimer, the expectation value of $\bW$ has the following property:
\begin{align}
\langle \bW \rangle = \langle \bS_1 \times \bS_2 \rangle \neq \langle \bS_1 \rangle \times \langle \bS_2 \rangle.
\end{align}
This is because the two spins are strongly coupled in a dimer and their wavefunction cannot be decoupled.
The observed magnetic field dependence of the induced electric polarization in \Tl~cannot be explained by
$\langle \bS_1 \rangle \times \langle \bS_2 \rangle$.
For the quantitive explanation, we emphasize that the expectation value must be taken as
$\langle \bS_1 \times \bS_2 \rangle$ on the basis of the wavefunction of a dimer.
\cite{Kimura-2016}

\begin{figure}[t]
\begin{center}
\includegraphics[width=8cm,clip]{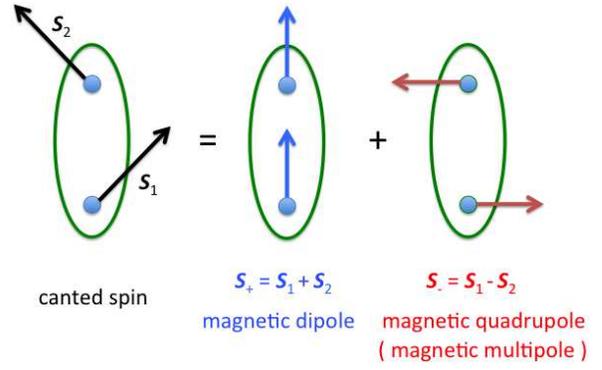}
\end{center}
\caption{
(Color online)
Schematic of the canted magnetic structure in the field-induced magnetic ordered phase.
It can be divided into uniform and staggered components,
which are represented by the $\bS_+=\bS_1+\bS_2$ and $\bS_-=\bS_1-\bS_2$ operators, respectively.
The latter can be interpreted as a magnetic quadrupole (or multipole) (see text).
Only in the presence of both moments, the vector spin chirality, $\bW=\bS_1\times\bS_2$, can be finite.
This induces a finite electric dipole moment.
}
\label{fig:multipole}
\end{figure}

In the conventional classification of multipole operators,
a magnetic (electric) operator has an odd (even) character with respect to the $T$ transformation.
The multipole characters, such as those of the dipole, quadrupole, octupole, etc.,
are determined by the rank of the tensor operator.
In the case of spin-dependent operators, they correspond to
$S^\alpha$, $S^\alpha S^\beta$, $S^\alpha S^\beta S^\gamma$, etc., respectively.
This classification of the multipoles is applied to cases of a single-spin site.
In the spin dimer case, there are two kinds of spin operators ($\bS_1$ and $\bS_2$),
reflecting the two spin sites.
The interesting point of a dimer is that the two spins are strongly coupled by a strong intradimer interaction.
Thus, a dimer can be regarded as a single atom having a wide spatial distribution.
The spin-dependent operators for a dimer are given by combinations of $S_1^\alpha$ and $S_2^\beta$.
Under a uniform external magnetic field, for instance, the interaction is expressed as $-\bm{h}_{\rm ex}\cdot\bS_+$.
Therefore, $\bS_+=\bS_1+\bS_2$ can be understood as a magnetic dipole operator for a dimer.
In the presence of a staggered magnetic field, the interaction is expressed as $-\bm{h}_{\rm AF}\cdot\bS_-$.
Thus, $\bS_-=\bS_1-\bS_2$ is related to higher-rank magnetic multipole operators.
As shown in Table \ref{table:dimer}, it has an odd character for both $I$ and $T$ transformations.
A finite expectation value of the $\bS_-$ operator generates finite staggered magnetic moments
on the left and right sides of the dimer.
Close to one of the spin sites, a finite magnetic dipole moment can be observed at each spin site.
On the other hand, far from the dimer,
the staggered components cancel out and the magnetic field behaves as that from a magnetic multipole moment.
In this sense, $\bS_-=\bS_1-\bS_2$ can be interpreted as an ``extended magnetic quadrupole''.
\cite{Spaldin-2008,Hitomi-2014,Hayami-2016}
Precisely speaking, it contains pseudoscalar, quadrupole, and toroidal magnetic components in general.
\cite{Spaldin-2008}
The present result indicates that the electric dipole moment
is owing to both magnetic dipole and quadrupole moments,
because the canted magnetic moments at the two sites give rise to a finite value of the vector spin chirality,
$\bW=\bS_1\times\bS_2$, as shown in Fig. \ref{fig:multipole}.
As a similar effect caused by magnetic multipoles,
magnetodielectric effect has recently been observed in Ba(TiO)Cu$_4$(PO$_4$)$_4$,
where a magnetic quadrupole moment is present in the magnetic ordered phase.
\cite{Kimura-K-2016}

\subsection{In the absence of inversion symmetry}
\label{sec:absence}

When there is no inversion center between the two spins,
the even-odd parity mixing allows both symmetric ($\bm{p}_{\rm S}$)
and antisymmetric ($\bm{p}_{\rm A}$) spin-dependent components
in the electric dipole operator, as shown in Eq. (\ref{eqn:P2}).
The spin dependences of the electric dipole operator are listed in Table \ref{table:PS} for various symmetries.
Unlike $\bm{p}_{\rm A}$, $\bm{p}_{\rm S}$ does not survive in the high-symmetry case,
since it is not allowed in the presence of the inversion symmetry.
In this sense, $\bm{p}_{\rm S}$ is not robust compared with $\bm{p}_{\rm A}$.

\subsubsection{Antiferromagnetic spin dimer}

Let us consider the electric dipole operator derived from the spin products at different sites,
i.e., the $A^\alpha_{\beta\gamma}$ term in Eq. (\ref{eqn:P2}).
When the two spins have the $C_2(z)$ and $C_2(x)$ symmetries, for instance,
the electric dipole operator is given by (see Tables \ref{table:PS} and \ref{table:PA})
\begin{align}
&(p_{\rm S}^x, p_{\rm S}^y, p_{\rm S}^z) = ( A^x F_{yz}, A^y F_{zx}, A^z F_{xy}), \cr
&(p_{\rm A}^x, p_{\rm A}^y, p_{\rm A}^z) = ( C^x W_y, C^y W_x, 0).
\label{eqn:p-F}
\end{align}
Here, $A^x$, $A^y$, $A^z$, $C^x$, and $C^y$ are arbitrary constants.
As in the caption of Table \ref{table:PS}, $F_{\alpha\beta}$ are defined as
\begin{align}
F_{\alpha\beta} = S_1^\alpha S_2^\beta + S_1^\beta S_2^\alpha.
\end{align}
It is interesting to see the difference from the vector spin chirality $\bW=\bS_1\times \bS_2$.
As listed in Table \ref{table:dimer}, $F_{\alpha\beta}$ has an even parity.
It is classified as an electric quadrupole operator induced by two spins.
In the absence of the inversion symmetry,
$\bm{p}_{\rm S}$ is described by $F_{\alpha\beta}$ owing to the even-odd parity mixing.

Since $F_{\alpha\beta}$ has an even parity, it only has matrix elements between the triplet states given as
\begin{align}
F_{\alpha\beta} = -\frac{1}{2} [ (|t_\alpha\rangle \langle t_\beta|) + (|t_\beta\rangle \langle t_\alpha|) ].
\end{align}
In the pressure-induced ordered phase, the local groundstate is expressed in the following form:
\begin{align}
|{\rm GS}\rangle = a_s |s\rangle + a_t \sin\theta ( \cos\phi |t_x\rangle + \sin\phi |t_y\rangle )
                                 + a_t \cos\theta |t_z\rangle,
\label{eqn:GS-pressure}
\end{align}
where $a_s$ and $a_t$ are real constants.
The expectation values of $\bS_-$ and $\bm{p}_{\rm S}$ are respectively given by
\begin{align}
\begin{pmatrix}
\langle S_-^x \rangle \cr
\langle S_-^y \rangle \cr
\langle S_-^z \rangle
\end{pmatrix}
=  2 a_s a_t
\begin{pmatrix}
x \cr
y \cr
z
\end{pmatrix},~
\begin{pmatrix}
\langle p_{\rm S}^x \rangle \cr
\langle p_{\rm S}^y \rangle \cr
\langle p_{\rm S}^z \rangle
\end{pmatrix}
=  - a_t^2
\begin{pmatrix}
A^x yz \cr
A^y zx\cr
A^z xy
\end{pmatrix},
\label{eqn:p}
\end{align}
where $(x,y,z)=(\sin\theta\cos\phi,\sin\theta\sin\phi,\cos\theta)$.
The direction of the staggered moment is expressed by the angles $\theta$ and $\phi$.
Although there is an antisymmetric operator ($\bm{p}_{\rm A}$) in the absence of the inversion symmetry,
its expectation value is zero, i.e., $\langle \bm{p}_{\rm A}\rangle=0$ (or $\langle \bm{W}\rangle=0$),
as discussed below Eq. (\ref{eqn:W}).
\cite{note-W}
We can see in Eq. (\ref{eqn:p})
that both a magnetic quadrupole ($\bm{S}_-$) and an electric quadrupole ($F_{\alpha\beta}$) can coexist
since both have an even character for the $IT$ transformation, as shown in Table \ref{table:dimer}.
In the absence of the inversion symmetry between the two spins,
a finite electric dipole moment $\langle \bm{p}_{\rm S}\rangle$ can be induced through the electric quadrupole.
The induced electric dipole moment is proportional to the triplet amplitude ($a_t^2$)
and it can be large for a strong interdimer interaction.

In a field-induced ordered phase with $\bm{H}\parallel z$-direction,
a staggered magnetic moment aligns in the $xy$-plane.
The expectation values of $\bm{p}_{\rm S}$ and $\bm{p}_{\rm A}$ given by Eq. (\ref{eqn:p-F}) are respectively expressed as
\begin{align}
&
\begin{pmatrix}
\langle p_{\rm S}^x \rangle \cr
\langle p_{\rm S}^y \rangle \cr
\langle p_{\rm S}^z \rangle
\end{pmatrix} =
\begin{pmatrix}
0 \cr
0 \cr
- 2A^z a_1 a_{-1} xy
\end{pmatrix}, \cr
&
\begin{pmatrix}
\langle p_{\rm A}^x \rangle \cr
\langle p_{\rm A}^y \rangle \cr
\langle p_{\rm A}^z \rangle
\end{pmatrix}
= \frac{1}{\sqrt{2}} a_s ( a_1 - a_{-1} )
\begin{pmatrix}
-C^x x \cr
C^y y \cr
0
\end{pmatrix},
\end{align}
with $(x,y)=(\cos\phi,\sin\phi)$.
Here, the expectation value was taken with the groundstate given by Eq. (\ref{eqn:GS-H})
for the field-induced ordered phase, and Eq. (\ref{eqn:WS2}) was used for $\langle \bm{p}_{\rm A}\rangle$.

The interesting point of the electric dipole moment by the symmetric $F_{\alpha\beta}$ operator
is that it can be induced even in collinear magnetic structures, as shown by Eq. (\ref{eqn:p}).
This is in contrast to the dipole moment induced by the vector spin chirality,
where the canted spin structure on a dimer is essential.

\subsubsection{Ferromagnetic spin dimer}

The $F_{\alpha\beta}$ operator having matrix elements between the triplet states
can be active for a ferromagnetic interdimer interaction.
In this case, the triplet state is stabilized against the singlet state and the spin dimer behaves as an $S=1$ spin.
The local groundstate in the magnetic ordered phase at low temperatures is expressed as
\begin{align}
|{\rm GS}\rangle &=  \frac{1}{\sqrt{2}} \left[
  ( - \cos\theta \cos\phi + i \sin\phi ) |t_x\rangle \right. \cr
&~\left.
    -  ( \cos\theta \sin\phi + i \cos\phi ) |t_y\rangle
       + \sin\theta |t_z\rangle \right].
\end{align}
The expectation values of $\bS_+$ and $\bm{p}_{\rm S}$ are expressed as
\begin{align}
\begin{pmatrix}
\langle S_+^x \rangle \cr
\langle S_+^y \rangle \cr
\langle S_+^z \rangle
\end{pmatrix}
=
\begin{pmatrix}
x \cr
y \cr
z
\end{pmatrix},~~~
\begin{pmatrix}
\langle p_{\rm S}^x \rangle \cr
\langle p_{\rm S}^y \rangle \cr
\langle p_{\rm S}^z \rangle
\end{pmatrix}
=  \frac{1}{2}
\begin{pmatrix}
A^x yz \cr
A^y zx\cr
A^z xy
\end{pmatrix},
\label{eqn:F-F}
\end{align}
where $(x,y,z)=(\sin\theta\cos\phi,\sin\theta\sin\phi,\cos\theta)$.
The angles $\theta$ and $\phi$ express the direction of the ordered magnetic moment.
Note that $\langle \bm{S}_-\rangle=\langle \bm{p}_{\rm A}\rangle=0$.
For a ferromagnetic dimer, only the symmetric component is active
and the electric dipole moment is only induced in the absence of the inversion center.
This is essentially the same as the single-spin case, since the ferromagnetic dimer behaves as an $S=1$ single spin.

\section{Summary and Discussion}

\begin{table*}[p]
\caption{
Classification of magnetic dipole, electric dipole, and electric quadrupole operators
induced by a single spin in point groups compatible with the space group without the inversion symmetry.
$S^\alpha$ and $p_{\rm S}^\alpha$ represent the $\alpha~(=x,y,z)$ component of the spin operator
and the symmetric spin-dependent electric dipole operator, respectively.
The quadruple operators (products of spin operators) are defined as
$O_{\alpha^2} = (S^\alpha)^2$,
$O_{x^2\pm y^2} = (S^x)^2 \pm (S^y)^2$, and
$O_{\alpha\beta} = S^\alpha S^\beta + S^\beta S^\alpha~{\rm for}~\alpha\neq\beta$.
They have an even parity with respect to the inversion operation.
In the absence of the inversion symmetry, the electric dipole is expressed
by a linear combination of the quadrupole operators in the same representation by parity mixing,
since both of them (electric dipole and quadrupoles) are invariant under the time-reversal transformation.
The coefficients of the linear combination are listed in Table \ref{table:P}.
Here, we only show the quadrupole operators in the same representation of the electric dipole.
In $f$-electron systems with a strong spin-orbit interaction,
the magnetic dipole and electric quadrupole operators are obtained by replacing $S^\alpha\rightarrow J^\alpha$,
where $J^\alpha$ represents the operator of the total angular momentum.
}
\begin{tabular}{cccccc}
\hline
Point group & Label & Magnetic dipole & Electric dipole & Electric quadrupole \cr
\hline
$C_2$ & $A$ $\Gamma_1$ & $S^z$        & $p_{\rm S}^z$        & $O_{x^2}$, $O_{y^2}$, $O_{z^2}$, $O_{xy}$ \cr
      & $B$ $\Gamma_2$ & $S^x,S^y$ & $p_{\rm S}^x,p_{\rm S}^y$ & $O_{yz}$, $O_{zx}$ \cr
\hline
$C_s$ & $A'$~~$\Gamma_1$ & $S^z$        & $p_{\rm S}^x$, $p_{\rm S}^y$ & $O_{x^2}$, $O_{y^2}$, $O_{z^2}$, $O_{xy}$ \cr
      & $A''$ $\Gamma_2$ & $S^x,S^y$ & $p_{\rm S}^z$        & $O_{yz}$, $O_{zx}$ \cr
\hline
$D_2$ & $A$~~$\Gamma_1$  &       &       &          \cr
      & $B_1$ $\Gamma_2$ & $S^z$ & $p_{\rm S}^z$ & $O_{xy}$ \cr
      & $B_2$ $\Gamma_3$ & $S^y$ & $p_{\rm S}^y$ & $O_{zx}$ \cr
      & $B_3$ $\Gamma_4$ & $S^x$ & $p_{\rm S}^x$ & $O_{yz}$ \cr
\hline
$C_{2v}$ & $A_1$ $\Gamma_1$ &       & $p_{\rm S}^z$ & $O_{x^2}$, $O_{y^2}$, $O_{z^2}$ \cr
         & $A_2$ $\Gamma_2$ & $S^z$ &       &          \cr
         & $B_1$ $\Gamma_3$ & $S^y$ & $p_{\rm S}^x$ & $O_{zx}$ \cr
         & $B_2$ $\Gamma_4$ & $S^x$ & $p_{\rm S}^y$ & $O_{yz}$ \cr
\hline
$C_4$, $S_4$ & $A$ $\Gamma_1$ & $S^z$        & $p_{\rm S}^z(C_4)$   & $O_{x^2+y^2}$, $O_{z^2}$ \cr
             & $B$ $\Gamma_2$ &              & $p_{\rm S}^z(S_4)$   & $O_{x^2-y^2}$, $O_{xy}$ \cr
             & $E$ ($\Gamma_3,\Gamma_4$) & ($S^x,S^y$) & ($p_{\rm S}^x,p_{\rm S}^y$) & ($O_{yz}$, $O_{zx}$) \cr
\hline
$D_4$, $C_{4v}$, $D_{2d}$ & $A_1$ $\Gamma_1$ &              & $p_{\rm S}^z(C_{4v})$ & $O_{x^2+y^2}$, $O_{z^2}$ \cr
                          & $A_2$ $\Gamma_2$ & $S^z$        & $p_{\rm S}^z(D_4)$    & $-$ \cr
                          & $B_1$ $\Gamma_3$ &              &               & \cr
                          & $B_2$ $\Gamma_4$ &              & $p_{\rm S}^z(D_{2d})$ & $O_{xy}$ \cr
                          & $E$~~$\Gamma_5$  & ($S^x,S^y$) & ($p_{\rm S}^x,p_{\rm S}^y$)  & ($O_{yz}$, $O_{zx}$) \cr
\hline
$C_3$ & $A$ $\Gamma_1$ & $S^z$        & $p_{\rm S}^z$        & $O_{x^2+y^2}$, $O_{z^2}$ \cr
      & $E$ ($\Gamma_2,\Gamma_3$) & ($S^x,S^y$) & ($p_{\rm S}^x,p_{\rm S}^y$) & ($O_{x^2-y^2}$, $O_{xy}$), ($O_{yz}$, $O_{zx}$) \cr
\hline
$D_3$, $C_{3v}$ & $A_1$ $\Gamma_1$ &              & $p_{\rm S}^z(C_{3v})$ & $O_{x^2+y^2}$, $O_{z^2}$ \cr
                & $A_2$ $\Gamma_2$ & $S^z$        & $p_{\rm S}^z(D_3)$    & $-$ \cr
                & $E$~  $\Gamma_3$ & ($S^x,S^y$) & ($p_{\rm S}^x,p_{\rm S}^y$)  & ($O_{x^2-y^2}$, $O_{xy}$), ($O_{yz}$, $O_{zx}$) \cr
\hline
$C_6$ & $A$~  $\Gamma_1$ & $S^z$        & $p_{\rm S}^z$         & $O_{x^2+y^2}$, $O_{z^2}$ \cr
      & $B$~  $\Gamma_4$ & \cr
      & $E_1$ ($\Gamma_5,\Gamma_6$) & ($S^x,S^y$) & ($p_{\rm S}^x,p_{\rm S}^y$)  & ($O_{yz}$, $O_{zx}$) \cr
      & $E_2$ ($\Gamma_2,\Gamma_3$) &              &               &                    \cr
\hline
$C_{3h}$ & $A'$~ $\Gamma_1$ & $S^z$        &              &                         \cr
         & $A''$ $\Gamma_4$ &              & $p_{\rm S}^z$        & $-$ \cr
         & $E'$~ ($\Gamma_2,\Gamma_3$) &              & ($p_{\rm S}^x,p_{\rm S}^y$) & ($O_{x^2-y^2}$, $O_{xy}$) \cr
         & $E''$ ($\Gamma_5,\Gamma_6$) & ($S^x,S^y$) &              &                         \cr
\hline
$D_6$, $C_{6v}$ & $A_1$ $\Gamma_1$ &              & $p_{\rm S}^z(C_{6v})$ & $O_{x^2+y^2}$, $O_{z^2}$ \cr
                & $A_2$ $\Gamma_2$ & $S^z$        & $p_{\rm S}^z(D_6)$    & $-$ \cr
                & $B_1$ $\Gamma_3$ \cr
                & $B_2$ $\Gamma_4$ \cr
                & $E_1$ $\Gamma_5$ & ($S^x,S^y$) & ($p_{\rm S}^x,p_{\rm S}^y$)  & ($O_{yz}$, $O_{zx}$),      \cr
                & $E_2$ $\Gamma_6$ &              &               &                          \cr
\hline
$D_{3h}$ & $A_1'$~ $\Gamma_1$ &              &              &                         \cr
         & $A_2'$~ $\Gamma_2$ & $S^z$ \cr
         & $A_1''$ $\Gamma_3$ \cr
         & $A_2''$ $\Gamma_4$ &              & $p_{\rm S}^z$        & $-$ \cr
         & $E'$~   $\Gamma_5$ &              & ($p_{\rm S}^x,p_{\rm S}^y$) & ($O_{x^2-y^2}$, $O_{xy}$) \cr
         & $E''$   $\Gamma_6$ & ($S^x,S^y$) &              &                           \cr
\hline
$T$ & $A$ $\Gamma_1$ &                     &                     &                              \cr
    & $E$ ($\Gamma_2,\Gamma_3$) &                     &                     &                   \cr
    & $T$ $\Gamma_4$ & ($S^x,S^y,S^z$) & ($p_{\rm S}^x,p_{\rm S}^y,p_{\rm S}^z$) & ($O_{yz}$, $O_{zx}$, $O_{xy}$) \cr
\hline
$O$ & $A_1$ $\Gamma_1$ & \cr
    & $A_2$ $\Gamma_2$ & \cr
    & $E$~  $\Gamma_3$ & \cr
    & $T_1$ $\Gamma_4$ & ($S^x,S^y,S^z$) & ($p_{\rm S}^x,p_{\rm S}^y,p_{\rm S}^z$) & $-$ \cr
    & $T_2$ $\Gamma_5$ & \cr
\hline
$T_d$ & $A_1$ $\Gamma_1$ & \cr
      & $A_2$ $\Gamma_2$ & \cr
      & $E$~  $\Gamma_3$ & \cr
      & $T_1$ $\Gamma_4$ & ($S^x,S^y,S^z$) \cr
      & $T_2$ $\Gamma_5$ &                     & ($p_{\rm S}^x,p_{\rm S}^y,p_{\rm S}^z$) & ($O_{yz}$, $O_{zx}$, $O_{xy}$) \cr
\hline
\end{tabular}
\label{table:list}
\end{table*}


We investigated the spin-dependent electric dipole operator in both single-spin and two-spin cases.
For a single spin, Table \ref{table:list} shows that a magnetic dipole,
electric dipole, and electric quadrupole (product of spin operators)
can be classified in irreducible representations for 20 point groups without the inversion symmetry.
In the absence of the inversion symmetry, the even and odd parities are indistinguishable, namely,
the electric dipole and quadrupole are classified in the same irreducible representation.
As listed in Table I, the electric dipole operator can be written with the electric quadrupole operators by parity mixing
and the coefficients between them are obtained so as to satisfy the point-group symmetry.
The characteristic point of the group theoretical analysis is that the possible spin dependences can be determined
without having to consider the microscopic origin.
The results summarized in Tables \ref{table:P} and \ref{table:list} will be useful
for future works investigating magnetoelectric effects in magnetic materials in various point-group symmetries.

One of the typical examples of magnetoelectric effects is an induced static electric dipole moment
in magnetically ordered states.
This leads to a cross-correlation that allows the magnetic (electric) field to control the electric (magnetic) dipole
and related electromagnon excitation.
We discussed the selection rule for light absorption, focusing on tetragonal point groups.
The results are summarized in Fig. \ref{fig:e-magnon} with Tables \ref{table:list-tetra} and \ref{table:spin-tetra}.
When a magnon can be excited by both electric and magnetic fields,
its cross-correlation appears as directional dichroism.
In this case, the directional dependence of the transparency of light
can be controlled by the direction of an external magnetic field,
as summarized in Fig. \ref{fig:dichroism}.
This phenomenon can be used as an optical diode, as in the case of BiFeO$_3$.
\cite{Kezsmarki-2015}

The above results can be applied not only to a spin in crystals, such as in \Ba,
but also to a metal ion embedded in molecules.
An attractive example of this is a heme protein, where there is no inversion symmetry at the metal ion site.
High-frequency and high-field measurements of electron paramagnetic resonance
were carried out to identify the electric state of the metal ion in a heme protein
\cite{Andersson-2003,Miyajima-2004}
The present study implies that a heme protein is a possible soft material exhibiting multiferroic behavior,
such as light absorption caused by the electric field component.
It should be kept in mind that the conventional selection rule of the transition
caused by a magnetic dipole is not simply applicable
when the inversion symmetry is broken at the metal-ion site.
The study of multiferroic properties in a heme protein and related molecules with a metal ion is left for future work.

For a clear observation of multiferroic behavior and for practical applications to multiferroic devices,
a strong spin-orbit interaction is required.
Therefore, heavy ions or $f$-electron systems are more promising for these purpose
when the inversion symmetry is broken at the magnetic-ion site.
In this case, the present study can be applied straightforwardly:
the spin operator $\bm{S}$ in Eq. (\ref{eqn:P}) is simply replaced by the total angular momentum $\bm{J}$.
Since the electric dipole moment is screened in metals, insulating systems are favorable.
An $f$-electron ion embedded in a molecule at a site without the inversion symmetry is also interesting.
Giant electromagnetic effects, such as electric-field-controlled magnetic moments,
are expected to be more promising for applications than quantum spin systems with a transition-metal ion.
The study of the multiferroic properties of heavy ions or $f$-electron systems is also left for future work.

In the case of an electric dipole operator induced by two spins, there are both symmetric and antisymmetric spin dependences
with respect to the spatial inversion at the center of the two spins.
Beyond the work by Kaplan and Mahanti,
\cite{Kaplan-2011}
we considered the symmetric components of the electric dipole operators and summarized them in Table \ref{table:PS},
as well as the antisymmetric ones listed in Table \ref{table:PA}.
The result was applied to spin dimer systems.
In the presence of an inversion center between the two spins,
the electric dipole operator is described by the vector spin chirality $\bS_1\times \bS_2$.
\cite{Katsura-2005,Kaplan-2011}
We discussed an electric dipole moment from a symmetrical viewpoint
and found that it is induced when the $IT$ symmetry is broken,
where $I$ and $T$ represent the inversion and time-reversal symmetries, respectively.
This is realized in the field-induced ordered phase of spin dimer systems with a canted spin structure
and was recently revealed by Kimura et al. by observing the induced electric polarization in \Tl.
\cite{Kimura-2016}
Thus, spin dimer systems can be intriguing playgrounds to search for magnetoelectric effects.

Finally, we discuss an application to spin-nematic ordered phases.
\cite{Blume-1969,Chen-1971,Andreev-1984,Papanicolaou-1988}
It is necessary to have a suitable probe for observing spin-nematic and bond-nematic ordered states.
In our study, we demonstrated that an electric dipole moment can be induced by both a single spin and two spins.
For the former and latter, the electric dipole can be described by the local ($O$ operators in Table \ref{table:P})
and intersite ($F$ operators in Table \ref{table:PS}) electric quadruple operators, respectively.
Note that they correspond to spin-nematic and bond-nematic operators, respectively.
In a spin-nematic phase, the order parameter is a single-site electric quadrupole.
Note that it can induce an electric dipole moment
in the absence of the inversion symmetry at the magnetic-ion site.
In the same way, in a bond-nematic phase, the order parameter is a quadrupole induced by two spins.
It can induce an electric dipole moment in the absence of the inversion center between the bond spins.
Thus, Tables \ref{table:P} and \ref{table:PS} are also useful for detecting spin-nematic and bond-nematic ordered states
through the magnetoelectric effect.

\acknowledgments

The authors express their sincere thanks to S. Kimura for fruitful discussions
on the spin-dependent electric polarization in spin dimer systems.
They also thank Y. Inagaki, T. Kawae, K. Kimura, and R. Shiina for useful discussions.
This work was supported by JSPS KAKENHI Grant Numbers 23540390 and 26400332.
One of the authors (M. K.) was supported by JSPS KAKENHI Grant Number 16H01070 (J-Physics).

\appendix
\setcounter{equation}{0}

\section{Coefficient Tensor for Basal Symmetry Transformation}
\label{appendix:polarization}

The symmetric spin-dependent electric dipole operator for a single spin is expressed as
\begin{align*}
p_{\rm S}^\alpha = K^\alpha_{\beta\gamma} S^\beta S^\gamma.
\end{align*}
In this appendix, we present the coefficient tensor $K^\alpha_{\beta\gamma}$ under various basal symmetries.
Since the tensor is symmetric as
{\footnotesize
\begin{align*}
K^\alpha =
\begin{pmatrix}
K^\alpha_{xx} & K^\alpha_{xy} & K^\alpha_{zx} \cr
K^\alpha_{xy} & K^\alpha_{yy} & K^\alpha_{yz} \cr
K^\alpha_{zx} & K^\alpha_{yz} & K^\alpha_{zz}
\end{pmatrix},
\end{align*}
}
we represent it in the following form:
{\footnotesize
\begin{align*}
K \rightarrow
\begin{pmatrix}
K^x_{xx} & K^x_{yy} & K^x_{zz} & K^x_{yz} & K^x_{zx} & K^x_{xy} \cr
K^y_{xx} & K^y_{yy} & K^y_{zz} & K^y_{yz} & K^y_{zx} & K^y_{xy} \cr
K^z_{xx} & K^z_{yy} & K^z_{zz} & K^z_{yz} & K^z_{zx} & K^z_{xy}
\end{pmatrix}.
\end{align*}
}
There are $18~(=6\times 3)$ degrees of freedom.
The symmetry transformations reduces the number of free parameters.
We can obtain the coefficient tensor for a point group
by retaining the common coefficients over all possible symmetry transformations in the point group.

\subsection{$C_2(z), C_2(x), C_2(y)$}

$C_2(\alpha)$ represents the $\pi$ rotation around the $\alpha~(=x,y,z)$-axis.
The coefficient tensors invariant under these transformations are expressed as
{\footnotesize
\begin{align*}
&K[C_2(z)] =
\begin{pmatrix}
0        & 0        & 0        & K^x_{yz} & K^x_{zx} & 0 \cr
0        & 0        & 0        & K^y_{yz} & K^y_{zx} & 0 \cr
K^z_{xx} & K^z_{yy} & K^z_{zz} & 0        & 0        & K^z_{xy}
\end{pmatrix}, \cr
&K[C_2(x)] =
\begin{pmatrix}
K^x_{xx} & K^x_{yy} & K^x_{zz} & K^x_{yz} & 0        & 0 \cr
0        & 0        & 0        & 0        & K^y_{zx} & K^y_{xy} \cr
0        & 0        & 0        & 0        & K^z_{zx} & K^z_{xy}
\end{pmatrix}, \cr
&K[C_2(y)] =
\begin{pmatrix}
0        & 0        & 0        & K^x_{yz} & 0        & K^x_{xy} \cr
K^y_{xx} & K^y_{yy} & K^y_{zz} & 0        & K^y_{zx} & 0 \cr
0        & 0        & 0        & K^z_{yz} & 0        & K^z_{xy} \cr
\end{pmatrix}.
\end{align*}
}
For instance, the $D_2$ point group has the above three symmetry transformations.
By retaining the common coefficients over the three transformations,
we obtain the following coefficient tensor for $D_2$:
{\footnotesize
\begin{align*}
&K[D_2] =
\begin{pmatrix}
0        & 0        & 0        & K^x_{yz} & 0 & 0 \cr
0        & 0        & 0        & 0        & K^y_{zx} & 0 \cr
0        & 0        & 0        & 0        & 0        & K^z_{xy}
\end{pmatrix}.
\end{align*}
}
The symmetric spin-dependent electric dipole operator is then expressed as
\begin{align*}
(p_{\rm S}^x,p_{\rm S}^y,p_{\rm S}^z) = (K^x_{yz}O_{yz},K^y_{zx}O_{zx},K^z_{xy}O_{xy}).
\end{align*}
Here, the quadrupole operators, $O_{\alpha\beta}$, are defined by Eq. (\ref{eqn:Gamma}).
In the same way, we can obtain the electric dipole operators for other point groups.

\subsection{$\sigma(z), \sigma(x), \sigma(y)$}

$\sigma(\alpha)$ represents the mirror transformation
with respect to the plane whose normal vector is along the $\alpha~(=x,y,z)$-axis.
For the mirror transformations, the coefficient tensors are expressed as
{\footnotesize
\begin{align*}
&K[\sigma(z)] =
\begin{pmatrix}
K^x_{xx} & K^x_{yy} & K^x_{zz} & 0        & 0        & K^x_{xy} \cr
K^y_{xx} & K^y_{yy} & K^y_{zz} & 0        & 0        & K^y_{xy} \cr
0        & 0        & 0        & K^z_{yz} & K^z_{zx} & 0
\end{pmatrix}, \cr
&K[\sigma(x)] =
\begin{pmatrix}
0        & 0        & 0        & 0        & K^x_{zx} & K^x_{xy} \cr
K^y_{xx} & K^y_{yy} & K^y_{zz} & K^y_{yz} & 0        & 0 \cr
K^z_{xx} & K^z_{yy} & K^z_{zz} & K^z_{yz} & 0        & 0
\end{pmatrix}, \cr
&K[\sigma(y)] =
\begin{pmatrix}
K^x_{xx} & K^x_{yy} & K^x_{zz} & 0        & K^x_{zx} & 0 \cr
0        & 0        & 0        & K^y_{yz} & 0        & K^y_{xy} \cr
K^z_{xx} & K^z_{yy} & K^z_{zz} & 0        & K^z_{zx} & 0
\end{pmatrix}.
\end{align*}
}

\subsection{$C_3(z), C_4(z), C_6(z), S_4$}

$C_n(z)$ represents the $2\pi/n$ rotation around the $z$-axis.
$S_4$ represents the $\pi/2$ rotary reflection with respect to the $z$-axis.
The coefficient tensors are expressed as
{\footnotesize
\begin{align*}
&K[C_3(z)] =
\begin{pmatrix}
K^x_{xx} & -K^x_{xx} & 0        & K^x_{yz} &  K^x_{zx} &  K^x_{xy} \cr
K^x_{xy} & -K^x_{xy} & 0        & K^x_{zx} & -K^x_{yz} & -K^x_{xx} \cr
K^z_{xx} &  K^z_{xx} & K^z_{zz} & 0        &  0        &  0 \cr
\end{pmatrix}, \cr
&K[C_4(z)] =
\begin{pmatrix}
0        & 0        & 0        & K^x_{yz} &  K^x_{zx} & 0 \cr
0        & 0        & 0        & K^x_{zx} & -K^x_{yz} & 0 \cr
K^z_{xx} & K^z_{xx} & K^z_{zz} & 0        &  0        & 0
\end{pmatrix}, \cr
&K[C_6(z)] =
\begin{pmatrix}
0        & 0        & 0        & K^x_{yz} &  K^x_{zx} & 0 \cr
0        & 0        & 0        & K^x_{zx} & -K^x_{yz} & 0 \cr
K^z_{xx} & K^z_{xx} & K^z_{zz} & 0        &  0        & 0
\end{pmatrix}, \cr
&K[S_4] =
\begin{pmatrix}
0 & 0 & 0 & K^x_{yz} & K^x_{zx} & 0 \cr
0 & 0 & 0 & -K^x_{zx} & K^x_{yz} & 0 \cr
K^z_{xx} & -K^z_{xx} & 0 & 0 & 0 & K^z_{xy}
\end{pmatrix}.
\end{align*}
}

\subsection{$C_2(\frac{\pi}{4}), C_2(-\frac{\pi}{4}), \sigma_v(\frac{\pi}{4}), \sigma_v(-\frac{\pi}{4})$}

\begin{figure}[t]
\begin{center}
\includegraphics[width=8cm,clip]{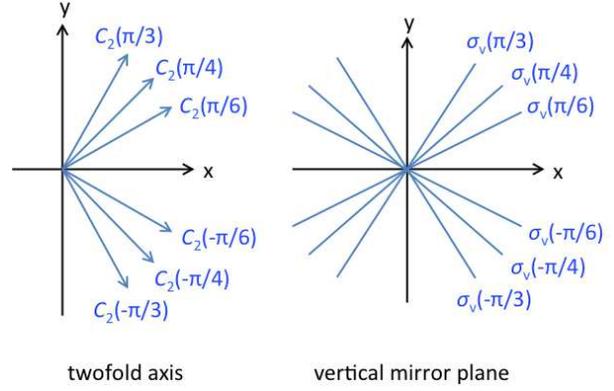}
\end{center}
\caption{
(Color online)
Schematic of various twofold axes and vertical mirror planes.
The twofold axes are in the $xy$-plane.
They are represented by the angle measured from the $x$-axis.
The vertical mirrors are represented in the same way.
}
\label{fig:symmetry}
\end{figure}

$C_2(\pm\frac{\pi}{4})$ represents the $\pi$ rotation around the axes shown in Fig. \ref{fig:symmetry}.
$\sigma_v(\pm\frac{\pi}{4})$ represents the mirror transformation (see Fig. \ref{fig:symmetry}).
The coefficient tensors are expressed as
{\scriptsize
\begin{align*}
&K[C_2(\frac{\pi}{4})] =
\begin{pmatrix}
K^x_{xx} &  K^x_{yy} & K^x_{zz} &  K^x_{yz} &  K^x_{zx} & K^x_{xy} \cr
K^x_{yy} &  K^x_{xx} & K^x_{zz} & -K^x_{zx} & -K^x_{yz} & K^x_{xy} \cr
K^z_{xx} & -K^z_{xx} & 0        &  K^z_{zx} &  K^z_{zx} & 0
\end{pmatrix}, \cr
&K[C_2(-\frac{\pi}{4})] =
\begin{pmatrix}
 K^x_{xx} &  K^x_{yy} &  K^x_{zz} &  K^x_{yz} &  K^x_{zx} &  K^x_{xy} \cr
-K^x_{yy} & -K^x_{xx} & -K^x_{zz} & -K^x_{zx} & -K^x_{yz} & -K^x_{xy} \cr
 K^z_{xx} & -K^z_{xx} &  0        & -K^z_{zx} &  K^z_{zx} &  0
\end{pmatrix}, \cr
&K[\sigma_v(\frac{\pi}{4})] =
\begin{pmatrix}
K^x_{xx} & K^x_{yy} & K^x_{zz} & K^x_{yz} & K^x_{zx} & K^x_{xy} \cr
K^x_{yy} & K^x_{xx} & K^x_{zz} & K^x_{zx} & K^x_{yz} & K^x_{xy} \cr
K^z_{xx} & K^z_{xx} & K^z_{zz} & K^z_{zx} & K^z_{zx} & K^z_{xy}
\end{pmatrix}, \cr
&K[\sigma_v(-\frac{\pi}{4})] =
\begin{pmatrix}
K^x_{xx} & K^x_{yy} & K^x_{zz} & K^x_{yz} & K^x_{zx} & K^x_{xy} \cr
-K^x_{yy} & -K^x_{xx} & -K^x_{zz} & K^x_{zx} & K^x_{yz} & -K^x_{xy} \cr
K^z_{xx} & K^z_{xx} & K^z_{zz} & -K^z_{zx} & K^z_{zx} & K^z_{xy}
\end{pmatrix}.
\end{align*}
}

\subsection{$C_2(\frac{\pi}{6}), C_2(-\frac{\pi}{6}), C_2(\frac{\pi}{3}), C_2(-\frac{\pi}{3})$}

We show the coefficient tensors that are invariant
under the $C_2(\pm\frac{\pi}{6})$ and $C_2(\pm\frac{\pi}{3})$ transformations (see Fig. \ref{fig:symmetry}).
The matrix elements in the $K^y$ component surrounded by squares and double squares are written by the $K^x$ component.
They are shown separately below the tensor.
{\tiny
\begin{align*}
&K[C_2(\frac{\pi}{6})] \cr
&=
\begin{pmatrix}
K^x_{xx} &  K^x_{yy} & K^x_{zz} &  K^x_{yz} &  K^x_{zx} & K^x_{xy} \cr
\fbox{$K^y_{xx}$} & \fbox{$K^y_{yy}$} & \frac{1}{\sqrt{3}} K^x_{zz} & -K^x_{zx} & \doublebox{$K^y_{zx}$} & \fbox{$K^y_{xy}$} \cr
K^z_{xx} & -K^z_{xx} & 0        &  \frac{1}{\sqrt{3}} K^z_{zx} & K^z_{zx} & -\frac{1}{\sqrt{3}} K^z_{xx}
\end{pmatrix}, \cr
&~~~\doublebox{$K^y_{zx}$} = -\frac{2}{\sqrt{3}} K^x_{zx} - K^x_{yz}, \cr
&~~~
\fbox{$
\begin{pmatrix}
K^y_{xx} \cr
K^y_{xy} \cr
K^y_{yy}
\end{pmatrix}$}
=
\begin{pmatrix}
-\frac{1}{2\sqrt{3}} & 1 & \frac{\sqrt{3}}{2} \cr
\frac{1}{2} & 0 & -\frac{1}{2} \cr
\frac{\sqrt{3}}{2} & -1 & -\frac{1}{2\sqrt{3}}
\end{pmatrix}
\begin{pmatrix}
K^x_{xx} \cr
K^x_{xy} \cr
K^x_{yy}
\end{pmatrix}, \cr
&K[C_2(-\frac{\pi}{6})] \cr
&=
\begin{pmatrix}
K^x_{xx} &  K^x_{yy} & K^x_{zz} &  K^x_{yz} &  K^x_{zx} & K^x_{xy} \cr
\fbox{$K^y_{xx}$} & \fbox{$K^y_{yy}$} & -\frac{1}{\sqrt{3}} K^x_{zz} & -K^x_{zx} & \doublebox{$K^y_{zx}$} & \fbox{$K^y_{xy}$} \cr
K^z_{xx} & -K^z_{xx} & 0        & -\frac{1}{\sqrt{3}} K^z_{zx} & K^z_{zx} & \frac{1}{\sqrt{3}} K^z_{xx}
\end{pmatrix}, \cr
&~~~\doublebox{$K^y_{zx}$} = \frac{2}{\sqrt{3}} K^x_{zx} - K^x_{yz}, \cr
&~~~
\fbox{$
\begin{pmatrix}
K^y_{xx} \cr
K^y_{xy} \cr
K^y_{yy}
\end{pmatrix}$}
=
\begin{pmatrix}
\frac{1}{2\sqrt{3}} & 1 & -\frac{\sqrt{3}}{2} \cr
\frac{1}{2} & 0 & -\frac{1}{2} \cr
-\frac{\sqrt{3}}{2} & -1 & \frac{1}{2\sqrt{3}}
\end{pmatrix}
\begin{pmatrix}
K^x_{xx} \cr
K^x_{xy} \cr
K^x_{yy}
\end{pmatrix}, \cr
&K[C_2(\frac{\pi}{3})] \cr
&=
\begin{pmatrix}
K^x_{xx} &  K^x_{yy} & K^x_{zz} &  K^x_{yz} &  K^x_{zx} & K^x_{xy} \cr
\fbox{$K^y_{xx}$} & \fbox{$K^y_{yy}$} & \sqrt{3} K^x_{zz} & -K^x_{zx} & \doublebox{$K^y_{zx}$} & \fbox{$K^y_{xy}$} \cr
K^z_{xx} & -K^z_{xx} & 0        &  \sqrt{3} K^z_{zx} &  K^z_{zx} & \frac{1}{\sqrt{3}} K^z_{xx}
\end{pmatrix}, \cr
&~~~\doublebox{$K^y_{zx}$} = \frac{2}{\sqrt{3}} K^x_{zx} - K^x_{yz}, \cr
&~~~
\fbox{$
\begin{pmatrix}
K^y_{xx} \cr
K^y_{xy} \cr
K^y_{yy}
\end{pmatrix}$}
=
\begin{pmatrix}
\frac{\sqrt{3}}{2} & -1 & \frac{\sqrt{3}}{2} \cr
-\frac{1}{2} & \frac{\sqrt{3}}{2} & \frac{1}{2} \cr
\frac{\sqrt{3}}{2} & 1 & \frac{\sqrt{3}}{2}
\end{pmatrix}
\begin{pmatrix}
K^x_{xx} \cr
K^x_{xy} \cr
K^x_{yy}
\end{pmatrix}, \cr
&K[C_2(-\frac{\pi}{3})] \cr
&=
\begin{pmatrix}
K^x_{xx} &  K^x_{yy} & K^x_{zz} &  K^x_{yz} &  K^x_{zx} & K^x_{xy} \cr
\fbox{$K^y_{xx}$} & \fbox{$K^y_{yy}$} & -\sqrt{3} K^x_{zz} & -K^x_{zx} & \doublebox{$K^y_{zx}$} & \fbox{$K^y_{xy}$} \cr
K^z_{xx} & -K^z_{xx} & 0        &  -\sqrt{3} K^z_{zx} &  K^z_{zx} &  -\frac{1}{\sqrt{3}} K^z_{xx}
\end{pmatrix}, \cr
&~~~\doublebox{$K^y_{zx}$} = -\frac{2}{\sqrt{3}} K^x_{zx} - K^x_{yz}, \cr
&~~~
\fbox{$
\begin{pmatrix}
K^y_{xx} \cr
K^y_{xy} \cr
K^y_{yy}
\end{pmatrix}$}
=
\begin{pmatrix}
-\frac{\sqrt{3}}{2} & -1 & -\frac{\sqrt{3}}{2} \cr
-\frac{1}{2} & -\frac{\sqrt{3}}{2} & \frac{1}{2} \cr
-\frac{\sqrt{3}}{2} & 1 & -\frac{\sqrt{3}}{2}
\end{pmatrix}
\begin{pmatrix}
K^x_{xx} \cr
K^x_{xy} \cr
K^x_{yy}
\end{pmatrix}.
\end{align*}
}

\subsection{$\sigma_v(\frac{\pi}{6}), \sigma_v(-\frac{\pi}{6}), \sigma_v(\frac{\pi}{3}), \sigma_v(-\frac{\pi}{3})$}

The coefficient tensors for the $\sigma_v(\pm\frac{\pi}{6})$ and $\sigma_v(\pm\frac{\pi}{3})$ transformations
(see Fig. \ref{fig:symmetry}) are expressed as
{\tiny
\begin{align*}
&K[\sigma_v(\frac{\pi}{6})] \cr
&=
\begin{pmatrix}
K^x_{xx} &  K^x_{yy} & K^x_{zz} &  K^x_{yz} &  K^x_{zx} & K^x_{xy} \cr
\fbox{$K^y_{xx}$} & \fbox{$K^y_{yy}$} & \frac{1}{\sqrt{3}} K^x_{zz} & \doublebox{$K^y_{yz}$} & K^x_{yz} & \fbox{$K^y_{xy}$} \cr
K^z_{xx} & K^z_{yy} & K^z_{zz} & \frac{1}{\sqrt{3}} K^z_{zx} &  K^z_{zx} & \frac{\sqrt{3}}{2} ( K^z_{xx} - K^z_{yy} )
\end{pmatrix}, \cr
&~~~\doublebox{$K^y_{yz}$} = K^x_{zx} - \frac{2}{\sqrt{3}} K^x_{yz}, \cr
&~~~
\fbox{$
\begin{pmatrix}
K^y_{xx} \cr
K^y_{xy} \cr
K^y_{yy}
\end{pmatrix}$}
=
\begin{pmatrix}
-\frac{1}{2\sqrt{3}} & 1 & \frac{\sqrt{3}}{2} \cr
\frac{1}{2} & 0 & -\frac{1}{2} \cr
\frac{\sqrt{3}}{2} & -1 & -\frac{1}{2\sqrt{3}}
\end{pmatrix}
\begin{pmatrix}
K^x_{xx} \cr
K^x_{xy} \cr
K^x_{yy}
\end{pmatrix}, \cr
&K[\sigma_v(-\frac{\pi}{6})] \cr
&=
\begin{pmatrix}
K^x_{xx} &  K^x_{yy} & K^x_{zz} &  K^x_{yz} &  K^x_{zx} & K^x_{xy} \cr
\fbox{$K^y_{xx}$} & \fbox{$K^y_{yy}$} & -\frac{1}{\sqrt{3}} K^x_{zz} & \doublebox{$K^y_{yz}$} & K^x_{yz} & \fbox{$K^y_{xy}$} \cr
K^z_{xx} & K^z_{yy} & K^z_{zz} & -\frac{1}{\sqrt{3}} K^z_{zx} &  K^z_{zx} & \frac{\sqrt{3}}{2} ( K^z_{yy} - K^z_{xx} )
\end{pmatrix}, \cr
&~~~\doublebox{$K^y_{yz}$} = K^x_{zx} + \frac{2}{\sqrt{3}} K^x_{yz}, \cr
&~~~
\fbox{$
\begin{pmatrix}
K^y_{xx} \cr
K^y_{xy} \cr
K^y_{yy}
\end{pmatrix}$}
=
\begin{pmatrix}
\frac{1}{2\sqrt{3}} & 1 & -\frac{\sqrt{3}}{2} \cr
\frac{1}{2} & 0 & -\frac{1}{2} \cr
-\frac{\sqrt{3}}{2} & -1 & \frac{1}{2\sqrt{3}}
\end{pmatrix}
\begin{pmatrix}
K^x_{xx} \cr
K^x_{xy} \cr
K^x_{yy}
\end{pmatrix}, \cr
&K[\sigma_v(\frac{\pi}{3})] \cr
&=
\begin{pmatrix}
K^x_{xx} &  K^x_{yy} & K^x_{zz} &  K^x_{yz} &  K^x_{zx} & K^x_{xy} \cr
\fbox{$K^y_{xx}$} & \fbox{$K^y_{yy}$} & \sqrt{3} K^x_{zz} & \doublebox{$K^y_{yz}$} & K^x_{yz} & \fbox{$K^y_{xy}$} \cr
K^z_{xx} & K^z_{yy} & K^z_{zz} & \sqrt{3} K^z_{zx} &  K^z_{zx} & \frac{\sqrt{3}}{2} ( K^z_{yy} - K^z_{xx} )
\end{pmatrix}, \cr
&~~~\doublebox{$K^y_{yz}$} = K^x_{zx} + \frac{2}{\sqrt{3}} K^x_{yz}, \cr
&~~~
\fbox{$
\begin{pmatrix}
K^y_{xx} \cr
K^y_{xy} \cr
K^y_{yy}
\end{pmatrix}$}
=
\begin{pmatrix}
\frac{\sqrt{3}}{2} & -1 & \frac{\sqrt{3}}{2} \cr
-\frac{1}{2} & \frac{2}{\sqrt{3}} & \frac{1}{2} \cr
\frac{\sqrt{3}}{2} & 1 & \frac{\sqrt{3}}{2}
\end{pmatrix}
\begin{pmatrix}
K^x_{xx} \cr
K^x_{xy} \cr
K^x_{yy}
\end{pmatrix}, \cr
&K[\sigma_v(-\frac{\pi}{3})] \cr
&=
\begin{pmatrix}
K^x_{xx} &  K^x_{yy} & K^x_{zz} &  K^x_{yz} &  K^x_{zx} & K^x_{xy} \cr
\fbox{$K^y_{xx}$} & \fbox{$K^y_{yy}$} & -\sqrt{3} K^x_{zz} & \doublebox{$K^y_{yz}$} & K^x_{yz} & \fbox{$K^y_{xy}$} \cr
K^z_{xx} & K^z_{yy} & K^z_{zz} & -\sqrt{3} K^z_{zx} &  K^z_{zx} & \frac{\sqrt{3}}{2} ( K^z_{xx} - K^z_{yy} )
\end{pmatrix}, \cr
&~~~\doublebox{$K^y_{yz}$} = K^x_{zx} - \frac{2}{\sqrt{3}} K^x_{yz}, \cr
&~~~
\fbox{$
\begin{pmatrix}
K^y_{xx} \cr
K^y_{xy} \cr
K^y_{yy}
\end{pmatrix}$}
=
\begin{pmatrix}
-\frac{\sqrt{3}}{2} & -1 & -\frac{\sqrt{3}}{2} \cr
-\frac{1}{2} & -\frac{2}{\sqrt{3}} & \frac{1}{2} \cr
-\frac{\sqrt{3}}{2} & 1 & -\frac{\sqrt{3}}{2}
\end{pmatrix}
\begin{pmatrix}
K^x_{xx} \cr
K^x_{xy} \cr
K^x_{yy}
\end{pmatrix}.
\end{align*}
}

\section{Electric Dipole Induced by Parity Mixing under Broken Inversion Symmetry}
\label{appendix:parity-mix}

\subsection{Electric dipole induced by $d$-$p$ hybridization}

\begin{figure}[t]
\begin{center}
\includegraphics[width=6.3cm,clip]{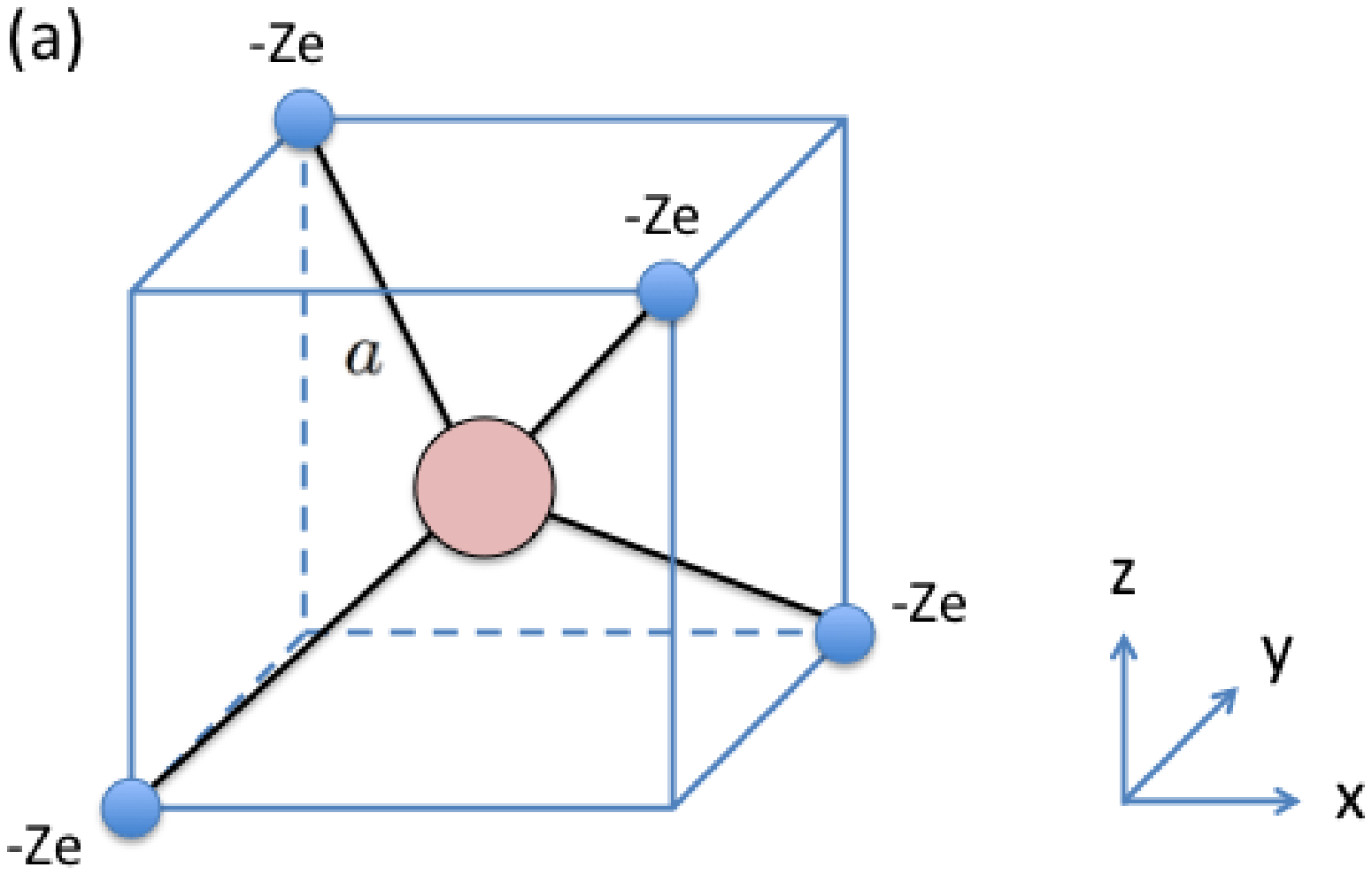}
\includegraphics[width=6.5cm,clip]{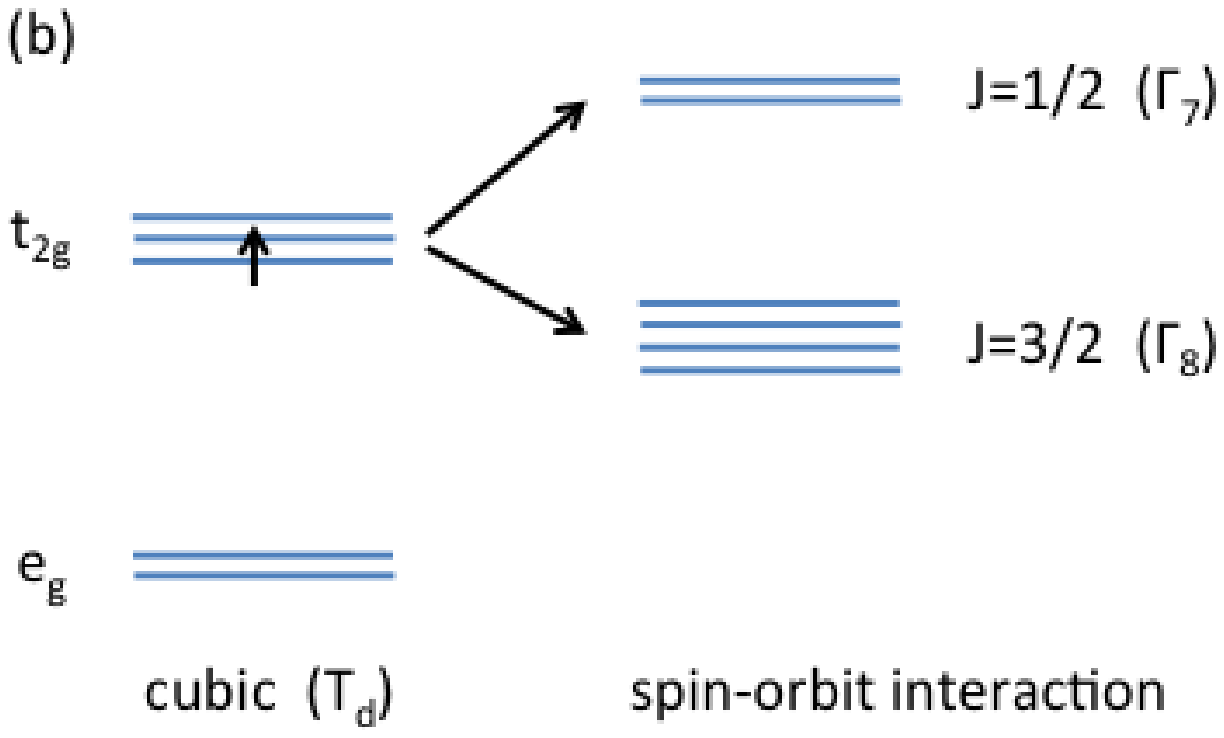}
\end{center}
\caption{
(Color online)
(a) Schematic of a magnetic ion surrounded by ligand ions at the apices of a regular tetrahedron.
$-Ze$ represents the electric charge of the ligand ion.
$a$ is the distance between the magnetic ion and ligand ion.
(b) Schematic of energy levels of $d$ orbitals in $T_d$ point-group symmetry,
where the fivefold-degenerate $d$ orbitals split into $e_g$ and $T_{2g}$ orbitals.
We assume that an electron occupies one of the $t_{2g}$ orbitals.
The angular momentum of the $t_{2g}$ state can be mapped on an $L=1$ model
and the $t_{2g}$ states split into $J=3/2$ ($\Gamma_8$) and $J=1/2$ ($\Gamma_7$) states
in the presence of the spin-orbit interaction.
}
\label{fig:Td}
\end{figure}

In this appendix, we consider a microscopic model
and show how an electric dipole moment is induced by the broken inversion symmetry.
This helps us understand the group theoretical results summarized in Table \ref{table:P}.
For this purpose, we focus on a $d$-electron in the $T_d$ point-group symmetry,
where the $d$ orbitals are split into $t_{2g}$ and $e_g$ orbitals, as shown in Fig. \ref{fig:Td}.
We assume that an electron occupies one of the $t_{2g}$ orbitals.
They are defined as
\begin{align}
&|d_{yz}\rangle = i\frac{1}{\sqrt{2}} |Y_{2,1}\rangle + i\frac{1}{\sqrt{2}} |Y_{2,-1}\rangle, \cr
&|d_{zx}\rangle = -\frac{1}{\sqrt{2}} |Y_{2,1}\rangle + \frac{1}{\sqrt{2}} |Y_{2,-1}\rangle, \cr
&|d_{xy}\rangle = -i\frac{1}{\sqrt{2}} |Y_{2,2}\rangle + i\frac{1}{\sqrt{2}} |Y_{2,-2}\rangle.
\label{eqn:tg2}
\end{align}
Here, $|Y_{l,m}\rangle$ represents a harmonic spherical function.
We omitted the function for the radius component.
Note that the wavefunctions in Eq. (\ref{eqn:tg2}) take real values.
There are six degenerated states including the spin states.
The orbital angular momentum is active among the $t_{2g}$ orbitals.
It can be mapped on a pseudo-$L=1$ model within the three basal states.
In the presence of the spin-orbit interaction,
the sixfold states split into four lower pseudo-$J=3/2$ ($\Gamma_8$) states
and two higher pseudo-$J=1/2$ ($\Gamma_7$) states [see Fig. \ref{fig:Td}(b)].
We consider the lower $J=3/2$ states in the following discussion.
They are expressed as
\begin{align}
&\left|\frac{3}{2}\right\rangle = - \frac{1}{\sqrt{2}} |d_{yz\uparrow}\rangle - i \frac{1}{\sqrt{2}} |d_{zx\uparrow}\rangle, \cr
&\left|\frac{1}{2}\right\rangle = \sqrt{\frac{2}{3}} |d_{xy\uparrow}\rangle - \frac{1}{\sqrt{6}} |d_{yz\downarrow}\rangle
                                                                       - i \frac{1}{\sqrt{6}} |d_{zx\downarrow}\rangle, \cr
&\left|-\frac{1}{2}\right\rangle = \frac{1}{\sqrt{6}} |d_{yz\uparrow}\rangle - i \frac{1}{\sqrt{6}} |d_{zx\uparrow}\rangle
                                                                           +  \sqrt{\frac{2}{3}} |d_{xy\downarrow}\rangle, \cr
&\left|-\frac{3}{2}\right\rangle = \frac{1}{\sqrt{2}} |d_{yz\downarrow}\rangle - i \frac{1}{\sqrt{2}} |d_{zx\downarrow}\rangle.
\end{align}
Here, $|m\rangle$ ($m=\frac{3}{2},\frac{1}{2},-\frac{1}{2},-\frac{3}{2}$) represents the $J^z=m$ state for $J=3/2$.
We introduce a unitary transformation $U$ whose matrix elements are defined by $(U)_{nm}=\langle d_n|m\rangle$.
Here, $n=d_{yz\uparrow},d_{zx\uparrow},d_{xy\uparrow},d_{yz\downarrow},d_{zx\downarrow},d_{xy\downarrow}$
and $m=\frac{3}{2},\frac{1}{2},-\frac{1}{2},-\frac{3}{2}$.
Its matrix form is expressed as
\begin{align}
U =
\begin{pmatrix}
\frac{-1}{\sqrt{2}} & 0 & \frac{1}{\sqrt{6}} & 0 \cr
\frac{-i}{\sqrt{2}} & 0 & \frac{-i}{\sqrt{6}} & 0 \cr
0 & \sqrt{\frac{2}{3}} & 0 & 0 \cr
0 & \frac{-1}{\sqrt{6}} & 0 & \frac{1}{\sqrt{2}} \cr
0 & \frac{-i}{\sqrt{6}} & 0 & \frac{-i}{\sqrt{2}} \cr
0 & 0 & \sqrt{\frac{2}{3}} & 0
\end{pmatrix}.
\label{eqn:U}
\end{align}
This is used in the following discussions.

\subsubsection{Quadrupole operator}

For the $J=3/2$ states, quadrupole operators are defined by $O_{\alpha\beta}=J^\alpha J^\beta+J^\beta J^\alpha$,
where $J^\alpha$ represents the total angular moment of the $\alpha~(=x,y,z)$ component.
The quadrupole operators are expressed as
\begin{align}
&O_{yz} = \sqrt{3}
\begin{pmatrix}
0 & -i & 0 & 0 \cr
i & 0 & 0 & 0 \cr
0 & 0 & 0 & i \cr
0 & 0 & -i & 0
\end{pmatrix}, \cr
&O_{zx} = \sqrt{3}
\begin{pmatrix}
0 & 1 & 0 & 0 \cr
1 & 0 & 0 & 0 \cr
0 & 0 & 0 & -1 \cr
0 & 0 & -1 & 0
\end{pmatrix}, \cr
&O_{xy}= \sqrt{3}
\begin{pmatrix}
0 & 0 & -i & 0 \cr
0 & 0 & 0 & -i \cr
i & 0 & 0 & 0 \cr
0 & i & 0 & 0
\end{pmatrix}.
\label{eqn:O-dp}
\end{align}
We discuss next how the quadrupole operators are related to the electric dipole operators.

\subsubsection{$d$-$p$ hybridization induced by broken inversion symmetry}

Since the $d$ orbitals have an even parity, the expectation values of the $x$, $y$, and $z$ positions vanish.
When the inversion symmetry is broken, however, this is not the case.
This is owing to the fact that even-parity $d$ orbitals are mixed with odd-parity orbitals.
In usual cases, odd-rank crystal-field potentials are not considered
since the energy splitting of orbitals with the same $L$ value is focused on.
In the absence of the inversion symmetry, odd-rank crystal-field potentials exist in principle.
For this purpose, we consider the $T_d$ point-group symmetry in this appendix.
The crystal-field potential contains the following third-rank term for $T_d$:
\begin{align}
V_3(r,\theta,\phi) = V_3 r^3 xyz,
\label{eqn:V3}
\end{align}
where
\begin{align}
xyz = (\sin\theta\cos\phi) (\sin\theta\sin\phi) (\cos\theta).
\label{eqn:xyz}
\end{align}
In Eq. (\ref{eqn:V3}), $V_3$ represents the amplitude of the potential.
It is given by
\begin{align}
V_3 = - \frac{4\pi}{7}  \frac{-Ze}{a^4} \frac{105}{2\pi} \frac{2}{3\sqrt{3}}.
\end{align}
Here, $-Ze$ represents the charge of ligand ions on the apices of a regular tetrahedron surrounding the transition metal,
while $a$ represents the distance between the ligand ions and the metal ion, as shown in Fig. \ref{fig:Td}.
The $xyz$ symmetry is classified as an ``electric octupole",
and Eq. (\ref{eqn:V3}) indicates that there is an $xyz$ electric octupole in the crystal-field potential.
This is consistent with the fact that $xyz$ is invariant under the symmetry operations for the $T_d$ point group
and that it is a basis function of the $\Gamma_1$ representation.

The odd-parity crystal-field potential of the $xyz$ type leads to $d$-$p$ hybridization.
On the basis of first-order perturbation theory, $t_{2g}$ orbitals are mixed with $p$ orbitals as
\begin{align}
&|\tilde{d}_{yz}\rangle = |d_{yz}\rangle + v |p_x\rangle, \cr
&|\tilde{d}_{zx}\rangle = |d_{zx}\rangle + v |p_y\rangle, \cr
&|\tilde{d}_{xy}\rangle = |d_{xy}\rangle + v |p_z\rangle,
\label{eqn:dp}
\end{align}
with
\begin{align}
v= \frac{V_3 t}{E_d-E_p},~~~
t = \langle p_x|r^3xyz|d_{yz}\rangle.
\label{eqn:b}
\end{align}
Here, $t$ takes the same value under the cyclic $x\rightarrow y \rightarrow z \rightarrow x$ transformations.
In Eq. (\ref{eqn:b}), $E_d$ and $E_p$ represent the energy levels of the $d$ and $p$ orbitals, respectively.
\cite{note-dp}
Note that $t$ is a real value when we choose real wavefunctions for the $d$ and $p$ orbitals.
In the following discussion, we use the $\tilde{d}$ orbitals instead of the $d$ orbitals.

\subsubsection{Electric dipole operator}

The wavefunctions of the $\tilde{d}_{yz}$, $\tilde{d}_{zx}$, and $\tilde{d}_{xy}$ states are not invariant
under the inversion transformation.
This means that the charge distribution has an asymmetric component that can induce an electric dipole moment.
We next calculate the expectation values of the $x$, $y$, and $z$ positions and introduce matrices $X_\alpha$ ($\alpha=x,y,z$)
whose elements are defined by $\langle n|\alpha|n'\rangle$ with
$n,n'=\tilde{d}_{yz\uparrow},\tilde{d}_{zx\uparrow},\tilde{d}_{xy\uparrow},\tilde{d}_{yz\downarrow}, \tilde{d}_{zx\downarrow},\tilde{d}_{xy\downarrow}$.
They are expressed in the following matrix form:
\begin{align}
&X_\alpha =
\begin{pmatrix}
r_\alpha & 0 \cr
0 & r_\alpha
\end{pmatrix},
\label{eqn:X-alpha}
\end{align}
where
{\footnotesize
\begin{align}
r_x =
\begin{pmatrix}
0 & 0 & 0 \cr
0 & 0 & b \cr
0 & b & 0
\end{pmatrix},~
r_y =
\begin{pmatrix}
0 & 0 & b \cr
0 & 0 & 0 \cr
b & 0 & 0
\end{pmatrix},~
r_z =
\begin{pmatrix}
0 & b & 0 \cr
b & 0 & 0 \cr
0 & 0 & 0
\end{pmatrix}.
\label{eqn:r-x}
\end{align}
}
Here, $b=2t'v$ with $t'= \langle p_y|x|d_{xy}\rangle = \langle d_{zx}|x|p_z\rangle$.
$t'$ takes the same value under the cyclic $x\rightarrow y \rightarrow z \rightarrow x$ transformations.
$X_\alpha$ represents the matrix of the $x$, $y$, and $z$ positions
in a $6\times 6$ matrix form on the basis of the $d$-$p$-hybridized $\tilde{t}_{2g}$ orbitals.

The energy eigenstates are described by the four basal states of $J=3/2$.
The unitary transformation $U$ defined by Eq. (\ref{eqn:U})
enables us to calculate the expectation value of the position of an electron on the basis of the four $J=3/2$ states.
The transformed matrix $\tilde{X}_\alpha=U^\dagger X_a U$ is expressed in a $4\times 4$ matrix form.
Comparing this with the quadrupole operators, we can directly show that
\begin{align}
(\tilde{X}_x,\tilde{X}_y,\tilde{X}_z) = - \frac{1}{3} b (O_{yz},O_{zx},O_{xy}).
\label{eqn:X-O-dp}
\end{align}
Here, the matrix forms of the quadrupole operators are given by Eq. (\ref{eqn:O-dp}).
This indicates that the symmetric spin-dependent electric dipole operators are given by
\begin{align}
(p_{\rm S}^x,p_{\rm S}^y,p_{\rm S}^z) = - \frac{1}{3} (-e) b (O_{yz},O_{zx},O_{xy}),
\label{eqn:p-O}
\end{align}
where $-e$ represents the electron charge.
This is consistent with the result shown in Table \ref{table:P} for $T_d$.
The coefficient is determined as $K^x_{yz}=(1/3)eb$ within the present microscopic model.
Although the quadrupole operators $(O_{\alpha\beta})$ in Eq. (\ref{eqn:p-O}) have an even parity,
an odd-parity component can be induced by the quadrupoles
when the inversion symmetry is broken in the environment.

The result given by Eq. (\ref{eqn:X-O-dp}) can be understood from another point of view.
We consider the matrix forms of $yz$, $zx$, and $xy$
on the basis of the three $\tilde{d}_{yz}$, $\tilde{d}_{zx}$, and $\tilde{d}_{xy}$ orbitals,
where the properties of the $T_d$ symmetry are taken into account by the $d$-$p$ hybridization.
The $3\times 3$ matrices for $yz$, $zx$, and $xy$ are expressed by the matrix $r_\alpha$ given by Eq. (\ref{eqn:r-x}) as
\begin{align}
(r_{yz},r_{zx},r_{xy}) = (r_x,r_y,r_z)_{b\rightarrow b'}
\end{align}
by replacing $b$ with $b'=v^2\langle d_{zx}|yz|d_{xy}\rangle$.
This indicates that $(X_x,X_y,X_z) \propto (X_{yz},X_{zx},X_{xy})$
and $(x,y,z) \propto (yz,zx,xy)$ in the $T_d$ point-group symmetry.
This is consistent with the fact that both are classified in the same $\Gamma_5$ representations,
as shown in Table \ref{table:list}.
\cite{note-Gamma5}
Since $(yz,zx,xy)$ can be replaced by the equivalent operators as
$(yz,zx,xy) \rightarrow (O_{yz},O_{zx},O_{xy})$, we can obtain $(x,y,z) \propto (O_{yz},O_{zx},O_{xy})$.

\subsubsection{In the presence of a quadrupole field}

To show that an electric dipole moment is induced by the quadruple operators,
we consider a quadrupole ordered phase.
As an example, we focus on the $O_{xy}$ type here.
In the ordered phase, the local Hamiltonian for the $J=3/2$ state can be expressed as
\begin{align}
\H = \lambda \langle O_{xy} \rangle O_{xy},
\label{eqn:H-Oxy}
\end{align}
where $\langle O_{xy} \rangle$ is the quadrupole moment at the neighboring sites
and $\lambda$ represents the coupling constant.
In Eq. (\ref{eqn:H-Oxy}), $\lambda\langle O_{xy} \rangle$ plays the role of a quadruple field
coupling to the $O_{xy}$ operator.
Under the quadrupole field, the four degenerate states split
into two degenerate ground ($E_g=-\sqrt{3}\lambda\langle O_{xy}\rangle$)
and excited ($E_e=\sqrt{3}\lambda\langle O_{xy}\rangle$) states.
They are expressed as
\begin{align}
&|{\rm GS}_1\rangle = \frac{1}{\sqrt{2}} \left|\frac{3}{2}\right\rangle
                   - i\frac{1}{\sqrt{2}} \left|\frac{-1}{2}\right\rangle, \cr
&|{\rm GS}_2\rangle = i\frac{1}{\sqrt{2}} \left|\frac{1}{2}\right\rangle
                     + \frac{1}{\sqrt{2}}\left|\frac{-3}{2}\right\rangle, \cr
&|{\rm ES}_1\rangle = \frac{1}{\sqrt{2}} \left|\frac{3}{2}\right\rangle
                  + i \frac{1}{\sqrt{2}} \left|\frac{-1}{2}\right\rangle, \cr
&|{\rm ES}_2\rangle = -i\frac{1}{\sqrt{2}} \left|\frac{1}{2}\right\rangle
                      + \frac{1}{\sqrt{2}}\left|\frac{-3}{2}\right\rangle.
\label{eqn:GS-ES}
\end{align}
Here $|{\rm GS}\rangle$ and $|{\rm ES}\rangle$ denote the ground and excited states, respectively.
These states have the following expectation values of the quadruple operator:
$\langle O_{xy}\rangle = \mp \sqrt{3}$ for $|{\rm GS}\rangle$ and $|{\rm ES}\rangle$, respectively.
The expectation values are zero for $O_{yz}$ and $O_{zx}$.
The induced electric dipole moment is calculated as
\begin{align}
\langle p_{\rm S}^z \rangle_\psi = \mp \frac{1}{\sqrt{3}}eb,
\label{eqn:z-av}
\end{align}
where the $\pm$ signs are for the ground ($|\psi\rangle=|{\rm GS}\rangle$)
and excited ($|\psi\rangle=|{\rm ES}\rangle$) states, respectively.
Note that $\langle p_{\rm S}^x \rangle_\psi = \langle p_{\rm S}^y \rangle_\psi = 0$.
We can see in Eq. (\ref{eqn:z-av}) that an electric dipole moment along the $z$-direction is induced
through the energy splitting of the four degenerate $J=3/2$ states caused by the quadrupole field.

To see the microscopic origin of the induced electric dipole moment,
we discuss the absolute value of the wavefunction given by Eq. (\ref{eqn:GS-ES}).
It contains both even- and odd-parity components;
$|\psi(\br)|^2 = \rho_{\rm even}(\br) + \rho_{\rm odd}(\br)$.
The latter has the following form:
\begin{align}
\rho_{\rm odd}(\br) &= \frac{2}{3} v [ d_{yz}(\br) p_x(\br) + d_{zx}(\br) p_y(\br) + d_{xy}(\br) p_z(\br) ] \cr
&~\pm \frac{1}{\sqrt{3}} v [ d_{zx}(\br) p_x(\br) + d_{yz}(\br) p_y(\br) ],
\label{eqn:do-product}
\end{align}
where the $\pm$ signs in the second term are for $|{\rm GS}\rangle$ and $|{\rm ES}\rangle$, respectively.
Here, $d_{\alpha\beta}(\br)$ and $p_\alpha(\br)$ represent the wavefunctions of the $d$ and $p$ orbitals, respectively.
The first term in Eq. (\ref{eqn:do-product}) has an $xyz$ symmetry,
which is the same as the crystal-field potential $V_3(\br)$
given by Eq. (\ref{eqn:V3}), while the second term has an $(x^2+y^2)z$ symmetry.
Both are classified as an ``electric octupole''.
The latter leads to a charge polarization in the $z$ component and results in a finite electric dipole moment.
The result in Eq. (\ref{eqn:z-av}) can be reproduced by using
$v\int d \br z [ d_{zx}(\br) p_x(\br) + d_{yz}(\br) p_y(\br) ] = b$.
The combination of the $d$ and $p$ orbitals in Eq. (\ref{eqn:do-product}) is a consequence of the $d$-$p$ hybridization
induced by the lack of the inversion symmetry.
To induce an electric dipole moment, the spin-orbit interaction and broken inversion symmetry are essential
in the present microscopic model.

The induced electric dipole moment can be simply understood as follows.
In the $T_d$ point-group symmetry, the $xyz$ symmetry is classified as the $\Gamma_1$ representation.
When we apply the $xy$-type quadrupole field, for instance, we obtain $(xyz)\times(xy)\rightarrow x^2 y^2 z$.
This indicates that the $z$ component of a dipole is induced by the quadrupole field.
This situation can also be realized when we apply a magnetic field along the $(1,1,0)$ direction,
where the magnetic moment is aligned along the $(1,1,0)$ direction
and a quadrupole moment of the $O_{xy}$ type is induced ($\langle O_{xy}\rangle \neq 0$).
We discuss this point in the following subsection.

\subsubsection{In the presence of a magnetic field}

To induce a finite magnetic moment, we next study the $J=3/2$ state under an external magnetic field.
For this purpose, we consider the following form of the Hamiltonian:
\begin{align}
\H = -h [ \sin\theta ( \cos\phi J^x + \sin\phi J^y ) + \cos\theta J^z ],
\end{align}
where $\theta$ and $\phi$ represent the angles of the field measured from the $z$- and $x$-axes, respectively.
The groundstate is expressed as
\begin{align}
|{\rm GS}\rangle &= e^{-i\frac{3}{2}\phi} \cos^3\left(\frac{\theta}{2}\right) \left|\frac{3}{2}\right\rangle
+ e^{i\frac{3}{2}\phi} \sin^3\left(\frac{\theta}{2}\right) \left|-\frac{3}{2}\right\rangle \cr
&~+ e^{-i\frac{1}{2}\phi} \frac{\sqrt{3}}{2} \cos\left(\frac{\theta}{2}\right) \sin(\theta) \left|\frac{1}{2}\right\rangle \cr
&~+ e^{i\frac{1}{2}\phi} \frac{\sqrt{3}}{2} \sin\left(\frac{\theta}{2}\right) \sin(\theta) \left|-\frac{1}{2}\right\rangle.
\label{eqn:GS-H}
\end{align}
The expectation values of the $J^x$, $J^y$, and $J^z$ operators are
\begin{align}
\langle{\rm GS}|
\begin{pmatrix}
J^x \cr
J^y \cr
J^z
\end{pmatrix}
|{\rm GS}\rangle
= \frac{3}{2}
\begin{pmatrix}
\sin\theta\cos\phi \cr
\sin\theta\sin\phi \cr
\cos\theta
\end{pmatrix}.
\label{eqn:J-d}
\end{align}
A finite magnetic moment is extracted in parallel with the external field, as expected.
Under a finite magnetic field, quadrupole moments are induced as
\begin{align}
&\langle {\rm GS}|
\begin{pmatrix}
O_{yz} \cr
O_{zx} \cr
O_{xy}
\end{pmatrix}
|{\rm GS}\rangle
= 3
\begin{pmatrix}
(\sin\theta\sin\phi)(\cos\theta) \cr
(\cos\theta)(\sin\theta\cos\phi) \cr
(\sin\theta\cos\phi)(\sin\theta\sin\phi)
\end{pmatrix},
\label{eqn:O-av}
\end{align}
where $O_{\alpha\beta}=J^\alpha J^\beta + J^\beta J^\alpha$.
Note that
$\langle O_{\alpha\beta}\rangle \neq \langle J_\alpha\rangle \langle J_\beta\rangle + \langle J_\beta\rangle \langle J_\alpha\rangle$, although $\langle O_{\alpha\beta}\rangle \propto \langle J^\alpha\rangle \langle J^\beta\rangle$.
The expectation value of the electric polarization is given by
\begin{align}
\langle
\begin{pmatrix}
p_{\rm S}^x \cr
p_{\rm S}^y \cr
p_{\rm S}^z
\end{pmatrix}
\rangle_{\rm GS}
&= eb
\begin{pmatrix}
(\sin\theta\sin\phi)(\cos\theta) \cr
(\cos\theta)(\sin\theta\cos\phi) \cr
(\sin\theta\cos\phi)(\sin\theta\sin\phi)
\end{pmatrix}.
\label{eqn:XYZ-av}
\end{align}

To see the microscopic origin of the induced electric dipole moment,
we examine the electron density of the groundstate wavefunction.
There is the following odd-parity component:
\begin{align}
&\rho_{\rm odd}(\br) = -v (\sin\theta\sin\phi) (\cos\theta) [ d_{xy}(\br) p_y(\br) + d_{zx}(\br) p_z(\br) ] \cr
&-v (\cos\theta)(\sin\theta\cos\phi) [ d_{yz}(\br) p_z(\br) + d_{xy}(\br) p_x(\br) ] \cr
&-v (\sin\theta\cos\phi)(\sin\theta\sin\phi) [ d_{zx}(\br) p_x(\br) + d_{yz}(\br) p_y(\br) ].
\end{align}
It has $(y^2+z^2)x$, $(z^2+x^2)y$, and $(x^2+y^2)z$ symmetries,
which induce a finite electric dipole moment along the $x$-, $y$-, and $z$-axes, respectively.
After the integration over the coordinate, the result in Eq. (\ref{eqn:XYZ-av}) is reproduced
by using $v\int d\br x[d_{xy}(\br) p_y(\br) + d_{zx}(\br) p_z(\br) ] = b$.

Finally, we comment on the twofold degenerate higher $J=1/2$ ($\Gamma_7$) state shown in Fig. \ref{fig:Td}.
The electric dipole moment is not induced by the $\Gamma_7$ state
since it is described by a $J=1/2$ model having no degrees of freedom of the quadrupoles.
In addition, the $e_g$ states are irrelevant since they do not mix with the $p$ orbitals
and they do not induce the electric dipole moment.

\subsection{Electric dipole induced by $f$-$d$ hybridization}

Induced electric dipole moments in $f$-electron systems can be discussed in parallel with the $d$-electron systems.
Let us consider an $f^1$ state, where one of the $f$ orbitals is occupied by an electron.
Under a strong spin-orbit interaction in metal ions having $f$ electrons,
the 14-fold ($L=3$ and $S=1/2$) degenerate energy levels split into lower $J=5/2$ and higher $J=7/2$ states.
We retain the former states.
They are expressed as
\begin{align}
&|n\rangle = - \sqrt{\frac{1}{2}-\frac{n}{7}}|Y_{3,n-\frac{1}{2}}\uparrow\rangle
             + \sqrt{\frac{1}{2}+\frac{n}{7}}|Y_{3,n+\frac{1}{2}}\downarrow\rangle.
\end{align}
Here, $|n\rangle$ and $|Y_{l,m}\rangle$ represent the $J^z=n$
($n=\frac{5}{2},\frac{3}{2},\frac{1}{2},-\frac{1}{2},-\frac{3}{2},-\frac{5}{2}$) state for $J=5/2$
and the spherical harmonic function, respectively.
We introduce a unitary transformation $U_{\rm SO}$ whose matrix elements are defined by
$(U_{\rm SO})_{mn}=\langle Y_{3,m}\sigma|n\rangle$ ($\sigma=\uparrow,\downarrow$).
This will be used later.

\begin{figure}[t]
\begin{center}
\includegraphics[width=6.5cm,clip]{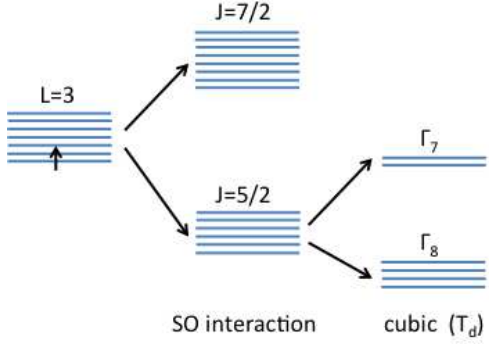}
\end{center}
\caption{
(Color online)
Schematic of energy levels of an $f^1$ state in a cubic ($T_d$) crystal-field potential.
The 14-fold degenerate $L=3$ states are split into $J=5/2$ and $J=7/2$ states by the spin-orbit (SO) interaction.
The lower $J=5/2$ states split into $\Gamma_8$ and $\Gamma_7$ states under a cubic crystal-field potential.
}
\label{fig:Td-f}
\end{figure}

In the presence of a cubic ($T_d$) crystal-field potential, the $J=5/2$ states
split into fourfold $\Gamma_8$ and twofold $\Gamma_7$ states.
When the $\Gamma_7$ is lower than $\Gamma_8$, we can restrict ourselves to the $\Gamma_7$ states.
In this case, an electric dipole moment is not induced
since the $\Gamma_7$ system can be mapped on a pseudo-$S=1/2$ spin model and it does not have quadrupole degrees of freedom.
Therefore, we consider the $\Gamma_8$ states with the lower energy in the following discussion.
The wavefunctions of the $\Gamma_8$ states are expressed as
\begin{align}
&|\Gamma_{8,1}\rangle = - \sqrt{\frac{1}{6}} \left|\frac{3}{2}\right\rangle
                        - \sqrt{\frac{5}{6}} \left|\frac{-5}{2}\right\rangle, \cr
&|\Gamma_{8,2}\rangle = \left|\frac{1}{2}\right\rangle,~~~
|\Gamma_{8,3}\rangle = - \left|\frac{-1}{2}\right\rangle, \cr
&|\Gamma_{8,4}\rangle = \sqrt{\frac{5}{6}} \left|\frac{5}{2}\right\rangle
                      + \sqrt{\frac{1}{6}} \left|\frac{-3}{2}\right\rangle.
\label{eqn:Gamma_8}
\end{align}
We introduce a unitary transformation $U_{\Gamma_8}$ whose matrix elements are defined by
$(U_{\Gamma_8})_{nm} = \langle n|\Gamma_{8,m}\rangle$.
Figure \ref{fig:Td-f} summarizes the energy levels of the $f^1$ state in a cubic crystal-field potential.
Within the $\Gamma_8$ states, the operators of the total angular momentum are expressed as
\begin{align}
&J^x =
\begin{pmatrix}
0 & \frac{-1}{\sqrt{3}} & 0 & \frac{-5}{6} \cr
\frac{-1}{\sqrt{3}} & 0 & \frac{-3}{2} & 0 \cr
0 & \frac{-3}{2} & 0 & \frac{-1}{\sqrt{3}} \cr
\frac{-5}{6} & 0 & \frac{-1}{\sqrt{3}} & 0
\end{pmatrix}, \cr
&J^y =
\begin{pmatrix}
0 & \frac{i}{\sqrt{3}} & 0 & \frac{-5i}{6} \cr
\frac{-i}{\sqrt{3}} & 0 & \frac{3i}{2} & 0 \cr
0 & \frac{-3i}{2} & 0 & \frac{i}{\sqrt{3}} \cr
\frac{5i}{6} & 0 & \frac{-i}{\sqrt{3}} & 0
\end{pmatrix}, \cr
&J^z =
\begin{pmatrix}
\frac{-11}{6} & 0 & 0 & 0 \cr
0 & \frac{1}{2} & 0 & 0 \cr
0 & 0 & \frac{-1}{2} & 0 \cr
0 & 0 & 0 & \frac{11}{6}
\end{pmatrix}.
\end{align}

\subsubsection{Quadrupole operator}

On the basis of the $\Gamma_8$ states, the quadrupole operators are expressed in the following matrix form:
\begin{align}
&O_{yz} = \frac{4}{3\sqrt{3}}
\begin{pmatrix}
0 & -i & 0 & 0 \cr
i & 0 & 0 & 0 \cr
0 & 0 & 0 & i \cr
0 & 0 & -i & 0
\end{pmatrix}, \cr
&O_{zx} = \frac{4}{3\sqrt{3}}
\begin{pmatrix}
0 & 1 & 0 & 0 \cr
1 & 0 & 0 & 0 \cr
0 & 0 & 0 & -1 \cr
0 & 0 & -1 & 0
\end{pmatrix}, \cr
&O_{xy}= \frac{4}{3\sqrt{3}}
\begin{pmatrix}
0 & 0 & -i & 0 \cr
0 & 0 & 0 & -i \cr
i & 0 & 0 & 0 \cr
0 & i & 0 & 0
\end{pmatrix}.
\label{eqn:O-fd}
\end{align}
The matrices in Eq. (\ref{eqn:O-dp}) for the $d$ orbital case appear here again
since the quadrupoles are classified in the same $\Gamma_5$ representation.

\subsubsection{$f$-$d$ hybridization induced by broken inversion symmetry}

As in the $d$-$p$ hybridization under the $T_d$ point-group symmetry,
$f$ (odd-parity) orbitals mix with even-parity ones.
To discuss the parity mixing, we introduce the following $f$ orbitals:
\begin{align}
&|f_{\Gamma_1}\rangle = \frac{-i}{\sqrt{2}} ( |Y_{3,2}\rangle - |Y_{3,-2}\rangle ), \cr
&|f_{\Gamma_{4x}}\rangle = \frac{1}{4} [ -\sqrt{5} ( |Y_{3,3}\rangle - |Y_{3,-3}\rangle )
                                        + \sqrt{3} ( |Y_{3,1}\rangle - |Y_{3,-1}\rangle ) ], \cr
&|f_{\Gamma_{4y}}\rangle = \frac{-i}{4} [ \sqrt{5} ( |Y_{3,3}\rangle + |Y_{3,-3}\rangle )
                                        + \sqrt{3} ( |Y_{3,1}\rangle + |Y_{3,-1}\rangle ) ], \cr
&|f_{\Gamma_{4z}}\rangle = |Y_{3,0}\rangle, \cr
&|f_{\Gamma_{5x}}\rangle = \frac{1}{4} [ \sqrt{3} ( |Y_{3,3}\rangle - |Y_{3,-3}\rangle )
                                       + \sqrt{5} ( |Y_{3,1}\rangle - |Y_{3,-1}\rangle ) ], \cr
&|f_{\Gamma_{5y}}\rangle = \frac{-i}{4} [ \sqrt{3} ( |Y_{3,3}\rangle + |Y_{3,-3}\rangle )
                                        - \sqrt{5} ( |Y_{3,1}\rangle + |Y_{3,-1}\rangle ) ], \cr
&|f_{\Gamma_{5z}}\rangle = \frac{1}{\sqrt{2}} ( |Y_{3,2}\rangle + |Y_{3,-2}\rangle ).
\label{eqn:f-orbital}
\end{align}
Here, the representations correspond to $\Gamma_1\rightarrow xyz$,
$\Gamma_{4x}\rightarrow x(5x^2-3r^2)$, $\Gamma_{4y}\rightarrow y(5y^2-3r^2)$, $\Gamma_{4z}\rightarrow z(5z^2-3r^2)$,
$\Gamma_{5x}\rightarrow x(y^2-z^2)$, $\Gamma_{5y}\rightarrow y(z^2-x^2)$, and $\Gamma_{5z}\rightarrow z(x^2-y^2)$.
Note that all the wavefunctions in Eq. (\ref{eqn:f-orbital}) take real values.
We introduce a unitary transformation $U_f$ whose matrix elements are defined by
$(U_f)_{mn} = \langle Y_{3,m}|f_{\Gamma_n}\rangle$.
The $xyz$-type crystal-field potential given by Eq. (\ref{eqn:V3}) leads to the following orbital mixing:
\begin{align}
&|\tilde{f}_{{\Gamma}_1}\rangle = |f_{{\Gamma}_1}\rangle + v_1 |s\rangle, \cr
&|\tilde{f}_{{\Gamma}_{4x}}\rangle = |f_{{\Gamma}_{4x}}\rangle + v_4 |d_{yz}\rangle, \cr
&|\tilde{f}_{{\Gamma}_{4y}}\rangle = |f_{{\Gamma}_{4y}}\rangle + v_4 |d_{zx}\rangle, \cr
&|\tilde{f}_{{\Gamma}_{4z}}\rangle = |f_{{\Gamma}_{4z}}\rangle + v_4 |d_{xy}\rangle.
\label{eqn:fd-mix}
\end{align}
Here, $|s\rangle$ represents an $s$ orbital.
Note that $f_{\Gamma_{5\alpha}}$ ($\alpha=x,y,z$) states do not mix with $d$ orbitals.
In Eq. (\ref{eqn:fd-mix}), $v_1$ and $v_4$ are hybridization parameters.
They are given by
\begin{align}
&v_1= \frac{V_3 t_1}{E_f-E_s},~~~~~~v_4= \frac{V_3 t_4}{E_f-E_d}, \label{eqn:v-f} \\
&t_1 = \langle s|r^3xyz|f_{\Gamma_1}\rangle,~~~
t_4 = \langle d_{yz}|r^3xyz|f_{\Gamma_{4x}}\rangle, \nonumber
\end{align}
where $t_4$ takes the same value under the cyclic $x\rightarrow y \rightarrow z \rightarrow x$ transformations.
In Eq. (\ref{eqn:v-f}), $E_s$, $E_d$, and $E_f$ are energy levels of the $s$, $d$, and $f$ orbitals, respectively.
In the following discussion, $\tilde{f}_{\Gamma_1}$ and $\tilde{f}_{\Gamma_{4\alpha}}$ ($\alpha=x,y,z$) orbitals are respectively used
instead of $f_{\Gamma_1}$ and $f_{\Gamma_{4\alpha}}$.

\subsubsection{Electric dipole operator}

To calculate the expectation value of the position of an electron,
we introduce matrices $X_\alpha$ ($\alpha=x,y,z$) whose elements are defined by $\langle n|\alpha|n'\rangle$.
Here, $n$ and $n'$ are
$\tilde{f}_{{\Gamma}_{1\uparrow}},\tilde{f}_{{\Gamma}_{4x\uparrow}},\tilde{f}_{{\Gamma}_{4y\uparrow}},\tilde{f}_{{\Gamma}_{4z\uparrow}},f_{\Gamma_{5x\uparrow}},f_{\Gamma_{5y\uparrow}},f_{\Gamma_{5z\uparrow}},\tilde{f}_{{\Gamma}_{1\downarrow}},\tilde{f}_{{\Gamma}_{4x\downarrow}},\tilde{f}_{{\Gamma}_{4y\downarrow}}$,
$\tilde{f}_{{\Gamma}_{4z\downarrow}},f_{\Gamma_{5x\downarrow}},f_{\Gamma_{5y\downarrow}}$, and $f_{\Gamma_{5z\downarrow}}$.
The matrix form is expressed as
\begin{align}
X_\alpha =
\begin{pmatrix}
r_\alpha & 0 \cr
0 & r_\alpha
\end{pmatrix}~~~(\alpha=x,y,z),
\label{eqn:X-alpha-f}
\end{align}
with
\begin{align}
&r_x =
\begin{pmatrix}
0 & b_1 & 0 & 0 \cr
b_1 & 0 & 0 & 0 \cr
0 & 0 & 0 & b_4 \cr
0 & 0 & b_4 & 0
\end{pmatrix}, \cr
&r_y =
\begin{pmatrix}
0 & 0 & b_1 & 0 \cr
0 & 0 & 0 & b_4 \cr
b_1 & 0 & 0 & 0 \cr
0 & b_4 & 0 & 0
\end{pmatrix}, \cr
&r_z =
\begin{pmatrix}
0 & 0 & 0 & b_1 \cr
0 & 0 & b_4 & 0 \cr
0 & b_4 & 0 & 0 \cr
b_1 & 0 & 0 & 0
\end{pmatrix}.
\label{eqn:r-x-f}
\end{align}
Here,
$b_1 = v_4 \langle d_{yz}|x|f_{\Gamma_1}\rangle + v_1 \langle f_{\Gamma_{4x}}|x|s\rangle$
and
$b_4 = v_4 \langle d_{xy}|x|f_{\Gamma_{4y}}\rangle + v_4 \langle f_{\Gamma_{4z}}|x|d_{zx}\rangle$.
They are real values and take the same values under the cyclic $x\rightarrow y \rightarrow z \rightarrow x$ transformations.
$X_\alpha$ is expressed in a $14\times 14$ matrix form.
Although $r_\alpha$ is a $7\times 7$ matrix, only the upper $4\times 4$ part is shown.
This is because the matrix elements are zero for the $f_{\Gamma_5}$ orbitals.
By using the unitary transformation $U = U_f^\dagger U_{\rm SO} U_{\Gamma_8}$,
the transformed matrix $\tilde{X}_\alpha=U^\dagger X_\alpha U$ is expressed in a $4\times 4$ $\Gamma_8$ basis form.
Comparing the matrix with that of the quadrupole operators, we can show that
\begin{align}
(\tilde{X}_x,\tilde{X}_y,\tilde{X}_z) = - \frac{27}{56} b_4 (O_{yz},O_{zx},O_{xy}),
\label{eqn:X-O-fd}
\end{align}
where the quadrupole operators are expressed by Eq. (\ref{eqn:O-fd}).
The symmetric spin-dependent electric dipole operators are then expressed as
\begin{align}
(p_{\rm S}^x,p_{\rm S}^y,p_{\rm S}^z) = \frac{27}{56} e b_4 (O_{yz},O_{zx},O_{xy}).
\label{eqn:ps-fd}
\end{align}
This is consistent with the result shown in Table \ref{table:P} for $T_d$.
The coefficient is determined as $K_{yz}^x=(27/56)eb_4$ within the present microscopic model.
The electric dipole operator is proportional to $b_4$, which indicates that the $f$-$d$ mixing is essential,
while the $f$-$s$ mixing ($b_1$ term) does not contribute to the dipole.

As in the $d$ orbital case, the result given by Eq. (\ref{eqn:X-O-fd}) can be understood as follows.
We consider matrix forms of $yz$, $zx$, and $xy$ on the basis of the
$\tilde{f}_{{\Gamma}_{1}},\tilde{f}_{{\Gamma}_{4x}},\tilde{f}_{{\Gamma}_{4y}},\tilde{f}_{{\Gamma}_{4z}},f_{\Gamma_{5x}},f_{\Gamma_{5y}}$, and $f_{\Gamma_{5z}}$ orbitals.
The $7\times 7$ matrices for $yz$, $zx$, and $xy$ are expressed by the $r_\alpha$ matrix given by Eq. (\ref{eqn:r-x-f}) as
\begin{align}
(r_{yz},r_{zx},r_{xy}) = (r_x,r_y,r_z)_{(b_1,b_4)\rightarrow (b_1',b_4')}
\end{align}
by replacing $(b_1,b_4)$ with $(b_1',b_4')$.
Here, $b_1'=v_1 v_4 \langle s|yz|d_{yz}\rangle$ and $b_4'=v_4^2 \langle d_{zx}|yz|d_{xy}\rangle$.
At this stage, $(X_{yz},X_{zx},X_{xy})$ is not proportional to $(X_{x},X_{y},X_{z})$
since the spin-orbit interaction and cubic symmetry are not taken into account.
Therefore, we perform the unitary transformation $U = U_f^\dagger U_{\rm SO} U_{\Gamma_8}$.
After the transformation, we obtain the following relation in a $4\times 4$ matrix form:
\begin{align}
(\tilde{X}_{yz},\tilde{X}_{zx},\tilde{X}_{xy}) &= U^\dagger (X_{yz},X_{zx},X_{xy})U \cr
&=(\tilde{X}_x,\tilde{X}_y,\tilde{X}_z)_{b_4\rightarrow b_4'},
\end{align}
where the matrices only depend on $b_4'$.
This means that $(x,y,z) \propto (yz,zx,xy)$,
which leads to $(x,y,z) \propto (O_{yz},O_{zx},O_{xy})$ through the equivalent operators, as in the $d$ orbital case.

\subsubsection{In the presence of a quadrupole field}

As in the $d$ orbital case, we focus on the $O_{xy}$-type quadrupole here.
In the quadrupole ordered phase, the local Hamiltonian can be expressed as
\begin{align}
\H = \lambda \langle O_{xy} \rangle O_{xy},
\label{eqn:H-Oxy-f}
\end{align}
where $O_{xy}=J^x J^y + J^y J^x$.
Under the quadrupole field, the fourfold degenerate $\Gamma_8$ states split
into twofold degenerate ground ($E_g=-\lambda \langle O_{xy} \rangle \frac{4}{3\sqrt{3}}$)
and excited ($E_e=\lambda \langle O_{xy} \rangle \frac{4}{3\sqrt{3}}$) states.
They are expressed as
\begin{align}
&|{\rm GS}_1\rangle = \frac{1}{\sqrt{2}} |\Gamma_{8,1}\rangle - i\frac{1}{\sqrt{2}} |\Gamma_{8,3}\rangle, \cr
&|{\rm GS}_2\rangle = i \frac{1}{\sqrt{2}} |\Gamma_{8,2}\rangle + \frac{1}{\sqrt{2}} |\Gamma_{8,4}\rangle, \cr
&|{\rm ES}_1\rangle = \frac{1}{\sqrt{2}} |\Gamma_{8,1}\rangle + i\frac{1}{\sqrt{2}} |\Gamma_{8,3}\rangle, \cr
&|{\rm ES}_2\rangle = - i \frac{1}{\sqrt{2}} |\Gamma_{8,2}\rangle + \frac{1}{\sqrt{2}} |\Gamma_{8,4}\rangle.
\label{eqn:GS-f}
\end{align}
Here, $|{\rm GS}\rangle$ and $|{\rm ES}\rangle$ denote the ground and excited states, respectively.
The expectation value of the symmetric spin-dependent electric dipole is given by
\begin{align}
\langle p_{\rm S}^z \rangle_\psi = \mp \frac{3\sqrt{3}}{14} e b_4
\label{eqn:z-av-f}
\end{align}
for the ground ($|\psi\rangle=|{\rm GS}\rangle$) and excited ($|\psi\rangle=|{\rm ES}\rangle$) states
given by Eq. (\ref{eqn:GS-f}), respectively.
Note that $\langle p_{\rm S}^x \rangle_\psi = \langle p_{\rm S}^y \rangle_\psi = 0$.
Thus, an electric dipole moment along the $z$-direction is induced
through the energy splitting of the fourfold degenerate $\Gamma_8$ state caused by the quadrupole field.
In other words, the quadrupole moment $\langle O_{xy}\rangle$ can be induced
by applying an external electric field in the $z$-direction.
This means that a quadrupole order can be controlled by an electric field in the absence of the inversion symmetry.

To understand the microscopic origin of the electric dipole moment, we analyze the electron density.
As in the $d$ orbital case in Eq. (\ref{eqn:do-product}), the electron density contains the following odd-parity component:
\begin{align}
\rho_{\rm odd}(\br) &= \pm v_4 \left\{
\frac{3\sqrt{3}}{14} [ d_{zx}(\br) f_{\Gamma_{4x}}(\br) + d_{yz}(\br) f_{\Gamma_{4y}}(\br) ] \right. \cr
&~~~
\left.+ \frac{\sqrt{5}}{14} [ d_{zx}(\br) f_{\Gamma_{5x}}(\br) - d_{yz}(\br) f_{\Gamma_{5y}}(\br) ]
\right\},
\label{eqn:rho-f}
\end{align}
where the signs $\pm$ are for the ground ($|{\rm GS}\rangle$) and excited ($|{\rm ES}\rangle$) states, respectively.
Here, the $xyz$ components are omitted.
Equation (\ref{eqn:rho-f}) leads to charge polarization in the $z$-direction.
After integrating over the coordinate, the results in Eq. (\ref{eqn:z-av-f}) are reproduced
by $v_4\int d\br z [ d_{zx}(\br) f_{\Gamma_{4x}}(\br) + d_{yz}(\br) f_{\Gamma_{4y}}(\br) ]=b_4$.
The terms with $f_{\Gamma_{5x}}(\br)$ and $f_{\Gamma_{5y}}(\br)$ cancel out after the integration.
Therefore, the electric dipole is induced by products of $t_{2g}$ and $f_{\Gamma_4}$ orbitals in the electron density,
which are caused by the parity mixing in the presence of the $xyz$-type electric octupole crystal-field potential.

\subsubsection{In the presence of a magnetic field}

\begin{figure}[t]
\begin{center}
\includegraphics[width=8cm,clip]{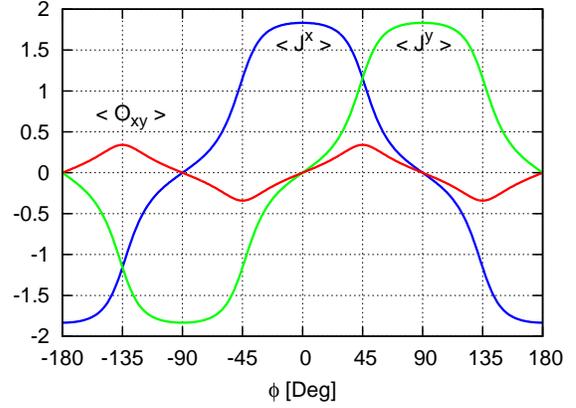}
\end{center}
\caption{
(Color online)
Magnetic dipole moment ($\langle J^x\rangle$ and $\langle J^y\rangle$) and quadrupole moment ($\langle O_{xy}\rangle$)
induced by magnetic field applied in the $xy$-plane.
The field direction is represented by $\phi$ measured from the $x$-axis.
}
\label{fig:JOz}
\end{figure}

We next discuss the $\Gamma_8$ states under a finite magnetic field.
We assume that the magnetic field is applied in the $xy$-plane and consider the following Hamiltonian:
\begin{align}
\H = - ( \cos\phi J^x + \sin\phi J^y ).
\end{align}
Here, $\phi$ represents the angle of the field measured from the $x$-axis.
Since the analytic form of the groundstate is complicated, we do not show the groundstate wavefunction here.
In Fig. \ref{fig:JOz}, we show $\phi$ dependences of the induced magnetic dipole moments
($\langle J^x\rangle$ and $\langle J^y\rangle$) and the quadrupole moment ($\langle O_{xy}\rangle$).
Note that $\langle J^z\rangle=\langle O_{yz}\rangle=\langle O_{zx}\rangle=0$ since the field is applied in the $xy$-plane.
Owing to the $|\frac{\mp 5}{2}\rangle$ components
in the $|\Gamma_{8,1}\rangle$ and $|\Gamma_{8,4}\rangle$ states, respectively,
the $\Gamma_8$ states expressed by Eq. (\ref{eqn:Gamma_8}) have no rotational invariance.
This appears as anisotropies in the angle dependences of $\langle J^x\rangle$ and $\langle J^y\rangle$.
As shown in Fig. \ref{fig:JOz}, they do not follow simple $\cos\phi$ and $\sin\phi$ functions, respectively,
differing from the $d$ orbital case represented by Eq. (\ref{eqn:J-d}) with $\theta=\pi/2$.
In the present $f$ orbital case, we can see that $\langle O_{xy}\rangle$ does not follow
a simple $\sin\phi\cos\phi \propto\sin2\phi$ function for the $d$ orbitals given by Eq. (\ref{eqn:O-av}), accordingly.
Even in the presence of such anisotropies, Eq. (\ref{eqn:ps-fd}) ensures that
$\langle p_{\rm S}^z\rangle = (27/56)eb_4 \langle O_{xy}\rangle$.

We emphasize that the results summarized in Tables \ref{table:P} and \ref{table:list} are universal
and that they do not depend on the circumstances of the microscopic models.
This is demonstrated by the microscopic models in Appendix \ref{appendix:parity-mix},
and the effectiveness of the symmetry analysis is confirmed.



\begin{thebibliography}{99}
\bibitem{note-sign}
\label{ref:note-sign}
There is another way to determine the sign of the coefficient.
For cyclic groups, such as $C_3$, $C_4$, and $C_6$,
the two-dimensional representation can be classified into two one-dimensional representations
whose basis functions are represented by $x\pm iy$.
As an example, we focus on the $C_3$ point group and explain how to determine the sign of the coefficient.
Since $z$ is a basis function in the A ($\Gamma_1$) representation,
$x \pm iy$ and $z(x \pm iy)$ are classified in the same representation.
Therefore, we can write $x+iy=a(zx+iyz)$ and $x-iy=b(zx-iyz)$ with arbitrary constants $a$ and $b$.
$x$ and $y$ are then expressed as $x=c_1 zx + c_2 yz$ and $y=-c_2 zx + c_1 yz$
with $c_1=\frac{1}{2}(a+b)$ and $c_2=i\frac{1}{2}(a-b)$.
Since ($x^2-y^2,-2xy$) transforms as ($x,y$) for the $C_3$ operation,
we obtain $x=d_1 (x^2-y^2) + d_2 (-2xy)$ and $y=-d_2 (x^2-y^2) zx + d_1 (-2xy)$ with arbitrary constants $d_1$ and $d_2$.
These lead to the following result:
\begin{align}
p_{\rm S}^x &= c_1 (S^zS^x+S^xS^z) + c_2 (S^yS^z+S^zS^y) \cr
    &~~~
    + d_1[(S^x)^2-(S^y)^2] + d_2 (S^xS^y+S^yS^x), \cr
p_{\rm S}^y &= - c_2 (S^zS^x+S^zS^x) + c_1 (S^yS^z+S^zS^y) \cr
    &~~~
     + d_2[(S^x)^2-(S^y)^2] + d_1 (-1)(S^xS^y+S^yS^x).
\nonumber
\end{align}
This is consistent with the result listed in Table \ref{table:P} for the $C_3$ point-group symmetry.
Here, we replaced $d_2$ with $-d_2$ for clarity of the comparison.
\bibitem{note-B3}
The $B_2$ and $B_3$ terms are identical since $O_{x^2+y^2,1} - O_{x^2+y^2,2} = - ( O_{z^2,1} - O_{z^2,2} )$.
This leads to $B_2 (O_{x^2+y^2,1} - O_{x^2+y^2,2}) + B_3 ( O_{z^2,1} - O_{z^2,2} ) = (B_3 - B_2) ( O_{z^2,1} - O_{z^2,2} )$.
\bibitem{note-W}
In the pressure-induced ordered phase, we assumed that
$\langle\bm{S}_-\rangle\neq 0$ and $\langle\bm{S}_+\rangle = 0$.
In this case, the mean-field Hamiltonian given by Eq. (\ref{eqn:H-dimer}) is invariant under the $IT$ transformation.
Since $\bm{W}$ has an odd character for the $IT$ transformation, its expectation value becomes zero,
i.e., $\langle\bm{W}\rangle=0$.
In other words, there is no canting spin structure in the pressure-induced ordered phase
and the vector spin chirality does not appear.
Precisely speaking, however, the left and right sides of a dimer are not equivalent in the absence of the inversion center
and we can expect that $\langle\bm{S}_\pm\rangle\neq 0$ in the ordered phase.
In this case, $\langle\bm{W}\rangle\neq 0$, which leads to $\langle p_{\rm A}\rangle\neq 0$.
We do not take this weak effect into account in Eq. (\ref{eqn:GS-pressure}).
\bibitem{note-dp}
In this discussion, the spin-orbit interaction is not taken into account in the $p$ orbitals.
When we consider the spin-orbit interaction in the $p$ orbitals,
the sixfold degenerate state splits into $J=3/2$ ($\Gamma_8$) and $J=1/2$ ($\Gamma_7$) states
under the $T_d$ point-group symmetry, as in the $t_{2g}$ orbital case.
The $\Gamma_8$ states from the $t_{2g}$ orbitals mix with the $p$ orbitals in the same $\Gamma_8$ states.
In this case, $E_d$ and $E_p$ in Eq. (\ref{eqn:b}) are written as
$E_d=E_d^{(0)} - \frac{1}{2}\lambda_d$ and $E_p=E_p^{(0)} + \frac{1}{2}\lambda_p$,
since the $\Gamma_8$ state is lower and higher for the $t_{2g}$ and $p$ orbitals, respectively.
Here, $E_d^{(0)}$ and $E_p^{(0)}$ represent the energies of the $d$ and $p$ orbitals
in the absence of the spin-orbit interaction, respectively.
$\lambda_d$ and $\lambda_p$ represent the coupling of the spin-orbit interaction for the $d$ and $p$ orbitals, respectively.
\bibitem{note-Gamma5}
In Table $\ref{table:list}$, only $(O_{yz},O_{zx},O_{xy})$ quadrupoles are shown.
Quadrupoles of $(yz,zx,xy)$ are also classified in the same $\Gamma_5$ representation.
\end{thebibliography}
\end{document}